\documentclass[11pt]{article}
\usepackage{cite}
\usepackage{amsmath}
\usepackage{amssymb}
\usepackage{accents} 
\usepackage{authblk}

\title{Minimum spacetime length and the thermodynamics of spacetime}
\author[1, 2]{Valeria Rossi}
\author[1, 2]{Sergio Cacciatori}
\author[3]{Alessandro Pesci}

\affil[1]{Dipartimento di Scienza e Alta Tecnologia,Università dell’Insubria,Via Valleggio 11, I-22100 Como, Italy}
\affil[2]{INFN, Sezione di Milano, Via Celoria 16,20133 Milano, Italy}
\affil[3]{INFN, Sezione di Bologna, Via Irnerio 46, I-40126 Bologna, Italy}

\begin{document}

	\maketitle

    \section*{Abstract}

	Theories of emergent gravity have established a deep connection between entropy and the geometry of spacetime by looking at the latter through a thermodynamic lens. In this framework, the macroscopic properties of gravity arise in a statistical way from an effective small scale discrete structure of spacetime and its information content.\\
	In this review we begin by outlining how theories of quantum gravity imply the existence of a minimum length of spacetime as a general feature. We then describe how such a structure can be implemented in a way that is independent from the details of the quantum fluctuations of spacetime via a bi-tensorial quantum metric $q_{\alpha\beta}(x, x')$ that yields a finite geodesic distance in the coincidence limit $x\rightarrow x'$. Finally, we discuss how the entropy encoded by these microscopic degrees of freedom can give rise to the field equations for gravity through a thermodynamic variational principle.

    \tableofcontents

	\section{Introduction}

	General relativity and quantum mechanics are highly successful and extensively experimentally tested in their respective domains. Trying to combine the two in a coherent framework, however, results in unphysical divergences and paradoxes that highlight an underlying clash of principles, a fundamental conflict between the way in which we describe the very small and the astronomical scale. There are two hints that may help guide the search for a theory of quantum gravity. First, quite general considerations suggest that considering gravitational effects modifies the Heisenberg uncertainty principle adding a term that increases the indeterminacy on the position linearly with that on the momentum. This would imply that, after a certain threshold, increasing the energy used to probe the position of an object would not result in an higher resolution, therefore setting a lower limit to the length that can be meaningfully measured. The second clue comes from black hole physics. Considering quantum effects in curved spacetime leads to interpreting black holes as thermodynamic systems, emitting a flux of particles according to a thermal spectrum. Further work suggests that thermodynamic properties should not be exclusive of black holes, but are rather associated to horizons and null surfaces in general.\\
	The existence of a minimum length and the thermodynamics of spacetime come together in a framework recently developed by Padmanabhan and colleagues: that of the q-metric. Rather than attempting a fundamental theory of quantum gravity, this approach focuses on the phenomenology of a minimum length scenario. It does so by constructing a semiclassical theory that still relies on the structures of a continuous geometry but replaces the metric with a bitensor $q_{\alpha\beta}(x, x')$ that has a non-vanishing zero point length $L$. Employing this q-metric allows to perform calculations in the familiar settings of a manifold with the tools of differential geometry while taking into account (at least some) effects of quantum gravity. Computing the equivalent of the Ricci scalar $R_q(x, x')$ in this framework gives a non trivial outcome: taking its coincidence limit $x\rightarrow x'$ followed by the local limit $L\rightarrow0$ does not result in the Ricci scalar of the usual metric theory, but rather in the entropy functional used in thermodynamic emergent theories of gravity. This suggests that the existence of a minimum length reflects a discrete, atomic-like structure of spacetime, with gravity being the result of the collective behaviour of microscopic degrees of freedom that evolve to maximize some entropy functional, instead of a fundamental force to be quantized.\\
	The aim of this work is to provide an introduction to the q-metric framework by outlining its motivations and consequences. Section \ref{Minimum spacetime length} briefly reviews some of the arguments for the existence of a minimum spacetime length from though experiments, entropy bounds and specific theories of quantum gravity. Section \ref{The q-metric} introduces the q-metric. Starting from bitensor theory, the section outlines the steps required to compute it for space-like, time-like, and null separations. The resulting expression is then used to explore some of its implications, including the calculation of the q-Ricci scalar. Section \ref{Thermodynamic emergent gravity} ties this last result with the landscape of theories of thermodynamic emergent gravity. The review ends with the conclusions of section \ref{Conclusions}.
	In section \ref{Minimum spacetime length} Planck units are used, but $G$ is kept explicit to emphasize the role of gravity in the emergence of the minimum length. In section \ref{Thermodynamic emergent gravity} the constants $\hbar$, $c$ and $k_B$ are reinstated as well to highlight the interplay between gravity, quantum physics, special relativity and thermodynamics.

    \section{Minimum spacetime length}\label{Minimum spacetime length}

	The elusive quest for quantum gravity has produced a variety of theories, each with a different viewpoint on the nature of spacetime, that lead to different predictions (see for example \cite{Tong2012_StringTheory} \cite{Mukhi2011_StringUpdates} on string theory, \cite{Rovelli2008_LoopQuantumGravity} \cite{Ashtekar2021_LQGUpdate} on loop quantum gravity,  \cite{Reuter2012_AsymptoticSafeGravity} on asymptotic safe gravity and \cite{Hinchliffe2004_NonCommutativePhenomenology} \cite{Chamseddine2023_NonCummutativeUpdate} on non commutative geometries as some of the main candidates). Even when working in such different frameworks, however, a common feature emerges: the existence of a minimum length scale. Indeed, the need for an extension of the Heisenberg uncertainty principle when considering both quantum and gravitational effects can be inferred on fairly general grounds. In quantum physics, objects exchange sharply defined position and momentum for conjugate probability clouds. On the other hand, the spacetime of general relativity is a dynamical background, morphed by the matter in it: the fuzziness of quantum objects translates to an uncertainty on the gravitational response to them, which in turn affects the position and momentum of the objects themselves. The resulting limitations are ultimately due to the cohabitation of the core principles of quantum mechanics, special relativity and general relativity: the Heisenberg principle swaps definite quantities for probability clouds, the finiteness of the speed of light introduces a lag in information gathering, and the equivalence principle provides an interplay between energy and spacetime geometry that amplifies quantum uncertainties. \\
	In this section, we shall outline some of the arguments that support this intuition in the form of Gedankenexperimente that show unavoidable limitations in the precision of measurements, as implied by entropy bounds and as results from specific quantum gravity frameworks. A more thorough discussion on the motivations and implications of a minimum length can be found in the well-recognised reviews by Garay \cite{Garay1995_QG} and by Hossenfelder \cite{Hossenfelder2013_MinLength}, which focus exclusively on the topic.

		\subsection{Gedankenexperimente for minimum length}
			
			\subsubsection{Heisenberg microscope with gravity}
			
			The Gedankenexperiment that introduced the uncertainty principle in physics is the Heisenberg microscope. In it, an experimental setup is considered to measure the position of a particle on the x-axis through scattering with a photon of energy $\nu$. The microscope that detects the photon after the collision can only identify its direction within the opening angle $\epsilon$, which, according to classical optics, translates into a limiting resolving power
			\begin{equation}\label{HeisenbergDeltax}
				\Delta x \gtrsim \frac{c}{2\pi \nu \sin\epsilon}.
			\end{equation} 
			The scattering used to locate the particle causes it to recoil, with a momentum transfer from the photon along the x-axis
			\begin{equation}
				\Delta p_x \gtrsim \frac{h}{c}\nu \sin\epsilon.
			\end{equation}
			Putting the two uncertainties together gives the Heisenberg principle
			\begin{equation}\label{Heisenberg principle}
				\Delta x \Delta p_x  \gtrsim \frac{h}{2\pi}.
			\end{equation}
			The nature of this equation is actually much deeper than a limitation for measurements on this specific setup: the particle does not have a well-defined position or momentum. Nevertheless, the capability of this intuitive picture to hint at the need for a profound modification of the underlying mathematical theory (the non-commutativity of the position and momentum operators) makes it a promising playground to explore the effects of gravity.
			Mead \cite{Mead1964_HeisenbergMicroscopeQG} does so by considering a sphere of radius $R$ centred on the particle as the region in which the gravitational pull of the photon is not negligible. As long as the photon travels close enough to the particle, the latter will experience an acceleration which, interpreting the photon energy as matter through $h \nu=m c^2$, can be evaluated simply with the Newtonian law as $a\approx Gh\nu/ c^2 R^2$. The particle will then gain an approximate velocity $v\approx aR/c\approx G h\nu / c^3R$ which results in a displacement $L\approx v R/c \approx G h\nu/c^4$ along an unknown direction. Projecting the latter on the x-axis gives
			\begin{equation}
				\Delta x \gtrsim \frac{ G h}{c^4}\nu\sin\epsilon
			\end{equation}
			which using relation (\ref{HeisenbergDeltax}) to get an expression for $\sin\epsilon$ reads
			\begin{equation}\label{HeisenbergMicroMinL}
				\Delta x \gtrsim \sqrt{\frac{G \hbar}{c^3}}=L_P.
			\end{equation}
			A slightly refined version of the argument takes into account momentum conservation. The photon gains momentum in the interaction with the particle, which results in an increased uncertainty
			\begin{equation}
				\Delta p_x \gtrsim \frac{h\nu}{c}\bigg(1+\frac{GM}{c^2 R}\bigg)\sin\epsilon,
			\end{equation}
			$M$ being the mass of the particle. The extra velocity translates into an additional uncertainty on the position
			\begin{equation}
				\Delta x \gtrsim \frac{R}{c} \frac{\Delta p_x}{M} \approx \frac{h\nu}{c^2}\bigg(\frac{R}{M}+\frac{G}{c^2}\bigg)\sin\epsilon\gtrsim \frac{hG}{c^4}\nu\sin\epsilon,
			\end{equation}
			from which the restriction (\ref{HeisenbergMicroMinL}) again follows.\\
			This approach suffers from all the limitations of a non-relativistic approximation dealing with a photon, but was chosen for its simplicity in arriving at a result which turns out to be quite general. A full relativistic treatment of the same setup can be found in the same reference \cite{Mead1964_HeisenbergMicroscopeQG} and arrives at a matching conclusion.

			\subsubsection{Resolution vs black holes}
			
			An alternative viewpoint for the same argument can be given \cite{Hossenfelder2013_MinLength}, noting that a particle with energy $\omega$ is associated with a Compton length $R\approx1/\omega$, which marks the limit for its localisation in quantum mechanics and general relativity because of pair production. Assuming that Thorne's hoop conjecture holds, an amount of energy $\omega$ squeezed in a region that is smaller than the Schwarzschild horizon circumference $4\pi G\omega$ in any direction will develop a black hole. Then, the Compton length has a lower limit when interpreted as the extension of a particle in the configuration
			\begin{equation}
				\frac{1}{\omega}\approx 4\pi G\omega
			\end{equation}
			which corresponds to $\omega\approx1/\sqrt{G}$. Instead of being point-like then the particle has a minimal extension
			\begin{equation}
				\Delta x \gtrsim \sqrt{G}=L_P
			\end{equation}

			\subsubsection{Limitations to the measurement of distances}

			Wigner and Salecker \cite{Salecker1958_LimitDistanceMeasurements} consider a setup for the measurement of distances based on the constancy of the speed of light. A non-relativistic clock of mass $m$ is triggered by the emission of a photon, which travels a distance $d$ at the end of which it is reflected by a mirror back to the clock, which registers the time at which the photon hits it. At time $t=0$ that marks the emission of the photon, the position of the clock is known with an uncertainty $\Delta x$, which gives it a velocity within $\Delta v = 1/ (2m\Delta x)$. As the photon returns to the clock after a time interval $2d$, the latter will have moved in an unknown direction due to its velocity, which makes the updated uncertainty on its position, and therefore on the measurement $d$, $\Delta x + (d/m\Delta x)$. The minimum value for this expression is for
			\begin{equation}
				\Delta x_{min} = \sqrt{\frac{d}{m}}
			\end{equation}
			and noting that we do not want the experiment to take place inside a black hole gives
			\begin{equation}
				\Delta x \gtrsim \sqrt{\frac{2Gm}{m}} \gtrsim \sqrt{G} \approx L_P.
			\end{equation}
			Meade \cite{Mead1964_HeisenbergMicroscopeQG} considers a similar setup to highlight a limit to clocks synchronisation as done in special relativity fashion through the exchange of a photon. The Heisenberg principle introduces an uncertainty on the time at which the photon interacts with the clock, related to the fluctuations in the energy of the particle. The latter, however, is also responsible for the gravitational field of the photon, which both accelerates the clock, introducing an additional uncertainty in its position and affects the flow of time. These effects result in an unavoidable limitation to the synchronisation of the clocks of an interval
			\begin{equation}
				\Delta t \gtrsim \sqrt{G}\approx L_P.
			\end{equation}

		\subsection{Minimum length from entropy bounds}

		The generalised covariant entropy bound sets a limit to the entropy of a gravitational system based on its geometric extension \cite{ Bousso1999_CovariantEntropy} \cite{Bousso2002_HolographicPrinciple} and can be considered a general expression of the holographic principle \cite{tHooft:1993dmi, Susskind:1994vu} in a semiclassical context. As it can be noted (see \cite{Pesci2022_InformationContent}), hidden in it is the fact that taking into account gravitational effects enforces an additional constraint to quantum indeterminacy that results in a lower extension limit that is independent from other variables.
		A 3-dimensional light sheet is constructed from the light rays that emanate orthogonally from a 2-dimensional space-like surface $B$ in non-expanding directions. The sheet terminates at a second 2-dimensional space-like surface $B'$, taken before any focal points occur. The generalised covariant entropy bound states that the entropy of a system on $L$ has to satisfy
		\begin{equation}\label{GCE bound}
			S(L)\leq \frac{A_B - A_{B'}}{4G}.
		\end{equation}
		Considering, for simplicity, some matter in the hydrodynamic regime in the configuration of a layer of thickness $l$, \cite{Pesci2007_BoussoBoundHydrodynamic} explores the limiting conditions that still conform to the bound. Choosing a homogeneous system of photons as a particularly entropic candidate while avoiding the complication of allowing for multiple species, the bound (\ref{GCE bound}) is satisfied as long as
		\begin{equation} \label{bound to l}
		l \geq \frac{1}{\pi T}.
		\end{equation}
		This condition ties strongly with other thermodynamic bounds that do not consider gravity. It is the same threshold enforced by the Hods' bound to relaxation times, which sets a universal lower limit to the time $\tau$ that a perturbed thermodynamic system takes to return to equilibrium as $1/\pi T$. The above photon layer with $c=1$ translates this requirement back to relation (\ref{bound to l}) \cite{Hod2007_RelaxationBound}.
		Limits of the same order on the extension of a homogeneous system are also given by the Bekenstein bound $S\leq2\pi E R$ for spherically symmetric systems of energy $E$ \cite{Pesci2022_InformationContent} and by the Kovtun-Son-Starinet bound to the viscosity to entropy density ratio $\eta/s \geq 1/4\pi$, when considering a layer of gluons (a better choice than photon to attempt to saturate the bound, as they are strongly coupled) \cite{Pesci2009_KSSBound}. These restrictions are ultimately protected by the Heisenberg indeterminacy principle (\ref{Heisenberg principle}), as the temperature is nothing but the average energy of the particles that make up the system. \\
		One can argue (see \cite{Pesci2022_InformationContent}) that the inclusion of gravity in the generalised covariant entropy bound (\ref{GCE bound}) gives rise to an additional restriction to (\ref{bound to l}). Consider the photon gas between $B$ and $B'$ as composed of a single layer of photons, such that $l=\lambda=1/\pi T$, $\lambda$ being the photon wavelength. The temperature $T$ is then gradually increased, and the surface $B'$ is moved to the new allowed position. This process increases the focusing of the light rays emanating from $B$ and diminishes the area $A_{B'}$: it therefore has a natural stopping point as the latter vanishes. Such a limiting case corresponds to the maximum entropy that the gas can have, all that is made available by $A_B$, which corresponds to a temperature $\overline{T}$ and a related wavelength $\overline{\lambda}$. This bound $S\leq A_B/4G$ in terms of the entropy density reads
		\begin{equation}
			s\leq \frac{1}{4G\overline{\lambda}},
		\end{equation}
		which, as for a thermal photon gas $s=(4\pi^2 /45)T^3$, implies
		\begin{equation}
			\overline{\lambda} = \frac{4\sqrt{G}}{\sqrt{45\pi}} \approx 0.3 \sqrt{G} = \mathcal{O}(L_P).
		\end{equation}
		This limit to the extension of photons, once again, crucially arises when considering gravitational effects alongside quantum indeterminacy.

		\subsection{Minimum length in quantum gravity theories}

		The existence of a limit minimum length is a general feature of many different frameworks for quantum gravity. It is, in fact, a natural way to introduce a cutoff for high energies that would help regularize divergences. This motivation also lead several physicists to consider the same hypothesis in the 1930s, with the hope that it would cure the infinities of quantum field theory, but the advent of renormalization avoided its need. The same solution is however not applicable to gravity and the possibility of a minimum length is gaining renewed attention. In this subsection we shall briefly point out in which sense this is the case for some of the leading approaches.\\
		One of the most well developed theories of quantum gravity is string theory. In it the fundamental entities are one-dimensional vibrating objects sweeping two-dimensional surfaces, known as world sheets, in $D$ dimensional spacetime. Upgrading from point-particles to strings introduces a characteristic length, the string length $l_s$, that is related to the Planck length $L_P$ through the string coupling constant $g_s$. The latter is fixed by the vacuum expectation value of the dilaton $\Phi$, one of the many scalar fields of the theory, as $g_s= e^{\langle\Phi\rangle}$. The extended nature of strings makes them unfit to probe lengths shorter than their own. The study of high energy scattering of strings at a fixed angle, as done by Amati, Ciafaloni and Veneziano in \cite{Amati1987_StringPlanckianScattering} and \cite{Amati1989_StringGUP}, suggests a modification to the Heisenberg uncertainty principle in this regime to account for the fact that, as the energy increases, more oscillatory modes are activated and the string spreads out. This generalized uncertainty principle (GUP) takes the form
		\begin{equation} \label{GUP}
			\Delta x \gtrsim \frac{1}{\Delta p} + l_s^2 \Delta p
		\end{equation}
		and has also been recovered with an approach based on path integrals and the renormalization group in \cite{Konishi1989_StringGUPRenormalizationGroup} and interpreted through T-duality, which connects small and large scales. The additional term for $\Delta x$, growing linearly in $\Delta p$, sets a lower bound to lengths of the order of the Planck scale that is independent from the energy.\\
		Loop quantum gravity (LQG) tackles quantum gravity from a radically different approach and yet it is compatible with expression (\ref{GUP}) (obviously with the proportionality factor $l_s$ suitably redefined for the new context, but still of order of the Planck length). It is a background independent, non-perturbative, formulation of gravity as a gauge theory. The quantization is done in the basis of spin networks, which are graphs constructed by the intersection of all possible closed loops. Closed graphs correspond to gravitational states.
		Given a physical surface, as can be that identified by a matter field, its area operator is diagonalized in the spin network basis and shows a discrete spectrum, which implies a minimum area \cite{Rovelli1995_LQGDiscreteArea}
		\begin{equation}
			A = 4\sqrt{3}\pi \gamma L_P^2
		\end{equation}
		where $\gamma$ is the Barbero-Immirzi constant, a dimensionless free parameter of the theory. Requiring that the black hole entropy computed as the number of microstates in LQG is equal to the Bekenstein entropy fixes $\gamma$ to be of order 1, which makes the above expression compatible with the existence of minimum length around the Planck scale. \\
		Asymptotically safe gravity attempts to make gravity renormalizable by postulating a fixed point for the renormalization group flow at the Planck scale. The theory is approached as an effective field theory with an explicit UV cutoff $k$, the couplings are renormalized and the limit $k\rightarrow\infty$ is studied. The value of the gravitational constant $G$ then becomes energy dependent, resulting in weaker gravitational attraction as one approaches the fixed point for the dimensionless coupling $\tilde{G}=k^2 G$.
		For a given value of the cutoff, the renormalized theory in $G(k)$ should describe well phenomena of energy scale $k$ and is capable of resolving lengths down to $1/k$. The possibility of taking the $k\rightarrow\infty$ limit could hint at a vanishing limit length, but the point is more subtle, as noted by Percacci and Vacca in \cite{Percacci2010_ASGMinimalLength}. 
		$k$ is a dimensionful parameter, used as a unit of mass for the other dimensionful quantities of the theory. To make physical sense of the limit $k\rightarrow\infty$, $k$ must be measured in some other units: in Planck units for example its value is $k\sqrt{G(k)}$. The latter however is equal to $\sqrt{\tilde{G}}$, which is required to be finite even in the $k\rightarrow\infty$ limit for the theory to be asymptotically safe. \\
		Finally, a quite general mathematical framework for quantum gravity is represented by non-commutative geometry. In it, the presence of a minimum length is explicit: it builds a generalization of spacetime that accounts for its expected discretization around the Planck scale by promoting coordinates to operators and imposing that they do not commute. This can be enforced via the commutator $[\hat{x^\alpha}, \hat{x^\beta}]=-i\theta^{\alpha\beta}$, where $\theta^{\alpha\beta}$ is a real valued, antisymmetric tensor to be interpreted as a minimal area.

    \section{The q-metric}  \label{The q-metric}
		
			\subsection{Bitensors}
			
			Ordinary local tensors find a non-local generalisation in n-tensors, sets of functions of n points of spacetime that under coordinate change transform in a way akin to tensors, with the difference that each subset of indices refers to a different point. 2-tensors, also known as bitensors, in particular offer an elegant way to perform non-local calculations while keeping general covariance manifest.
			DeWitt and Brehme in \cite{DeWitt1960_RadiationDamping} develop bitensor theory starting from the simplest example: the product $C^\alpha_{\alpha'}$ of two local vectors $A^\alpha$ and $B^{\alpha'}$ taken at two different points in spacetime, $x$ and $x'$ respectively
			\begin{equation} \label{bitensor example}
				C^{\alpha}_{\mu'} (x, x') = A^\alpha(x) B_{\mu'}(x').
			\end{equation}
			Following a common convention in the literature, we shall denote the indices that refer to $x$ with unprimed Greek letters and the indices that refer to $x'$ with primed ones. Under a coordinate change ${x}\rightarrow{y}$  this object transforms as
			\begin{equation}
				C^{\alpha}_{\mu'} (y, y') = \frac{\partial y^\alpha}{x^\beta} \frac{\partial x'^{\nu '}}{\partial y'^{\mu'}} C^\beta_{\nu'} (x, x').
			\end{equation}
			A bitensor can be covariantly differentiated with respect to either variable, taking care to ignore all the indices that do not refer to the considered variable. Explicitly, for our test bitensor of eq. (\ref{bitensor example}):
			\begin{equation}
				C^{\alpha}_{\mu';\beta} = C^\alpha_{\mu',\beta} + \Gamma^\alpha_{\gamma\beta} C^\gamma_{\mu'}
			\end{equation}
			\begin{equation}
				C^\alpha_{\mu';\nu'} = C^\alpha_{\mu',\nu'} - \Gamma^{\sigma'}_{\mu'\nu'} C^\alpha_{\sigma'}
			\end{equation}
			where the indices clarify where the Christoffel symbol, a local quantity, should be evaluated.\\
			The limit in which the point $x$ approaches the point $x'$ is known as the coincidence limit and is commonly denoted by the square bracket notation:
			\begin{equation}
				[C^\alpha_{\mu'}] := \lim_{x\rightarrow x'} C^\alpha_{\mu'}(x, x').
			\end{equation}
			The computation of the coincidence limit of primed derivatives can be significantly simplified using Synge's rule, which for  a generic bitensor $C^A_{B'} (x, x')$ reads 
			\begin{equation} \label{Synge's rule}
				[C_{AB'}]_{;\alpha'} = [C_{AB;\alpha'}] + [C_{AB;\alpha}]
			\end{equation}
			where $A$ is a hyperindex that encases all unprimed indices, while $B$ stands for the primed indices. 
			To prove the above statement, let us follow Poisson, Pound and Vega in \cite{Poisson2011_PointParticlesCurved} and consider two arbitrary tensors $P^M(z)$ and $Q^N(z)$ on the geodesic $\gamma$ that connects $x$ and $x'$. $M$ and $N$ are hyperindices containing as many indices as, respectively, $A$ and $B$. We can use these objects to build a biscalar defined on the geodesic
			\begin{equation}
				H(x, x') = C_{AB'}(x, x') P^A(x) Q^{B'}(x')
			\end{equation}
			and consider its expansions for $x$ and $x'$ being close to each other. Since the object we are considering is defined on a specified geodesic, we can equally as well interpret it as a function of the variable $\lambda$ that parametrises the geodesic itself and consider either $x':=z(\lambda_0)$ or $x:=z(\lambda_1)$ as the initial point for the expansion:
			\begin{equation}
				H(\lambda_1, \lambda_0) = H(\lambda_0, \lambda_0)+(\lambda_1-\lambda_0) \frac{\partial H}{\partial\lambda_1}\bigg\rvert_{\lambda_1=\lambda_0} + ...
			\end{equation}
			\begin{equation}
				H(\lambda_1, \lambda_0) = H(\lambda_1, \lambda_1)-(\lambda_1-\lambda_0) \frac{\partial H}{\partial\lambda_0}\bigg\rvert_{\lambda_0=\lambda_1} + ... .
			\end{equation}
			From these, we can compute the derivative of $H(\lambda_0, \lambda_0)$ with respect to $\lambda_0$ as
			\begin{equation}
				\frac{d}{d\lambda_0}H(\lambda_0,\lambda_0) = \lim_{\lambda_1\rightarrow\lambda_0} \frac{[H(\lambda_1,\lambda_1) - H(\lambda_0,\lambda_0)]}{(\lambda_1-\lambda_0)} = \lim_{\lambda_1\rightarrow\lambda_0} \big[ \frac{\partial H}{\partial\lambda_0}\bigg\rvert_{\lambda_0=\lambda_1} + \frac{\partial H}{\partial\lambda_1}\bigg\rvert_{\lambda_1=\lambda_0}\big].
			\end{equation}
			If we assume that the tensors $P^A$ and $Q^{B'}$ are parallel transported along the geodesic, so that
			\begin{equation}
				P^A_{;\alpha} t^\alpha = 0
			\end{equation}
			\begin{equation}
				Q^{B'}_{;\alpha'} t^{\alpha'} = 0
			\end{equation}
			where $t^\alpha$ is the vector tangent to $\gamma$, the above equation reduces to
			\begin{equation}
				[C_{AB'}]_{;\alpha'} t^{\alpha'} P^{A'} Q^{B'} = [C_{AB';\alpha'}] t^{\alpha'} P^{A'} Q^{B'} + [C_{AB';\alpha}] t^{\alpha'} P^{A'} Q^{B'}.
			\end{equation}
			Since $P^{A'}$, $Q^{B'}$ and the direction of the geodesic $\gamma$ are arbitrary, this proves Synge's rule of eq. (\ref{Synge's rule}).\\
			Before DeWitt and Brehme, Ruse \cite{Ruse1931_TaylorTheorem} and Synge \cite{Synge1931_Function} independently studied the biscalar that will later become known as Synge's world function in Riemannian spaces.
			Their interest in such a quantity stems from their effort to develop a Taylor expansion for tensorial quantities. The difficulties of the task lie in the fact that tensors defined at different points of spacetime belong to different spaces. We can try to avoid this issue by considering a bitensor instead. The most straightforward candidate would be the geodesic distance $\sigma$ between two spacetime points $x$ and $x'$, but computing its derivatives in the coincidence limit we find that $\lim_{x\rightarrow x'} \frac{\partial \sigma}{\partial x^i}$ depends on the path on which we take the limit and $\lim_{x\rightarrow x'} \frac{\partial^2 \sigma}{\partial x^{i 2}}$ diverges. 
			To build Synge's world function $\Omega$ let us take the point $x$ in the normal convex neighbourhood of $x'$, i.e. close enough so that the geodesic connecting the two points is unique. $x'$ is called the base point, and is to be thought of as fixed, while $x$ is denominated "field point" and is the one approaching $x'$ in the coincidence limit \cite{Poisson2011_PointParticlesCurved}. The geodesic $\gamma$ linking the two is parametrized through the variable $\lambda$ that runs from $\lambda_0$ (such that $z(\lambda_0) = x'$) to $\lambda_1$ (such that $z(\lambda_1) = x$). Then, on the geodesic, we define
			\begin{equation} \label{Synge' world function}
				\Omega(x, x') := \frac{1}{2} (\lambda_1-\lambda_0) \int_{\lambda_0}^{\lambda_1} g_{\mu\nu} (z) \frac{d z^\mu}{d\lambda} \frac{d z^\nu}{d\lambda} d\lambda.
			\end{equation}
			This quantity is independent of the way in which we choose to parametrise $\gamma$, since the above expression is invariant under a transformation $\lambda \rightarrow \tilde{\lambda} = a \lambda + b$, $a$ and $b$ being constants. Given a smooth metric $g_{\mu\nu}$, $\Omega(x',x)$ is an analytic function of the coordinates of the two points.
			The geodesic $\gamma$ is defined by the equation for its tangent vector $\frac{d z^\mu}{d\lambda} = t^\mu$
			\begin{equation}
				\frac{D t^\mu}{d\lambda} = t^\mu_{;\nu} t^\nu = 0,
			\end{equation}
			which means that the integrand
			\begin{equation}
				\epsilon := g_{\mu\nu}(z)\frac{d z^\mu}{d\lambda} \frac{d z^\nu}{d\lambda} = g_{\mu\nu} t^\mu t^\nu
			\end{equation}
			is actually constant in the domain of integration. Synge's world function is therefore numerically equal to half of the squared geodesic distance:
			\begin{equation}
				\Omega(x', x) = \frac{1}{2}\epsilon(\lambda_1 - \lambda_0)^2 = \frac{1}{2} \sigma^2(x', x).
			\end{equation}
			We can build additional bitensors starting from $\Omega$ and differentiating it. Since Synge's world function is a scalar both with respect to $x$ and $x'$, first derivatives are simple partial ones:
			\begin{equation}
				\Omega_\alpha := \frac{\partial\Omega}{\partial x}	\qquad	\Omega_\alpha' := \frac{\partial\Omega}{\partial x'}
			\end{equation}
			so that it's straightforward to claim
			\begin{equation}
				\Omega_{\alpha\beta'} = \frac{\partial^2 \Omega}{\partial x \partial x'} = \Omega_{\beta'\alpha }.
			\end{equation}
			To evaluate $\Omega_\alpha$ we shall consider the variation $\delta\Omega$ coming from a change of the field point $x\rightarrow x+\delta x$ as in \cite{Poisson2011_PointParticlesCurved}. $\Omega(x+\delta x, x)$ is then evaluated on the geodesic connecting $x+\delta x$ and $x'$ whose points are identified by the relation $z^\mu(\lambda) + \delta z^\mu (\lambda)$, with the affine parameter $\lambda\in [\lambda_0,\lambda_1]$. Then 
			\begin{equation}
				\delta\Omega := \Omega(x+\delta x, x') - \Omega(x,x')
			\end{equation}
			which, expanding the first term and only considering the first order approximation, gives
			\begin{equation}
				\delta\Omega = (\lambda_1-\lambda_0) \int_{\lambda_0}^{\lambda_1} g_{\mu\nu} \dot{z}^\mu \delta\dot{z}^\nu d\lambda + \frac{1}{2} (\lambda_1-\lambda_0) \int_{\lambda_0}^{\lambda_1} g_{\mu\nu,\sigma} \dot{z}^\mu \dot{z}^\nu \delta z^\sigma d\lambda.
			\end{equation}
			The first integral can be evaluated by parts as
			\begin{equation}
				\delta\Omega = (\lambda_1-\lambda_0) [g_{\mu\nu}\dot{z}^\mu \delta z^\nu]^{\lambda_1}_{\lambda_0} - (\lambda_1-\lambda_0) \int_{\lambda_0}^{\lambda_1} \big(g_{\mu\nu} \ddot{z}^\mu+ \big(g_{\mu\nu,\sigma} - \frac{1}{2}g_{\mu\sigma, \nu}\big)\dot{z}^\mu \dot{z}^\sigma\big) \delta z^\nu d\lambda.
			\end{equation}
			Writing $\dot{z}^\mu_{;\nu} \dot{z}^\nu$ in terms of the metric allows us to recognize the integrand in brackets as $g_{\sigma\mu}(\dot{z}^\mu_{;\nu} \dot{z}^\nu)$, which vanishes as $\dot{z}^\mu$ obeys the geodesic equation. To instead evaluate the first term, note that the change we are considering only affects the field point, so that $\delta z^\mu (\lambda_0) = 0$ while $\delta z^\mu (\lambda_1) = \delta x^\mu$:
			\begin{equation}
				\delta\Omega = (\lambda_1-\lambda_0) g_{\mu\nu}(x) t^\mu (x) \delta x^\nu.
			\end{equation}
			The first derivative of the Synge's world function with respect to the field point $x$ can therefore be understood as the tangent vector evaluated on the latter multiplied by the length of the geodesic itself
			\begin{equation} \label{OmegaAlpha}
				\Omega_\alpha = \frac{\delta\Omega}{\delta x^\alpha} = (\lambda_1-\lambda_0) t_\alpha.
			\end{equation}
			$\Omega_{\alpha'}$ can obviously be evaluated via the same steps, with the difference that considering a change of the base point $x'\rightarrow x'+\delta x'$ gives an extra minus sign:
			\begin{equation} \label{OmegaAlpha'}
				\Omega_{\alpha'} = -(\lambda_1-\lambda_0) t_{\alpha'}.
			\end{equation}
			Computing the norm of $\Omega_{\alpha}$
			\begin{equation}
				g^{\alpha\beta}\Omega_{\alpha}\Omega_{\beta} = (\lambda_1-\lambda_0)\epsilon = 2\Omega
			\end{equation}
			allows us to interpret $\Omega$ also as the solution to the above expression seen as a differential equation for fixed $x'$.
			$\Omega_{\alpha}$ and $\Omega_{\alpha'}$ are dual vectors as functions of their differentiated variable and scalars with respect to the other point, therefore an additional derivative with respect to the same variable needs to be a covariant one if we want to preserve the tensorial character of the quantity
			\begin{equation}
				\Omega_{\alpha \beta} := \nabla_\beta \Omega_{\alpha}.
			\end{equation}
			 Derivatives with respect to spacetime points are inherently local objects, so derivatives in unprimed indices will act independently from those in primed ones. The two groups can, in fact, be interchanged, while taking care to keep the internal order of the indices, without altering the value of the expression \cite{Synge1960_OpticalObservations}
			 \begin{equation}
			 	\Omega_{\alpha\beta...\gamma \alpha'\beta'...\gamma'} = \Omega_{ \alpha'\beta'...\gamma'\alpha\beta...\gamma}.
			 \end{equation}
			Another aspect of interest of Synge's world function, and crucial for its originally intended purpose of computing the Taylor expansion of tensorial objects, is the coincidence limit $x\rightarrow x'$ of $\Omega$ and its derivatives. Because of the assumed analyticity of the function, we expect such a limit to be independent of the path via which $x$ approaches $x'$. Then, from the definition (\ref{Synge' world function})  $[\Omega] = 0$ and from relations (\ref{OmegaAlpha}) and (\ref{OmegaAlpha'}) $[\Omega_\alpha] = 0 = [\Omega_{\alpha'}]$. Differentiating equations (\ref{OmegaAlpha}) and (\ref{OmegaAlpha'}) also tells us $[\Omega_{\alpha\beta}] = g_{\alpha'\beta'} = [\Omega_{\alpha' \beta'}]$. The coincidence limit of derivatives up to the sixth order is computed in \cite{Synge1931_Function}.\\
			Another biscalar of interest is the determinant introduced by Van Vleck in 1928 \cite{VanVleck1928_CorrespondencePrinciple} and developed by Morette \cite{Morette1951_ApproxFeynman}. It was first introduced to study the classical limit of quantum mechanics through the WKB approximation and later used in the context of point splitting techniques to compute the vacuum expectation value of the renormalised stress-energy tensor.
			For an arbitrary system with $n$ degrees of freedom, the Van Vleck determinant along a path $\gamma$ from the point $(q_i,t_i)$ to the point $(q_f,t_f)$ is defined as 
			\begin{equation}
				\Delta(q_f,t_f;q_i,t_i) := (-1)^n \det\big[\frac{\partial ^2 S_\gamma (q_f,t_f;q_i,t_i)}{\partial q_f \partial q_i}\big],
			\end{equation}
			where $S_\gamma$ is the action as computed along the mentioned path. For the specific case of a geodesic flow in a $D$-dimensional Lorentz space between the spacetime points $x'$ and $x$, the above relation specialises as 
			\begin{equation} \label{VanVleck definition}
				\Delta(x,x') := (-1)^{D-1} \frac{\det [-\Omega_{\alpha\beta'}]}{\sqrt{-g} \sqrt{-g'}}
			\end{equation}
			where $g$ and  $g'$ are the determinant of the metric as evaluated respectively in $x$ and in $x'$ \cite{Visser1993_VanVleck}. Definition (\ref{VanVleck definition}) can also be written in another way, setting up a local frame of orthonormal basis $e^\mu_a (z)$ that is parallel transported along the geodesic. Latin letters in $e^\mu_a (z)$ indicate frame indices, which are raised and lowered with the Minkowski metric $\eta_{ab}$ and are not to be confused with the Greek tensorial indices, that instead interact with the metric $g_{\alpha\beta}$. Keeping the primed notation we used earlier, we write the basis of the local frame in $x$ as $e^\alpha _a$ and that in $x'$ as $e^{\alpha'} _{a}$. The Van Vleck determinant then takes the form
			\begin{equation}\label{VanVleck definition alternative}
				\Delta(x,x') := \det[-g^{\alpha'}_\alpha \Omega^\alpha_{\beta'}(x, x')],
			\end{equation}
			where $g^{\alpha'}_\alpha := e^\alpha _a e^a _{\alpha'}$, known as the parallel propagator, is an object that takes a vector in $x'$ and parallel transports it to $x$ along the geodesic. The equivalence between expressions (\ref{VanVleck definition}) and (\ref{VanVleck definition alternative}) can be proven using the completeness relation in $x$ and in $x'$ to infer the determinants $e$ and $e'$ of the matrices made up of $e^\mu _a$ and $e^{\mu'} _{a}$:
			\begin{equation}
				g^{\alpha\beta} = \eta^{ab} e^\alpha_a e^\beta_b \qquad;\qquad g^{\alpha'\beta'} = \eta^{ab} e^{\alpha'}_a e^{\beta'}_b
			\end{equation}
			from which we get $e=1/\sqrt{-g}$ and $e'=1/\sqrt{-g'}$. The determinant of the parallel propagator $g^\alpha_{\alpha'}= \eta^{ab} g_{\alpha'\beta'} e^\alpha_a e^{\beta'}_b$ is then equal to $\sqrt{-g'}/\sqrt{-g}$, which proves the equivalence with expression (\ref{VanVleck definition}).

			\subsection{The q-metric for space-like and time-like intervals}
			
			The bitensors introduced above turn out to be particularly useful in implementing a quite general feature of quantum gravity theories: the existence of a minimal length in spacetime. As outlined in section \ref{Minimum spacetime length} of this review, there are several arguments for non-locality coming from Gedankenexperimente considering gravitational effects in the quantum regime, from the constraints imposed by entropy bounds and from a number of different theories of quantum gravity. Whichever its origin, this zero point length is to be understood as the overall effect of all possible quantum fluctuations of the background metric $g_{\alpha\beta}$ averaged over with a suitable prescription. The details of the actual form or nature of the fluctuations and how to perform the path integral to compute their expectation value are unknown, as they can only be specified by a complete theory of quantum gravity, but can be foregone for the time being in favour of a more practical approach that focuses on the phenomenology. To do so, we can consider the space generated by a modified squared geodesic distance that in the coincidence limit tends to a finite non-zero quantity $L_0$, which can possibly be of the order of the Planck length. The corresponding spacetime will not be a metric space strictly speaking, as any statement on distances is subject to the indeterminacy on the localisation of points; on the other hand, we have no clue how to describe a geometry without the notion of point or even whether the concept of spacetime still holds a physical meaning at the Planck length scale. In the spirit of practicality, what we can do is work in a semiclassical approximation, keeping the usual notion of manifold and the useful tools of differential geometry, but endowing spacetime with a metric-like bitensor $q_{\alpha\beta}(x, x')$ that gives us the desired finite geodesic distance in the coincidence limit. We can do so because the geodesic distance is able to describe a manifold's geometry equally as well as the metric tensor (that can in fact be recovered as the coincidence limit of $\Omega_{\alpha\beta}$ as seen before); this fact enables us to exchange the local degrees of freedom of $g_{\alpha\beta}(x')$ for the non local ones of $\sigma^2(x,x')$. Padmanabhan in \cite{Padmanabhan2020_GeodesicDistance} argues that the geodesic distance could actually be the more fundamental quantity in an emergent gravity framework, as it can be seen as the correlator of a pregeometric density of spacetime events, a quantity essentially akin to the density of molecules in a fluid.\\
			We modify the geodesic distance through a function $S_L(\sigma^2 (x, x'))$ that provides the desired limits: the usual squared geodesic distance $\sigma^2$ when the two points are far apart and a non-zero length $\epsilon L^2$ in the coincidence limit $x\rightarrow x'$, with $\epsilon = +1$ if the geodesic considered is space-like and $\epsilon=-1$ when it's time-like. We further require that under this modification equigeodesic surfaces keep their nature, that is, time-like (space-like) geodesics under $g_{\alpha\beta}$ are mapped into time-like (space-like) geodesics under the effective metric $q_{\alpha\beta}$ associated to $S_L(\sigma^2)$. The function $S_L$ could, in principle, be of any form that satisfies these demands. The relaxation of the requirement that the line element has to be just a quadratic function of the displacement $dx$ (while keeping the first degree homogeneity $f(x, \lambda dx)=\lambda f(x, dx)$) produces an extension of Riemmanian geometry known as Finsler geometry, in which the properties of the space depend not only on the position $x$ but also on an additional direction called element of support that can be taken as the geodesic congruence along which quantities evolve \cite{Kouretsis2013_RelativisticFinsler}. This geometry can be described by a metric-like tensor $\accentset{\ast}{g_{\alpha\beta}}$ that is related to the Riemmanian metric $g_{\alpha\beta}$ in a non trivial way by a disformal transformation, i.e. a local rescaling that depends on the direction identified by a scalar field $\phi$ defined on the manifold  \cite{Bekenstein1993_PhysicalAndGravitationalGeometry}: following Kothawala \cite{Kothawala2014_Finsleresque} we can write
			\begin{equation}\label{disformal metric _}
				\accentset{\ast}{g_{\alpha\beta}} = F(\phi) g_{\alpha\beta} -\epsilon G(\phi) t_\alpha t_\beta
			\end{equation}
			with
			\begin{equation}
				t_\alpha = \frac{\partial_\alpha \phi}{\sqrt{\epsilon g^{\mu\nu} \nabla_\mu \phi \nabla_\nu \phi}}.
			\end{equation}
			We can write the inverse metric imposing $\accentset{\ast}{g_{\alpha\gamma}} \accentset{\ast}{g^{\gamma\beta}} =\delta^\alpha_\beta$, from which we compute
			\begin{equation}\label{disformal metric ^}
				\accentset{\ast}{g^{\alpha\beta}} = F^{-1} g_{\alpha\beta} +\epsilon \big(\frac{G F^{-1}}{F-G}\big) t^\alpha t^\beta
			\end{equation}
			where the indices of $t^\alpha$ are raised with the usual metric $g_{\alpha\beta}$.
			The geometry induced on a level surface $\Sigma$ for the field $\phi$ turns out to be simply conformally related to its Riemannian counterpart, in fact, the induced metric on it $\accentset{\ast}{h_{\alpha\beta}}$ reads
			\begin{equation}
				\accentset{\ast}{h_{\alpha\beta}}=\accentset{\ast}{g_{\alpha\beta}}-\epsilon T_\alpha T_\beta=F h_{\alpha\beta}
			\end{equation}
			where
			\begin{equation}
				T_\alpha=\sqrt{F-G} t_\alpha
			\end{equation}
			is the vector tangent to $\phi$ normalized with respect to $\accentset{\ast}{g_{\alpha\beta}}$.\\
			Given these premises, we expect the metric-like bitensor $q_{\alpha\beta}$ to have a form akin to that of equation (\ref{disformal metric _}), with the role of the scalar field $\phi$ being played by the squared geodesic distance $\sigma^2$. The non-conformal term will be responsible for the zero point length in the coincidence limit and has to vanish in the large separation limit, where $q_{\alpha\beta} (x, x')\rightarrow g_{\alpha\beta}(x')$.
			We think of the bitensor $q_{\alpha\beta}$ as metric-like in the sense that it satisfies the same relations for $S_L$ that $g_{\alpha\beta}$ does for $\sigma^2$, so that following \cite{Kothawala2014_EntropySpacetime} from $g^{\alpha\beta} \Omega_\alpha \Omega_\beta = \Omega$ we can infer
			\begin{equation}\label{relation for alpha}
				q^{\alpha\beta} \partial_\alpha S_L \partial_\beta S_L = 4S_L.
			\end{equation}
			In order to satisfy this relation, it is convenient to recast (\ref{disformal metric _}) as
			\begin{equation} \label{q-metric _}
				q_{\alpha\beta} = A g_{\alpha\beta} + \epsilon \big(\frac{1}{\alpha}-A\big) t_\alpha t_\beta,
			\end{equation}
			and consequently (\ref{disformal metric ^}) as
			\begin{equation} \label{q-metric ^}
				q^{\alpha\beta} = \frac{1}{A} g^{\alpha\beta} + \epsilon \big(\alpha-\frac{1}{A}\big) t^\alpha t^\beta
			\end{equation}
			so that (\ref{relation for alpha}) can be exploited to determine the $\alpha$ term. Writing the q-metric in terms of the transversal metric $h^{\alpha\beta}= g^{\alpha\beta}-\epsilon t^\alpha t^\beta$ as $q^{\alpha\beta} = \frac{1}{A} h^{\alpha\beta}+\epsilon\alpha t^\alpha t^\beta$ in fact allows us to see the relation only depends on such a term. Plugging expression (\ref{disformal metric ^}) into (\ref{relation for alpha}) gives
			\begin{equation} \label{alpha}
				\alpha=\frac{1}{\sigma^2}\frac{S_L}{S_L^{'2}}
			\end{equation}
			where $'$ indicates a differentiation with respect to $\sigma^2$ ($S'_L:=d S_L/d(\sigma^2)$).\\
			In order to fix $A$, we can consider another aspect of the minimal length from quantum gravity theories, that is, we expect it to act as a regulating cut-off in quantum field theory. This leads \cite{Stargen2015_VanVleckDeterminant} to suggest that the q-metric equivalent of the Green function $G_q$, defined through the relation with the q-metric d'Alambertian $ \Box_q G_q(x, x') = \delta(x-x')$, should have the same form as the Green function $G$ for $g_{\alpha\beta}$, but with $S_L$ in place of $\sigma^2$:
			\begin{equation}
				G_q(\sigma^2) = G(S_L(\sigma^2)).
			\end{equation}
			Imposing this condition in D-dimensional maximally symmetric spaces, where the Green function only depends on the squared geodesic distance, allows Jaffino Stargen and Kothawala in \cite{Stargen2015_VanVleckDeterminant} to deduce
			\begin{equation} \label{A}
				A=\frac{S_L}{\sigma^2}\big(\frac{\Delta}{\tilde{\Delta}}\big)^{\frac{2}{D-1}},
			\end{equation}
			where $\Delta$ is the Van Vleck determinant introduced in the previous section and $\tilde{\Delta}(x, x')=\Delta(\tilde{x}, x')$ with $\tilde{x}$ such that $\sigma^2(\tilde{x}, x')=S_L(\sigma^2(x, x'))$.\\
			The resulting q-metric $q_{\alpha\beta}$ describes a spacetime which is essentially different from that of $g_{\alpha\beta}$, with different curvature invariants, as in general the two metrics are not related by a diffeomorphism unless the original spacetime is flat \cite{Kothawala2013_SmallScaleStructure}. The described approach moreover relies on geometrical considerations which are covariant in nature: if we consider the two points $x$ and $x'$ as being close enough that they can be described by a single local Lorentz frame, then $S_L$ won't depend on the frame chosen \cite{Pesci2022_InformationContent}.

			\subsection{The q-metric for null intervals}
			
			The generalisation of the above construction to null intervals is not straightforward: even the first relation (\ref{relation for alpha}) used in the space/time-like case here fails in giving any information about the q-metric. This is due to the fact that the geodesic distance $\sigma(x, x')$ (and therefore $S_L$) is identically zero on the light cone stemming from $x'$ and therefore of no use in localising the point $x$. 
			In order to solve this issue, one needs to identify a parameter that measures the length of a null geodesic. To do so, we can follow the construction that Visser uses in \cite{Visser1993_VanVleck} and on the null geodesic $\gamma$, affinely parametrised by $\lambda$, consider an observer with velocity $V^\alpha$ that is defined in $x'$ and then parallel transported. $V^\alpha$ is chosen such that $V^\alpha l^\beta g_{\alpha\beta}=-1$, where $l^\alpha=d x^\alpha/d\lambda$ is the vector tangent to the geodesic. The affine parameter $\lambda$ can then be interpreted as the distance along the geodesic from $x'$ (where $\lambda(x',x')=0$) as measured by the observer through
			\begin{equation}\label{lambda as length}
				l(x, x'):=\int_{0}^{\lambda} - V^\alpha l^\beta g_{\alpha\beta} d\lambda ' = \lambda(x, x').
			\end{equation}
			Following \cite{Pesci2018_SpacetimeAtomsInLorentz} and \cite{Pesci2019_QMetricForNull}, one can then impose the modification on distances as $l^2\rightarrow S_L(l^2)$ and reasonably assume that this implies a mapping of the affine parameter $\lambda$ into a quantum version $\lambda_q$ that encompasses the effect in the q-metric description. For consistency, we expect
			\begin{equation}
				\frac{\lambda_q^2}{\lambda^2} = \frac{S_L(l^2)}{l^2}
			\end{equation}
			and, similarly to the space-like and time-like case, we impose $\lambda_q (x,x')\rightarrow\lambda(x, x')$ for large separations, so that we recover the usual metric description in the regime where it is so well tested, and $\lambda_q(x, x')\rightarrow L$ in the coincidence limit to capture the effect of quantum fluctuations. Having identified the appropriate distance parameter, we can now look for the effective q-metric that can describe the desired geometry. The considerations outlined before on the connection with Finsler-like geometries still apply, so we expect the q-metric $q_{\alpha\beta}$ to be related to $g_{\alpha\beta}$ with a disformal transformation akin to (\ref{q-metric _}). Since the vector tangent to the geodesic $l^\alpha$ is null, however, we need an auxiliary null vector in order to write the transverse part of the metric. We can define it as \cite{Visser1993_VanVleck}
			\begin{equation}
				m^\alpha:=V^\alpha - \frac{1}{2} l^\alpha
			\end{equation}
			so that it's normalized as $m^\alpha l^\beta g_{\alpha\beta}=-1$ and $m^\alpha V^\beta g_{\alpha\beta}=1/2$ holds. We can then guess \cite{Pesci2022_InformationContent} 
			\begin{equation}\label{q-metric null ^}
				q^{\alpha\beta}=\frac{1}{A_\Gamma} g^{\alpha\beta}+\big(\frac{1}{A_\Gamma}-\alpha_\Gamma\big)(l^\alpha m^\beta + m^\alpha l^\beta)
			\end{equation}
			for some $A_\Gamma$ and $\alpha_\Gamma$ that depend on $\lambda(x, x')$. The advantage of this form is in separating the contributions to the transverse and longitudinal part of the metric with respect to the direction identified by the geodesic, i.e. that of the tangent vector $l^\alpha$. Separating the contribution of the transverse metric $h_{\alpha\beta}=g_{\alpha\beta}+l_\alpha m_\beta + m_\alpha l_\beta$ in fact allows us to write \cite{Pesci2018_SpacetimeAtomsInLorentz}
			\begin{equation}
				q^{\alpha\beta}= \frac{1}{A_\Gamma}h^{\alpha\beta}-\alpha_\Gamma (l^\alpha m^\beta + m^\alpha l^\beta).
			\end{equation}
			The inverse of the metric (\ref{q-metric null ^}) satisfies $q^{\alpha\gamma}q_{\gamma\beta}=\delta^\alpha_\beta$, therefore \cite{Pesci2022_InformationContent}
			\begin{equation} \label{q-metric null _}
				q_{\alpha\beta} = A_\Gamma g_{\alpha\beta} + \big(A_\Gamma -\frac{1}{\alpha_\Gamma}\big)(l_\alpha m_\beta + m_\alpha l_\beta).
			\end{equation}
			As already noted, we cannot follow the same route as before and apply (\ref{relation for alpha}) expecting it to fix $\alpha_\Gamma$, as it becomes a trivial identity in the null case. The physical sense of that equation was requiring that $S_L(\sigma^2)$ can be interpreted as a squared geodesic distance with respect to $q_{\alpha\beta}$ in the same way in which $\sigma^2$ is with respect to $g_{\alpha\beta}$. The spirit of this observation for the null case is requiring that the geodesic $\gamma$ of affine parameter $\lambda$ is mapped by $S_L(l^2)$ into a null geodesic of affine parameter $\gamma_q$, so that the geodesic equation \cite{Poisson2004_RelativistToolkit}
			\begin{equation}\label{geodesic equation}
				l^\beta \nabla_\beta l_\alpha = 0
			\end{equation}
			translates to \cite{Pesci2018_SpacetimeAtomsInLorentz}
			\begin{equation}\label{q-geodesic equation}
				l^\beta_q \nabla^q_\beta l^q_\alpha = 0,
			\end{equation}
			in the q-metric version, where the quantities are defined through the same relations as in the original metric case. That is, $l^\alpha_q$ is the vector tangent to the new geodesic parametrised by $\lambda_q$
			\begin{equation}\label{q-metric l^alpha}
				l^\alpha_q = \frac{d x^\alpha}{d \lambda_q} = l^\alpha \frac{d \lambda}{d \lambda_q},
			\end{equation}
			whose index gets lowered by $q_{\alpha\beta}$
			\begin{equation}
				l^q_\beta = q_{\alpha\beta} l^\alpha_q = \frac{d\lambda}{d \lambda_q} \frac{1}{\alpha_\Gamma}l_\beta
			\end{equation}
			and the covariant derivative with respect to the q-metric $\nabla^q_\beta$ acts on $l^q_\alpha$ as
			\begin{equation}
				\nabla^q_\beta l^q_\alpha = \partial_\beta l^q_\alpha - [\Gamma^\gamma_{\alpha\beta}]_q l^q_\gamma
			\end{equation}
			where the q-Christoffel symbol reads \cite{Kothawala2014_Finsleresque}
			\begin{equation}\label{q-metric Christoffel}
				\begin{aligned}
				\Gamma^{\gamma(q)}_{\alpha\beta} &= \frac{1}{2} q^{\gamma\delta}(\partial_\alpha q_{\delta\beta} + \partial_\beta q_{\alpha\delta}-\partial_\delta q_{\alpha\beta}) \\
				&=\frac{1}{2} q^{\gamma\delta} (-\nabla_\delta q_{\alpha\beta}+\nabla_{\alpha}q_{\beta\delta}+\nabla_{\beta}q_{\alpha\delta})+\Gamma^\gamma_{\alpha\beta},
			   \end{aligned}
			\end{equation}
			$\Gamma^\gamma_{\alpha\beta}$ being the Christoffel symbol of the usual metric $g_{\alpha\beta}$.
			Equation (\ref{q-geodesic equation}) can than be written in terms of the usual metric as
			\begin{equation}
				l^\alpha_q \nabla^q_\alpha l^q_\beta = \frac{d\lambda}{d\lambda_q} l_\beta \frac{d}{d\lambda} \bigg(\frac{d\lambda}{d\lambda_q}\frac{1}{\alpha_\Gamma}\bigg)-\bigg(\frac{d\lambda}{d\lambda_q}\bigg)^2 \bigg(\frac{1}{\alpha_\Gamma}-A_\Gamma\bigg)l^\alpha \nabla_\beta l_\alpha=0
			\end{equation}
			by substituting the above expressions and exploiting equation (\ref{geodesic equation}). The term $l^\alpha \nabla_\beta l_\alpha$ can be recast as $1/2\nabla_\beta(l^\alpha l_\alpha)$, which vanishes since $l^\alpha$ is null. What remains is an equation for $\alpha_\Gamma$ up to a multiplicative constant, which can be set considering that we want $q_{\alpha\beta}\rightarrow g_{\alpha\beta}$  as $\lambda\rightarrow\infty$. Then \cite{Pesci2019_QMetricForNull}
			\begin{equation} \label{null alpha}
				\alpha_\Gamma = \frac{d\lambda}{d\lambda_q}.
			\end{equation}
			As in the time/space-like case, the requirement that $S_L(l^2)$ maps geodesics into geodesics does not affect the $A_\Gamma$. In order to fix it one can turn to the Green function, imposing that the q-metric version can be obtained as $G(S_L)$ if $G(\sigma^2)$ is a solution for the ordinary metric. The application of this requirement is however troublesome in the null case once again as on the lightcone the Green function diverges: to obtain a relation functional to the definition of the q-metric, \cite{Pesci2019_QMetricForNull} considers the point $x''$ laying slightly off the null geodesic and then takes the limit bringing it back to $x$ on $\gamma$. In order to describe the position of this auxiliary point, we also need to consider the unique geodesic $\beta$ of affine parameter $\nu$ that connects $x$ (such that $\nu(x)=0$) to $x''$; then, in $x'$, we can write the derivative of the squared geodesic distance as
			\begin{equation}\label{sigma derivative in x''}
				\partial^\alpha \sigma^2 \vline_{x''} = 2\lambda l^\alpha\vline_{x''}+2\nu m^\alpha \vline_{x''}
			\end{equation}
			where the tangent vectors $l^\alpha$ and $m^\alpha$ are defined in $x$ and then parallel transported along $\beta$. We can exploit this expression in maximally symmetric spaces, where the Green function $G$ only depends on $\sigma^2$,  taking the limit in which $x''\rightarrow x$ (and therefore $\nu\rightarrow 0$) so that
			\begin{equation}
				\begin{aligned}
					0&= \Box G(\sigma^2)= (\nabla_\alpha \partial^\alpha \sigma^2) \frac{d G}{d\sigma^2} + (\partial^\alpha\sigma^2)(\partial_\alpha \sigma^2) \frac{d^2 G}{d(\sigma^2)^2}\\
					&= (\nabla_\alpha \partial^\alpha \sigma^2) \frac{d G}{d\sigma^2} \\
					&=(2\lambda \nabla_\alpha l^\alpha +4) \frac{dG}{d\sigma^2}
				\end{aligned}
			\end{equation}
			where in the second line we used the fact that $l^\alpha$ is a null vector to see that $(\partial^\alpha \sigma^2)(\partial_\partial\sigma^2)$ vanishes and in the last line we used (\ref{sigma derivative in x''}) \cite{Pesci2019_QMetricForNull}. In the q-metric realm, the considered geodesic remains null, so, assuming that the q-Green function keeps the same form as its $g_{\alpha\beta}$ metric counterpart with $S_L$ in place of $\sigma^2$, all the calculations above are still valid, and one can write
			\begin{equation}
				0=(2\lambda_q \nabla^q_\alpha l_q^\alpha +4) \frac{dG_q}{dS_L(\sigma^2)}.
			\end{equation}
			To satisfy this equation, the operator in brackets needs to vanish. Writing it in terms of the metric $g_{\alpha\beta}$ thanks to relations (\ref{q-metric l^alpha}) to (\ref{q-metric Christoffel}) and the found value of $\alpha_\Gamma$ (\ref{null alpha}) gives
			\begin{equation}
				-2\frac{d\lambda_q}{d\lambda}(\nabla_\alpha l^\alpha){\vline x''}+2\nabla_\alpha l^\alpha + (D-2) \frac{d}{d\lambda} \ln A_\Gamma =0.
			\end{equation}
			The first and second terms of the left-hand side can be expressed in terms of the Van Vleck determinant through \cite{Visser1993_VanVleck}
			\begin{equation} \label{cov derivative of l as van Vleck}
				\nabla_\alpha l^\alpha = \frac{D-2}{\lambda} + \frac{d}{d\lambda} \ln\Delta ^{-1}
			\end{equation}
			to obtain
			\begin{equation}
				-2\frac{d\lambda_q}{d\lambda}\frac{1}{\lambda_q} - \frac{2}{D-2}\frac{d\lambda_q}{d\lambda}\frac{d}{d\lambda_q}\ln\tilde\Delta^{-1} + \frac{2}{\lambda} + \frac{2}{D-2} \frac{d}{d\lambda}\ln\Delta^{-1} + \frac{d}{d\lambda} \ln A_\Gamma =0
			\end{equation}
			where $\tilde{\Delta}$ stands for the Van Vleck determinant taken in the points $x'$ and $\tilde{x}$ such that $\sigma^2(\tilde{x}, x')=S_L(\sigma^2(x, x'))$.
			The left-hand side can be recast as a total derivative with respect to $\lambda$
			\begin{equation}
				\frac{d}{d\lambda} \ln \bigg[\frac{\lambda^2}{\lambda_q^2} \bigg(\frac{\tilde{\Delta}}{\Delta}\bigg)^{\frac{2}{D-2}} A_\Gamma\bigg] = 0
			\end{equation}
			from which is easy to read \cite{Pesci2019_QMetricForNull}
			\begin{equation} \label{null A}
				A_\Gamma = \frac{\lambda^2_q}{\lambda^2} \bigg(\frac{\Delta}{\tilde{\Delta}}\bigg)^{\frac{2}{D-2}}
			\end{equation}
			further requiring that for large intervals we want to recover the usual metric (so $A_\Gamma\rightarrow 1$ as $\lambda\rightarrow\infty$).
			The q-metric (\ref{q-metric ^}) is then completely defined. \\
			The significant difference between the q-metric expression for the space-like and time-like case \ref{q-metric ^}  and that for the null case (\ref{q-metric null ^}) is the dependence in the latter on an observer through the auxiliary null vector $m^\alpha$. The building block that allowed the whole null-case construction was the interpretation of the affine parameter $\lambda$ as a measure of length of the geodesic through (\ref{lambda as length}), which only acquires meaning when an observer is set because otherwise an affine parameter could be arbitrarily rescaled by a multiplicative constant. With such premises, this dependence cannot be avoided. Even if the q-metric requires the definition of a specific frame, however, Lorentz invariance can be saved in the sense that any local observer will see the same minimal-length structure in terms of their space and time measurements \cite{Pesci2022_InformationContent}.

			\subsection{Lorentzian vs Euclidean metric}
			
			The effort just outlined obviously assumes that null intervals make sense in the region where the q-metric would be useful, that is, when quantum gravity effects are sensible. This seems very reasonable, since the observed metric for the world we live in is Lorentzian, and so far, there are no experimental results that clearly point towards a Euclidean metric. There are, however, a few lines of research, sparked by the papers of Hawking \cite{Hawking1993_NoBoundary} and Sakharov \cite{Sakharov1984_CosmologicalTransitions}, that consider the possibility of a signature change at very small scales or even as the beginning of the Universe, which are worth mentioning.\\
			The connection to the Euclidean realm can most immediately be made through the well-known Wick rotation, which is naively presented as an analytic continuation of time into imaginary values. Visser \cite{Visser2017_WickRotate} argues that thinking of the Wick rotation as being just about time is actually a simplification valid in flat spacetime, but that becomes ill-defined when generalised to arbitrary curvature, being coordinate dependent and not even always resulting in an Euclidean signature. This mapping can be better defined as a complex deformation of the metric that leaves the coordinate charts invariant and does not modify the topological structure of the manifold. The direct consequence is that, starting from a Lorentzian metric, not all possible Euclidean geometries are reachable, and this should be kept in consideration; first of all, when employing the Wick rotation as a mathematical tool to solve path integrals, but also in any Euclidean quantum gravity program. Another approach is that by Fernando Barbero \cite{Barbero1996_RealEuclidean}: starting from a modified action for Einstein's field equation, he finds as a solution a family of metrics whose signature is mediated by the value of two free parameters $\alpha$ and $\beta$. This not only provides a method to derive the Euclidean geometries which are compatible with a Lorentzian counterpart, but also opens the door to the possibility of endowing $\alpha$ and $\beta$ with a dynamic that would induce a change of signature. Greensite \cite{Greensite1993_DynamicalOriginLorentzian} goes a step further in applying this last idea and considers a generalized spacetime metric $g_{\mu\nu} = e^a_\mu \eta_{ab} e^b_\nu$, where $e^a_\mu$ are the usual tetrads but $\eta_{ab}$ instead of being the local frame Minkowski metric is defined as $\eta_{ab}=diag\{e^{i\theta}, 1, 1, 1\}$. The Euclidean theory can then be recovered for $\theta=0$ and the Lorentzian theory for $\theta=\pi$, but most importantly, $e^{i\theta(x)}$ can be thought of as a dynamical field with an effective potential whose expectation value fixes the signature. Considering a D-dimensional background populated by $n_B$ massless bosonic fields and $n_F$ massless fermionic fields, one can obtain a complex-valued effective $V(\theta)$ by integrating out all the fields' contributions, imposing a cut-off to deal with the non-renormalizability of gravity. At one-loop level, the minimisation of the real part of $V(\theta)$ and the stationarity of its imaginary part involves not only the value of $\theta$, but also the dimension of spacetime $D$ and the balance between the number of bosonic and fermionic fields: a solution that fixes $\theta$ ($\theta=\pm \pi$ as expected) can only be achieved if $n_F>n_B$ and $D=4$, meaning that, provided the field's asymmetry, the dynamic of the $\theta(x)$ field would not only select the Lorentzian signature, but the observed dimensionality of spacetime as well. Expanding the analysis to include massive fields as in \cite{Carlini1994_WhyLorentian} separates the point of minimisation for the real part and that of stationarity for the imaginary part, but allows an additional supersymmetric solution in $D=6$.\\
			Whichever the specific mechanism, however, the actual physical meaning, in any case, of a signature change event or of physics made in an Euclidean background is still controversial. Analogue models of gravity based on Bose-Einstein condensates provide an interesting setting to explore signature change events, which can be provoked modifying the nature of atomic interactions from repulsive to attractive through Feshbach resonance (see for example \cite{Weinfurtner2007_SignatureChangeBEC}); any experimental or simulation setting though needs to be aware of the subtleties regarding time that come with modelling the possible beginning of spacetime as we inhabit it. Before the signature change event that broke the symmetry among the coordinates and sparked the emergence of time, there could not be a time evolution, strictly speaking, only a foliation of spacetime in surfaces of constant $t$. A laboratory setting, moreover, introduces a background "real-world" time that breaks time parametrisation invariance; reintroducing it in a theoretical approach shows that a signature change event would ignite the production of an infinite number of particles with an infinite total energy \cite{White2010_SignatureChange}, so a viable theory should also take care of a regularization mechanism. Not being able to lean on a temporal structure also rules out the applicability of semiclassical approximations that implicitly rely on it, hindering any program that sees time as emerging in the larger scale limit of a timeless quantum gravity realm \cite{Chua2021_NoTime}.\\ 
			Despite the many aspects to be clarified, the possibility of interpreting euclideanization as more than just a useful mathematical trick to compute lorentzian path integrals and the concept of time as an emergent feature still provides an active ground for research. Even for the q-metric program, one could find a hint towards that direction in the work of Padmanabhan in \cite{Padmanabhan2021_WordlinePropagator}. Constructing a propagator that incorporates a minimal length, he shows that the evolution for time intervals under such a scale would no longer be unitary. If one were to extend this result from the mesoscopic scale to sub-Planckian lengths, this could point towards the emergence of time from a Euclidean effective quantum gravity metric.

			\subsection{The effects of a minimum length: minimum area, maximal acceleration, dimensional reduction at small scales and no focal points} \label{subsection: qmetric effects}
			
			The existence of a minimal length may naively lead to the conclusion that such a space should also display a minimal area and volume. This is not necessarily the case, and the point becomes more subtle when one realises that the q-metric identifies a direction along which it describes spacetime: the geodesic connecting the two points to which it refers. When talking about an area, we therefore first need to specify exactly what we are considering.
			Taking the field point $p'$ as fixed, let us consider the equigeodesic surface formed by all the points $p$ which are at a fixed geodesic distance from the former. In Euclidean space, such a surface would obviously be a sphere centred in $p'$, while in Minkowski space, as it's well known, we would distinguish between the two sheet hyperboloid of constant proper time for the time-like case, the one-sheet hyperboloid of constant proper space for the space-like case and the light cone for the null case. When talking about an infinitesimal area, we shall mean the equigeodesic surface element $d\Sigma$ around a point $p$. This quantity can be defined starting from the volume element $d V=\sqrt{g} d^D x$ in $D$ dimensional spacetime, where $g$ is the absolute value of the metric determinant and $x$ stands for the $D$ coordinates that identify a point, and constraining it on the one-dimensional equigeodesic surface to get for the non-null case $d\Sigma=\sqrt{h} d^{D-1}y$, where $h$ is the absolute value of the determinant of the induced metric $h_{\alpha\beta}$ and $y$ stands for the $D-1$ coordinates on the equigeodesic surface \cite{Poisson2004_RelativistToolkit}.
			In the q-metric space, the same relations apply, with the effective $q_{\alpha\beta}$ substituting $g_{\alpha\beta}$. According to expression (\ref{q-metric _}) the square root of the determinant of the q-metric $q$ will be related to $g$ as \cite{Kothawala2014_Finsleresque}
			\begin{equation}
				\sqrt{q} = \frac{A^{\frac{D-1}{2}}}{\sqrt{\alpha}} \sqrt{g}
			\end{equation}
			giving an effective q-volume element
			\begin{equation}\label{q-volume element}
				d V_q = \sqrt{q} d^D x = \frac{A^{\frac{D-1}{2}}}{\sqrt{\alpha}}  d V.
			\end{equation}
			Since the q-metric is related to the usual metric $g_{\alpha\beta}$ by a disformal transformation, we expect the geometry on the equigeodesic surface to be related to the usual one in a particularly simple way, through a conformal transformation of factor $A$ \cite{Kothawala2014_Finsleresque}. In fact, through $h^q_{\alpha\beta}= q_{\alpha\beta} - \epsilon T_\alpha T\beta= A h_{\alpha\beta}$, where $T_\alpha$ is the unit vector tangent to the geodesic connecting $p'$ and $p$ normalized with respect to the q-metric, one finds
			\begin{equation}
				\sqrt{h_q} = A^{\frac{D-1}{2}} \sqrt{h}
			\end{equation}
			and therefore
			\begin{equation}\label{q-area element}
				d \Sigma_q = \sqrt{h_q} d^{D-1}y = A^{\frac{D-1}{2}} d\Sigma.
			\end{equation}
			Now one can substitute the expressions for $A$ (\ref{A}) and consider the coincidence limit $p\rightarrow p'$. $S_L$ by construction takes a finite value $\pm L^2$, and since we expect that the usual area element scales like $(\sigma^2)^{\frac{D-1}{2}}$ we are left with
			\begin{equation}
				\lim_{p\rightarrow p'} d\Sigma_q=\frac{L^{D-1} }{\Delta_L} d^{D-1} y
			\end{equation}
			where $\Delta_L$ stands for $\Delta(\tilde{p}, p')$ with $\sigma^2(\tilde{p},p')=\epsilon L^2$ \cite{Padmanabhan2015_DistributionAtoms} \cite{Pesci2022_InformationContent}. Doing the same for the q-volume element with expression (\ref{alpha}) for $\alpha$ and assuming that the usual volume elements shrink as $(\sigma^2)^{\frac{D}{2}}$, we instead get an expression dominated by $\sigma^2$ that vanishes in the coincidence limit.\\
			The result is somewhat counterintuitive: a spacetime equipped with a minimum length provides a finite non-zero minimum area but no minimum volume. This fact can be understood by noting that you can recover the volume around $p$ by integrating the area along the geodesic direction, say from $q=p-dx$ to $s=p+dx$. The finite coincidence limit $\lim_{p\rightarrow p'}\sigma^2=\epsilon L^2$ prevents the equigeodesic surface from collapsing to a point, but along the geodesic direction an observer in $p'$ will see the distance between $s$ and $q$ as vanishing, because both $\sigma^2(p', s)$ and $\sigma^2(p', q)$ will tend to $\epsilon L^2$. One can convince oneself that this is indeed the case by computing explicitly the area and volume around $p$ in the simple case of 4-dimensional q-Minkowski spacetime as done in the thesis \cite{Perri2024_MinimumLength}. Setting $p'$ as the origin and choosing, e.g. time-like separations, one can introduce the coordinates $(\tau, \beta, \theta, \phi)$ as
			\begin{equation}
				\left\{
				\begin{array}{lll}
					x^0&=\tau \cosh(\beta)\\
					x^1&=\tau \sinh(\beta)\sin(\theta)\cos(\phi)\\
					x^2&=\tau \sinh(\beta)\sin(\theta)\sin(\phi)\\
					x^3&=\tau \sinh(\beta)\sin(\theta)
				\end{array}
				\right.
			\end{equation}
			with $\beta\in[0, \infty]$, $\theta\in[0, \pi]$ and $\phi\in[0, 2\pi]$, which are particularly convenient to identify points on the upper sheet of the hyperboloid of constant proper time $\tau_\Sigma$. The line element then reads
			\begin{equation}
				\begin{aligned}
					ds^2_q &= q_{\alpha\beta} dx^\alpha dx^\beta \\
					          &=-d\tilde{\tau}^2+\tilde{\tau}^2[\sinh^2 \beta(d\theta^2+\sin^2\theta d\phi^2)],
				\end{aligned}				
			\end{equation}
			where $\tilde{\tau}$ is proper time as modified through the q-metric prescription $\tilde{\tau}^2=-S_L$,
			and the line element as constrained on the equigeodesic hyperboloid $\Sigma$ of constant $\tilde{\tau}=\tilde{\tau}_\Sigma$ is
			\begin{equation}
				ds^2\vline_\Sigma = \tilde{\tau}_\Sigma^2[d\beta^2+\sinh^2\beta (d\theta^2+\sin^2\theta d\phi^2)].
			\end{equation}
			Then one can compute an area around $p$ by integrating (\ref{q-area element}) over e.g. the whole solid angle and an arbitrary finite range $[\beta_-, \beta_+]$ and obtain
			\begin{equation}
				\Sigma_q(p) = 4\pi\tilde{\tau}_\Sigma^3 \int_{\beta_-}^{\beta_+} \sinh^2(\beta) d\beta.
			\end{equation}
			As $p$ approaches $0$ along the geodesic, the transversal coordinates on the equigeodesic hypersurface remain constant, while the q-metric prescription bounds from below the value of $\tilde{\tau}$ to $L$. Doing the same for the q-volume around $p$ (\ref{q-volume element}) also requires integrating over a finite range $\tilde{\tau}_-<\tilde{\tau}<\tilde{\tau}_+$ and yields
			\begin{equation}
				V_q(p) = \pi(\tilde{\tau}^4_+ - \tilde{\tau}^4_-) \int_{\beta_-}^{\beta_+} \sinh^2(\beta) d\beta.
			\end{equation}
			In the coincidence limit the range $[\tilde{\tau}_-, \tilde{\tau}_+]$ collapses to the point $p$, at a finite proper time $L$ from $0$, and makes the volume vanish.\\
			The discussion for the area and volume around points on null equigeodesic surfaces is very similar and leads to the same conclusions in the coincidence limit, with the difference that the surface transverse to the light cone is actually $D-2$ dimensional, since a null vector is orthogonal also to itself.\\
            As mentioned before, in this framework the spacetime volume scales as $\sigma^2$ as $\sigma\rightarrow 0$. Such a feature suggests an effective dimensional reduction of spacetime at the small scales.
            One can introduce
            the following notion of
            effective dimension \cite{Padmanabhan2016_DimReduction},
			\begin{equation}
				D_{eff}=D+\frac{d}{d \ln \sigma} \bigg[\ln\bigg(\frac{V_D(\sigma, L)}{V_D(\sigma, {L=0})}\bigg)\bigg],
			\end{equation}
            where $V_D(\sigma, L)$ is the D-dimensional volume within a ordinary distance $\sigma$ if a limit length $L$ is present
            and $V_D(\sigma, L=0)$ is the same but for ordinary space,
            and where a correction is present, 
            $D - \frac{d}{d\ln\sigma} V_D(\sigma, L=0)$,
            to account for curvature effects
            (then assuring that $D_{eff} = D$ for ordinary space for finite $\sigma$).
            Well it turns out that, at least in the Euclidean case
            \cite{Padmanabhan2016_DimReduction},
            we have $D_{eff} \to 2$ when $\sigma \to 0$.\\
			The fact that a D-dimensional spacetime effectively reduces to a 2-dimensional one at the Planck scale seems to be a quite general prediction of theories of quantum gravity: a version of this effect can be recovered in the causal dynamical triangulation program, in asymptotically safe gravity, in loop quantum gravity and in high temperature strings \cite{Carlip2009_DimReduction}.\\
			The fact that the cross-sectional area of the geodesic congruence stemming from $p'$ does not collapse to a point in the coincidence limit suggests considering the possibility that a minimal length could prevent the formation of singularities and caustics in general, as has also been hinted at by specific theories of quantum gravity.
			The presence of singularities, in particular, is a problematic aspect in the description of spacetime, because it entails geodesic incompleteness. Geodesics are simply not defined past a singularity, and one would ideally expect a complete gravitational theory to clarify this disappearance. Caustics, while related to singularities, do not share this issue and simply signal the focusing of a bundle of geodesics. A measure of the change in the cross-sectional areas of a bundle of geodesics can be given by the trace of the deformation tensor $B_{\alpha\beta}=\nabla_\beta t_\alpha$, $\theta=\nabla_\alpha t^\alpha$, which is also known as the expansion scalar. A negative value of $\theta$ is associated with converging geodesics, up until the limit $\theta \rightarrow -\infty$, where caustics form. 
			The evolution of $\theta$ depends on the geometry of the spacetime considered, with shearing and the attractive nature of gravity promoting geodesic focusing and rotation opposing it, as is encapsulated by the Raychaudhuri equation, which for space-like and time-like geodesics reads
			\begin{equation}
				\frac{d\theta}{d\sigma} = -\frac{\theta^2}{D-1} - \sigma_{\alpha\beta} \sigma^{\alpha\beta}+\omega_{\alpha\beta}\omega^{\alpha\beta} - R_{\alpha\beta} t^\alpha t^\beta.
			\end{equation}
			The three terms reflect the decomposition of the deformation tensor
			\begin{equation}
				B_{\alpha\beta}= \frac{1}{D-1}\theta h_{\alpha\beta} + \sigma_{\alpha\beta} + \omega_{\alpha\beta}
			\end{equation}
			where $\sigma_{\alpha\beta}$ is the symmetric, traceless part known as the shear tensor that represents volume preserving deformations, and $\omega_{\alpha\beta}$ is the antisymmetric component that encodes rigid rotation.\\
			As $\theta$ describes how the tangent vector field changes, it can also be easily expressed in terms of the trace of the extrinsic curvature for the equigeodesic surface $K_{\alpha\beta}= -h^\mu_\alpha h^\nu_\beta \nabla_\mu t_\nu$, where $h_{\alpha\beta}$ is the induced metric on said surface, as $\theta=K=-h^{\alpha\beta} K_{\alpha\beta}$. This construction cannot be formally extended to the case of a null congruence and its equigeodesic surface, as the degeneracy induced by the light-like character of the geodesics implies that the vector normal to the surface is also tangent to it. Even though the extrinsic curvature of the latter cannot be properly defined, the expansion scalar can still be expressed via the null tangent vector $l^\alpha$ in analogy with the non-null case as $\theta=\nabla_{\alpha}l^\alpha$.\\
			Even though the expansion scalar is a local quantity, it can be expressed in a bitensorial framework leveraging the connection of the extrinsic curvature with the geodesic distance and the Van Vleck determinant $\Delta$. $K$ is a local quantity as well, defined on a point $x$ of the equigeodesic surface: considering $x'$ as a fixed point from which the geodesic $\gamma$ through $x$ originates, $\sigma(x, x')=\sqrt{\epsilon\sigma(x, x')^2}$ and $\Delta(x, x')$ while evaluated on $\gamma$ can be interpreted as functions of only the point of interest. In this sense, the extrinsic curvature of the equigeodesic surface can be written as
			\begin{equation} \label{Kbitensors}
				K(x)=\frac{D-1}{\sigma(x, x')}-\frac{d}{d\sigma}\ln \Delta (x, x'),
			\end{equation}
			where $\Delta(x, x')$ and $\sigma(x, x')$ are evaluated on the unique geodesic connecting $x'$ and $x$ and they are to be interpreted as functions of $x$ \cite{Stargen2015_VanVleckDeterminant} \cite{Pesci2019_SpacetimeAtomsExtrinsic}. This expression allows us to evaluate $K$, and therefore the expansion scalar, in the coincidence limit: as the geodesic distance approaches zero, $K$ diverges, signalling the convergence of the geodesics of the congruence, and the Raychaudhuri equation becomes ill-defined.\\
			This geometric construction can be brought over to the spacetime described by the q-metric using the relations elaborated by Kothawala in \cite{Kothawala2014_Finsleresque} that link two disformally coupled geometries. The extrinsic curvature becomes
			\begin{equation} \label{Kq}
				K_q=\sqrt{\alpha}\big[K+(D-1) \frac{d}{d\sigma} \ln \sqrt{A}\big]
			\end{equation}
			with $\alpha$ and $A$ as defined by equation (\ref{alpha}) and (\ref{A}) respectively. 
			Then the evolution of $K_q$ along the geodesic can be evaluated as
			\begin{equation}\label{Kq variation}
				\begin{aligned}
					\bigg(\frac{d K}{d\sigma}\bigg)_q &= \frac{K_q}{d\sqrt{\epsilon S_L}}  \\
					& = \frac{d\sigma}{d\sqrt{\epsilon S_L}} \frac{d}{d\sigma} \big[\sqrt{\alpha} K + (D-1) \sqrt{\alpha} \frac{d}{d\sigma} \ln \sqrt{A}\big]  \\
					& = \alpha \frac{d K}{d\sigma} + (D-1) \alpha \frac{d^2}{d\sigma^2}\ln\sqrt{A} + \frac{1}{2} \frac{d\alpha}{d\sigma} \big[K + (D-1) \frac{d}{d\sigma}\ln \sqrt{A}\big]
				\end{aligned}				
			\end{equation}
			where in the second line expression (\ref{Kq}) for the q-extrinsic curvature was used alongside a change in differentiation variable, which could be expressed through $\alpha$ as definition (\ref{alpha}) can be recast as $\alpha=\bigg(\frac{d\sqrt{\epsilon S_L}}{d\sigma}\bigg)^{-2}$. 
			These expressions can also be evaluated in terms of bitensors through (\ref{Kbitensors}), in order to be able to evaluate the coincidence limit that was problematic earlier, as done in \cite{Chakraborty2019_ZeroPointRaychaudhuri}. Substituting in (\ref{Kq}) $K$, $\alpha$ and $A$ with respectively relations (\ref{Kbitensors}), (\ref{alpha}) and (\ref{A}) $K_q$ becomes
			\begin{equation}\label{Kq_bitensor}
				\begin{aligned}
					K_q &= \frac{\sqrt{\epsilon S_L}}{\sigma S'_L}\bigg[\frac{(D-1)}{\sigma}-\frac{d\ln\Delta}{d\sigma}+(D-1)\frac{d}{d\sigma}\ln \bigg[\frac{\sqrt{\epsilon S_L}}{\sigma}\bigg(\frac{\Delta}{\tilde{\Delta}}\bigg)^{\frac{1}{D-1}}\bigg]\bigg]\\ 
					&=  \frac{(D-1)}{\sqrt{\epsilon S_L}} -\frac{d}{d\sqrt{\epsilon S_L}} \ln\tilde{\Delta}
				\end{aligned}
			\end{equation}
			with $S'_L:= d S_L/d\sigma^2$, where the second line was obtained by changing the differentiation variable from $\sigma$ to $\sigma^2$ and then to $\sqrt(\epsilon S_L)$ for the second term. Incidentally, this relation is the same as (\ref{Kbitensors}), but extended to the q-metric space using the equivalent minimum length quantities.
			$(dK/d\sigma)_q$ can then be readily obtained in terms of bitensors differentiating (\ref{Kq_bitensor}) by the modified geodesic distance $\sqrt{\epsilon S_L}$:
			\begin{equation}\label{dKq_bitensor}
					\bigg(\frac{d K}{d\sigma}\bigg)_q =  -\frac{(D-1)}{(\sqrt{\epsilon S_L})^2} - \frac{d^2 \ln\tilde{\Delta}}{d(\sqrt{\epsilon S_L})^2}.
			\end{equation}
			A general evaluation of the above quantities presents significant obstacles due to the fact that the Van Vleck determinant $\tilde{\Delta}$ is difficult to assess.  In the coincidence limit $x\rightarrow x'$, however, it has a known expansion in $\sqrt{\epsilon S_L}$ as
			\begin{equation}
				\tilde{\Delta} = 1 + \frac{\mathcal{S}(x')}{6}(\sqrt{\epsilon S_L})^2 + \frac{\dot{\mathcal{S}}(x')}{12}(\sqrt{\epsilon S_L})^3 + \mathcal{O}((\sqrt{\epsilon S_L})^4)
			\end{equation}
			with $\mathcal{S}= R_{\alpha\beta} t^\alpha t^\beta$ and $\dot{\mathcal{S}}=t^\alpha \partial_\alpha \mathcal{S}$. Then, supposing that there isn't an already formed singularity in $x'$, this quantity remains finite thanks to the minimum length prescription that was at the basis of the definition of $S_L$. The coincidence limit of $K_q$ can then be evaluated as
			\begin{equation}
				\begin{aligned}
					\lim_{x\rightarrow x'} K_q&= \lim_{x\rightarrow x'} \biggl\{\frac{(D-1)}{\sqrt{\epsilon S_L}} - \frac{d\ln\tilde{\Delta}}{d\sqrt{\epsilon S_L}} \biggr\} \\
					&= \frac{(D-1)}{L} - \frac{\mathcal{S}}{3} L - \frac{\dot{\mathcal{S}}}{4} L^2 +\mathcal{O}(L^3)
				\end{aligned}				
			\end{equation}
			and its variation $(dK/d\sigma)_q$ as
			\begin{equation}
				\begin{aligned}
					\lim_{x\rightarrow x'} \bigg(\frac{dK}{d\sigma}\bigg)_q &= \lim_{x\rightarrow x'}\biggl\{-\frac{(D-1)}{(\sqrt{\epsilon S_L})^2} -  \frac{d^2 \ln\tilde{\Delta}}{d(\sqrt{\epsilon S_L})^2}\biggr\}\\
					&= -\frac{(D-1)}{L^3} - \frac{\mathcal{S}}{3} - \frac{\dot{\mathcal{S}}}{2} L + \mathcal{O}(L^2).
				\end{aligned}
			\end{equation}
			Both of these quantities, contrary to their usual spacetime counterparts, do not diverge, signalling the possibility that a minimum spacetime length could, in fact, prevent the formation of caustics.\\
			This construction cannot be directly applied to the congruence of null geodesics because, as was briefly mentioned, the degeneracy of the space orthogonal to the equigeodesic surface prevents a meaningful definition of the extrinsic curvature. Still, one can work on the original definition for the expansion scalar $\theta = \nabla_{\alpha}l^\alpha$ as done in \cite{Pesci2018_NullRaychaudhuri}. The null tangent vector $l^\alpha$, when defined in the q-metric space, becomes
			\begin{equation}
				l^\alpha_q = \frac{d x^\alpha}{d\lambda_q} = \frac{d\lambda}{d\lambda_q} \frac{d x^\alpha}{d\lambda} = \alpha_\Gamma l^\alpha.
			\end{equation}
			Assuming that the covariant derivative maintains the same relation to the q-metric $q_{\alpha\beta}$ as it did to the usual metric $g_{\alpha\beta}$, the q-Christoffel symbol reads
			\begin{equation}
				\Gamma^{\beta (q)}_{\alpha\beta} = \frac{1}{2} q^{\beta\gamma} \nabla_\alpha q_{\beta\gamma} + \Gamma^\beta_{\alpha\beta}
			\end{equation}
			from which
			\begin{equation}\label{theta}
				\begin{aligned}
					\theta &= \nabla_{\alpha}^q l^\alpha_q = (\partial_\alpha + \frac{1}{2} q^{\beta\gamma} \nabla_\alpha q_{\beta\gamma} + \Gamma^\beta_{\alpha\beta}) \bigg(\frac{d\lambda}{d\lambda_q} l^\alpha\bigg) \\
					&= \nabla_\alpha \bigg(\frac{d\lambda}{d\lambda_q} l^\alpha\bigg) + \frac{1}{2} \frac{d\lambda}{d\lambda_q} q^{\beta\gamma} (\nabla_\alpha q_{\beta\gamma}) l^\alpha \\
					&= \frac{d}{d\lambda} \bigg(\frac{d\lambda}{d\lambda_q}\bigg) + \frac{d\lambda}{d\lambda_q} \nabla_\alpha l^\alpha + \frac{1}{2} \frac{d\lambda}{d\lambda_q} \biggl\{(D-2) \frac{d}{d\lambda} \ln A_\Gamma - 2 \frac{d}{d\lambda}\ln\alpha_\Gamma\biggr\}\\
					&= \alpha_\Gamma \nabla_\alpha l^\alpha + \frac{1}{2} (D-2) \alpha_\Gamma \frac{d}{d\lambda} \ln A_\Gamma \\
					&=  \alpha_\Gamma \bigg[ \theta + (D-2) \alpha_\Gamma \frac{d}{d\lambda} \ln \sqrt{A_\Gamma} \bigg]
				\end{aligned}
			\end{equation}
			where the third line was obtained using expressions (\ref{q-metric null ^}) and (\ref{q-metric null _}) for the null q-metric and the fourth line was the result of expression (\ref{null alpha}) for $\alpha_\Gamma$. The variation of the expansion scalar can then be computed as
			\begin{equation}
				\begin{aligned}
					\bigg( \frac{d\theta}{d\lambda}\bigg)_q &= \frac{d\theta_q}{d\lambda_q} = \alpha_\Gamma \frac{d\theta_q}{d\lambda}\\
					&= \alpha_\Gamma^2 \frac{d\theta}{d\lambda} + \frac{1}{2} (D-2) \alpha^2_\Gamma \frac{d^2 \ln A_\Gamma}{d\lambda ^2} + \frac{d\alpha_\Gamma}{d\lambda} \bigg[\alpha_\Gamma \theta + \frac{1}{2} \alpha_\Gamma (D-2) \frac{d\ln A_\Gamma}{d\lambda}\bigg]\\
					&= \alpha_\Gamma^2 \frac{d\theta}{d\lambda} + (D-2) \alpha_\Gamma^2\frac{d^2\ln\sqrt{A_\Gamma}}{d\lambda^2} + \frac{1}{2}\frac{d (\alpha_\Gamma^2)}{d\lambda}\bigg[\theta+(D-2) \frac{d\ln\sqrt{A_\Gamma}}{d\lambda}\bigg].
				\end{aligned}
			\end{equation}
			It is worth noting that these last two expressions maintain the relations (\ref{Kq}) and (\ref{Kq variation}) found in the non null case between the involved quantities: $D-2$ substitutes $D-1$ because of the degeneracy of the space orthogonal to the null equigeodesic surface, $\sqrt{\alpha_\Gamma}$ takes the place of $\alpha$ (where the square root is justified by the fact that we are working with $\lambda$ as a measure of the length of the geodesic instead of $\sigma^2$) and $A_\Gamma$ replaces $A$. \\
			Following the same path as before to get to an assessment of the above quantities in the coincidence limit, $\theta$ can be written in terms of bitensors, exploiting relation (\ref{cov derivative of l as van Vleck}) as was done in the evaluation of the $A_\Gamma$ term of the q-metric. The relation in terms of q-quantities reads
			\begin{equation}
				\theta_q = \frac{(D-2)}{\lambda_q} - \frac{d}{d\lambda_q} \ln\tilde{\Delta},
			\end{equation}
			an expression that can also be confirmed by replacing $A_\Gamma$ as (\ref{null A}) in the last line of equation (\ref{theta}). The variation on the geodesic in terms of bitensors can then be readily obtained by differentiating by $\lambda_q$
			\begin{equation}
				\bigg( \frac{d\theta}{d\lambda}\bigg)_q = \frac{d\theta}{d\lambda_q} = -\frac{(D-2)}{\lambda_q^2} - \frac{d^2\ln\tilde{\Delta}}{d\lambda_q}.
			\end{equation}
			These equalities once again reflect the same relations as in the non-null case, with the replacements $(D-1)\rightarrow(D-2)$ and $\sqrt{\epsilon S_L}\rightarrow\lambda_q$. The same expansion for the Van Vleck determinant in small spacetime intervals can then be used to evaluate the coincidence limit of these expressions as
			\begin{equation}
				\tilde{\Delta} = 1 + \frac{\mathcal{S}(x')}{6} \lambda_q^2 + \frac{\dot{\mathcal{S}}(x')}{12} \lambda_q^3 + \mathcal{O}(\lambda_q^4),
			\end{equation}
			with $\mathcal{S}=R_{\alpha\beta}l^\alpha l^\beta$, resulting in equivalent results as those found in the non-null case:
			\begin{equation}
				\lim_{x\rightarrow x'} \theta_q = \frac{(D-2)}{L} - \frac{\mathcal{S}}{3} L - \frac{\dot{\mathcal{S}}}{4} L^2 + \mathcal{O}(L^3)
			\end{equation}
			\begin{equation}
				\lim_{x\rightarrow x'} \bigg(\frac{d\theta_q}{d\lambda_q}\bigg)_q = -\frac{(D-2)}{L^2} - \frac{\mathcal{S}}{3} - \frac{\dot{\mathcal{S}}}{2} L + \mathcal{O}(L^2).
			\end{equation}

			\subsection{The Ricci scalar}
			
			Having an explicit form of the effective q-metric opens up the possibility to study quantum effects on general relativity by rewriting the Einstein-Hilbert Lagrangian, and to do so, the q-metric equivalent of the Ricci scalar must be computed. In general, the evaluation of curvature invariants in this setting is not an easy task, as the extra terms of the q-Christoffel symbol in the covariant derivative significantly complicate the calculations. 
			However, the geometries induced on equigeodesic surfaces by $g_{\alpha\beta}$ and $q_{\alpha\beta}$ are simply conformally related as $h^q_{\alpha\beta} = A h_{\alpha\beta}$, so that, when constrained on them, the Ricci (bi)scalar can be easily calculated as $R^q_\Sigma= A^{-1} R_\Sigma$. The Ricci biscalar for the whole spacetime manifold can then be reconstructed exploiting the Gauss-Codazzi equation as done by Padmanbhan and Kothawala in \cite{Kothawala2014_EntropySpacetime}:
			\begin{equation}
				R_q = R^q_\Sigma - \epsilon(K_q^2 + K^{\alpha\beta}_q K_{\alpha\beta}^q) -2\epsilon T^\alpha \nabla^q_\alpha K_q + 2\epsilon \nabla^q_\alpha (T^\beta \nabla_\beta^q T_\alpha).
			\end{equation}
			This relation can be recast in terms of quantities of the usual metric spacetime using the following identities first laid out by Kothawala in \cite{Kothawala2014_Finsleresque}
			\begin{equation}
				\begin{aligned}
					K_q^{\alpha\beta} K^q_{\alpha\beta} =& \alpha [K^{\alpha\beta}K_{\alpha\beta} + K t^\alpha\nabla_\alpha \ln A + (D-1)^2 (t^\alpha \nabla_\alpha \ln A)^2] \\
					 K_q^2 =& \alpha[K^2 + (D-1) K t^\alpha\nabla_\alpha \ln A + (D-1)^2 (t^\alpha \nabla_\alpha \ln A)^2] \\
					 T^\alpha \nabla^q_\alpha K_q =& \alpha t^\alpha \nabla_\alpha K + \frac{1}{2} (K + (D-1) t^\alpha \nabla_\alpha \ln \sqrt{A}) t^\alpha \nabla_\alpha \alpha \\
					 &+ (D-1) \alpha t^\alpha \nabla_\alpha (t^\alpha \nabla_\alpha \ln \sqrt{A}) \\
					 \nabla^q_\alpha (T^\beta \nabla_\beta^q T^\alpha) =& \frac{1}{A} \nabla_\alpha (t^\beta \nabla_\beta t^\alpha) \\
					 t^\alpha \nabla_\alpha K =& -K^{\alpha\beta}K_{\alpha\beta} - R_{\alpha\beta} t^\alpha t^\beta + \nabla_\alpha (t^\beta \nabla_\beta t^\alpha).
				\end{aligned}
			\end{equation}
			The induced Ricci scalar $R_\Sigma$ can then be written, exploiting the Gauss-Codazzi equation and the last one of the above relations as
			\begin{equation}
				R_\Sigma = R + \epsilon [-K^{\alpha\beta} K_{\alpha\beta} -2R_{\alpha\beta}t^\alpha t^\beta + K^2 + 2\nabla_\alpha(t^\beta \nabla_\beta t^\alpha)] + 2\epsilon\nabla_\alpha(t^\beta \nabla_\beta t^\alpha).
			\end{equation}
		Putting everything together results in \cite{Kothawala2014_Finsleresque}
		\begin{equation}
			\begin{aligned}
				R_q =& \frac{R}{A} - \alpha \bigg[2(D-1) \frac{1}{\sqrt{A}} \square \sqrt{A} + (D-1)(D-4) \frac{1}{A} (\nabla \sqrt{A})^2\bigg] \\
				& - \epsilon \alpha [(K+(D-1)t^\alpha \nabla_\alpha \ln\sqrt{A}) t^\alpha\nabla_\alpha \ln (\alpha A)] \\
				&+ \epsilon \bigg(\alpha - \frac{1}{A}\bigg) (2 R_{\alpha\beta}t^\alpha t^\beta + K^{\alpha\beta}K_{\alpha\beta} - K^2 -2\nabla_\alpha(t^\beta\nabla_\beta t^\alpha)).
			\end{aligned}
		\end{equation}
		Using the explicit expressions (\ref{A}) for $A$ and (\ref{alpha}) for $\alpha$ and the following identities for the Van Vleck determinant 
		\begin{equation}
			\begin{aligned}
				t^\alpha \nabla_\alpha \ln\Delta &= \frac{D-1}{\sigma} - K \\
				t^\alpha\nabla_\alpha(t^\alpha\nabla_\alpha \ln\Delta) &= -\frac{D-1}{\epsilon\sigma^2} + K^{\alpha\beta} K_{\alpha\beta} + R_{\alpha\beta} t^\alpha t^\beta
			\end{aligned}
		\end{equation}
		the above relation can be further recast as 
		\begin{equation}\label{q-Ricci biscalar}
			\begin{aligned}
				R_q = &\bigg[\frac{\sigma^2}{S_L}\bigg(\frac{\Delta}{\tilde\Delta}\bigg)^{-\frac{2}{D-1}} R_\Sigma - \frac{(D-1)(D-2)}{S_L} + 4(D-1) \frac{d\ln\tilde{\Delta}}{d S_L}\bigg] \\
				&- \frac{S_L}{\epsilon\sigma^2 S^{'2}_L} \bigg[K_{\alpha\beta}K^{\alpha\beta} - \frac{1}{D-1}K^2\bigg] \\
				&+ 4S_L \bigg[-\frac{D}{D-1} \bigg(\frac{d\ln\tilde{\Delta}}{d S_L}\bigg)^2 + 2\bigg(\frac{d^2\ln\tilde{\Delta}}{d S_L^2}\bigg)\bigg]
			\end{aligned}
		\end{equation}
		which is still an exact expression, as done in \cite{Stargen2015_VanVleckDeterminant}. Surprisingly, there is no dependence on higher derivatives of $S_L$: such contributions from the conformal and disformal terms of the q-metric cancel out.\\
		This quantity is obviously a biscalar, so in order to compare the outcome with classic general relativity, the coincidence limit $x\rightarrow x'$ of $R_q(x', x)$ should be considered. Assuming a smooth background, the involved geometric quantities can be replaced by their Taylor expansions for small geodesic distances, considering
		\begin{equation}
			K_{\alpha\beta} = \frac{1}{\sigma}\bigg[\nabla_\alpha \nabla_\beta\frac{\sigma^2}{2} - \epsilon t^\alpha t^\beta\bigg]
		\end{equation}
		and the expansion
		\begin{equation}
			\Omega_{\alpha\beta} = \nabla_\alpha \nabla_\beta\frac{\sigma^2}{2} = g_{\alpha\beta} - \frac{\epsilon\sigma^2}{3}\mathcal{S}_{\alpha\beta} + \frac{\sigma^3}{12} t^\mu \nabla_\mu \mathcal{S}_{\alpha\beta} + \mathcal{O}(\sigma^4),
		\end{equation}
		where $\mathcal{S}_{\alpha\beta}=R_{\alpha\mu\beta\nu}t^\mu t^\nu$.
		Then \cite{Stargen2015_VanVleckDeterminant}
		\begin{equation}\label{Kab,K, R espansions}
			\begin{aligned}
				K_{\alpha\beta} &= \frac{1}{\sigma} h_{\alpha\beta} - \frac{1}{3}\sigma\mathcal{S}_{\alpha\beta} + \frac{1}{12}\epsilon\sigma^2 t^\alpha\nabla_\alpha \mathcal{S}_{\alpha\beta} + \mathcal{O}(\sigma^3)\\
				K &= \frac{D-1}{\sigma}-\frac{1}{3}\sigma\mathcal{S} + \frac{1}{12}\epsilon\sigma^2 t^\alpha\nabla_\alpha \mathcal{S} + \mathcal{O}(\sigma^3) \\
				R_\Sigma &= \frac{(D-1)(D-2)}{\epsilon\sigma^2} + R - \frac{2}{3} \epsilon(D+1)\mathcal{S} + \mathcal{O}(\sigma)
			\end{aligned}
		\end{equation}
		 with $\mathcal{S}=g^{\alpha\beta}\mathcal{S}_{\alpha\beta}=R_{\alpha\beta}t^\alpha t^\beta$. These relations, alongside the previously used
		\begin{equation}\label{VVD espansions}
			\begin{aligned}
				\Delta &= 1 + \frac{1}{6} \epsilon\sigma^2 \mathcal{S} + \mathcal{O}(\sigma^3) \\
				\tilde{\Delta} &= 1 + \frac{1}{6} S_L \mathcal{S} + \mathcal{O}(S_L^{3/2})
			\end{aligned}
		\end{equation}
		allow an evaluation of the q-Ricci biscalar not only in the coincidence limit, but, once that has been taken, in the $L\rightarrow 0$ limit as well to assess any non-perturbative effect of the presence of a minimum spacetime length that could be sensible outside of the realm of quantum gravity measurements. This limit must be taken rigorously after considering $x\rightarrow x'$, as inverting the two would result in a levelling of the hypothesised mesoscopic structure that would obliterate any quantum effect; much like a strict $\hbar\rightarrow 0$ limit would make the atomic structure of solids a continuum that could no longer account for, e.g, the Ohm law.
		Writing the q-Ricci biscalar as in equation (\ref{q-Ricci biscalar}) highlights how the L-independent contributions that would survive the $L\rightarrow 0$ limit can only come from the first term: the $S_L$ factors that precede the second and third line result in contributions at least of order $L^2$. The term that makes up the second line could actually raise concerns about possible divergences, given the $\sigma$ at the denominator, but using the first two expansions of (\ref{Kab,K, R espansions}) it can be verified that it vanishes at order $L^2$. The last relation of (\ref{Kab,K, R espansions}) alongside (\ref{VVD espansions}) can be used to compute the coincidence limits
		\begin{equation}
			\begin{aligned}
				\lim_{x\rightarrow x'}& \bigg[\frac{\sigma^2}{S_L}\bigg(\frac{\Delta}{\tilde\Delta}\bigg)^{-\frac{2}{D-1}} R_\Sigma \bigg] = \frac{(D-1)(D-2)}{\epsilon L^2} \bigg(1+\frac{1}{6} \epsilon L^2 R_{\alpha\beta}(x')t^\alpha(x') t^\beta(x')\bigg)^\frac{2}{D-1} + \mathcal{O}(L) \\
				\lim_{x\rightarrow x'}& \bigg[- \frac{(D-1)(D-2)}{S_L} \bigg] = -\frac{(D-1)(D-2)}{\epsilon L^2} \\
				\lim_{x\rightarrow x'}& \bigg[4(D-1) \frac{d\ln\tilde{\Delta}}{d S_L}\bigg] = 
				4(D+1) \bigg[\frac{1}{6} \epsilon R_{\alpha\beta}(x')t^\alpha(x') t^\beta(x')\bigg] + \mathcal{O}(L).
			\end{aligned}
		\end{equation}
		To further proceed with the evaluation of the $L\rightarrow0$ limit, the de l'H\^opital's rule can be exploited to give
		\begin{equation}
			\begin{aligned}
				\lim_{L\rightarrow 0}\lim_{x\rightarrow x'}& \bigg[\frac{\sigma^2}{S_L}\bigg(\frac{\Delta}{\tilde\Delta}\bigg)^{-\frac{2}{D-1}} R_\Sigma - \frac{(D-1)(D-2)}{S_L}  \bigg] = \frac{(D-2)}{3}\epsilon [R_{\alpha\beta}(x')t^\alpha(x')t^\beta(x')] \\
				\lim_{L\rightarrow 0}\lim_{x\rightarrow x'}& \bigg[4(D-1) \frac{d\ln\tilde{\Delta}}{d S_L}\bigg] = 
				4(D+1) \bigg[\frac{1}{6} \epsilon R_{\alpha\beta}(x')t^\alpha(x') t^\beta(x')\bigg].
			\end{aligned}
		\end{equation}
		Putting these results and considerations together and recalling expression (\ref{q-Ricci biscalar}), the classical limit of the Ricci scalar in the effective q-metric framework turns out to be \cite{Stargen2015_VanVleckDeterminant}
		\begin{equation}
			\lim_{L\rightarrow 0} \lim_{x\rightarrow x'} R_q (x', x) = \epsilon D R_{\alpha\beta}(x') t^\alpha(x') t^\beta(x') = \epsilon D \mathcal{S}(x').
		\end{equation}
		While this result could numerically be equal to the usual metric Ricci scalar $R$, it encodes additional degrees of freedom through the normalised vectors $t^\alpha$, which became arbitrary once the coincidence limit was taken. \\
		The motivation for computing the Ricci scalar arose from its significance not merely as a curvature invariant, but as the quantity defining the Einstein-Hilbert Lagrangian: the semiclassical limit of $R_q$, $\epsilon D R_{\alpha\beta}(x') t^\alpha(x') t^\beta(x')$, then ought to be interpreted as the effective Einstein-Hilbert Lagrangian in the q-metric framework. What in general relativity was a variational principle for the metric naturally translates to a variational principle for the vectors $t^\alpha$, changing the interpretation of the metric from a dynamical variable to an emergent quantity. Gravity in the q-metric framework, therefore, assumes a thermodynamic interpretation as an emergent phenomenon that forces a reflection on the meaning of its quantisation. Not only because there is no reason to believe that such a procedure would result in anything meaningful, much like the quantisation of elasticity gives phonons instead of atoms or molecules \cite{Kothawala2014_EntropySpacetime}; but also because gravity is now intrinsically quantum in nature, in the sense that its manifestation relies on the atomic-like structure of spacetime. 
		These observations place the q-metric framework in a literature of theories of emergent gravity, with which the connection is indeed strong. The functional identified with $\mathcal{S}$ is in fact well know in such a landscape as the entropy density of spacetime; the idea being that null surfaces, like horizons, can be seen as one-way membranes for information to which an entropy can be assigned, and that field equations for gravity can then be recovered by extremizing such entropy. To be precise, $\mathcal{S}$ in this interpretation requires null vectors, so it would read $R_{\alpha\beta}l^\alpha l^\beta$. The connection to this quantity, the computation of $R_q$, should be carried out using the null q-metric (\ref{q-metric null _}). This is done in \cite{Pesci2020_RicciForNull}. The calculation follows the same steps as the space-like and time-like case, but with two additional complications. First of all, the light-like case requires an auxiliary null vector $m^\alpha$, as was the case for the construction of the q-metric itself. Secondly, what ensured the success of the previous approach was the fact that the $ D$-dimensional non-null congruence stemming from $x'$ does carry all the information to reconstruct the Ricci scalar in $x'$ via the Gauss-Codazzi equation. In the null case, however, the congruence of geodesics is $D-1$ dimensional, and this is in general no longer true, even for the usual metric spacetime, meaning that one can't avoid the need to compute some components of the Riemann tensor to get $R$. One can solve this issue by deforming the spacetime $\mathcal{M}$ outside of the null congruence into a spacetime $\mathcal{M}^*$ where this useful property is restored. The setup is the same built for the computation of the null q-metric: a congruence of null geodesics stemming from $x'$, affinely parametrized by $\lambda$, where a notion of distance can be given by introducing an observer with velocity $V^\alpha$ such that $V^\alpha l_\alpha = -1$ (that is defined in $x'$ and then parallel transported along the geodesic). The spacetime $\mathcal{M}^*$ can then be constructed by replacing the points arbitrarily close to the congruence with the congruence itself, parallel transported along $V^\alpha$. The geometric structure of $\mathcal{M}^*$ is inherited by that of the congruence, as parallel transport preserves it, and the covariant derivative in this new space is given by that in $\mathcal{M}$ restricted to the congruence. The Ricci scalar in $\mathcal{M}^*$ can then be written in terms of the null congruence in $\mathcal{M}$ starting from the null Gauss-Codazzi equation elaborated in \cite{Gemelli2002_nullGauss-Codazzi} as
		\begin{equation} \label{R* null}
			R^* = R_\Sigma + K \bar{K} + K^{\alpha\beta} \bar{K}_{\alpha\beta} + l^\alpha \partial_\alpha (\bar{K} - K)
		\end{equation}
		where $K_{\alpha\beta}= h^\gamma_\alpha h^\delta_\beta \nabla_\gamma l_\delta$ and $\bar{K}_{\alpha\beta}= h^\gamma_\alpha h^\delta_\beta \nabla_\gamma m_\delta$ are extrinsic curvature-like quantities with respect to $l^\alpha$ and $m^\alpha$ , $K=K^\alpha_\alpha$ and $\bar{K}=\bar{K}^\alpha_\alpha$. This construction can then be moved to the q-metric spacetime and, since both the deformation $\mathcal{M}\rightarrow\mathcal{M}^*$ and the disformal transformation $g_{\alpha\beta}\rightarrow q_{\alpha\beta}$ leave the null congruence untouched, it can be exploited for the calculation of $R_q$. Writing the q-metric equivalent of expression (\ref{R* null}) and substituting the relations for the q-metric quantities detailed in \cite{Pesci2020_RicciForNull}, allows the evaluation of $R_q$ in the coincidence limit and in the local limit in a way akin to what was done in the non-null case. The outcome is
		\begin{equation}
			\lim_{L\rightarrow 0} \lim_{x\rightarrow x'} R_q(x', x) = (D-1) R_{\alpha\beta}(x') l^\alpha(x') l^\beta(x').
		\end{equation}
		This result confirms the strong bond of the q-metric framework with the thermodynamic theories of gravity, linking it to the heat density of horizons $\mathcal{H} = -\frac{1}{L_P^2}R_{\alpha\beta}l^\alpha l^\beta$.

		\section{Thermodynamic emergent gravity} \label{Thermodynamic emergent gravity}

		\subsection{The thermodynamics of horizons}

		A first connection between gravity and thermodynamics famously arose with the 1974 paper by Bardeen, Carter, and Hawking, which began to establish an analogy between the laws governing black hole mechanics and the laws of thermodynamics for a system that includes a black hole. Indeed, black holes present a feature that directly associates them with the realm of thermodynamics, even at a classical level: the irreversibility of the path of the objects that fall into it, which introduces the concept of hidden information for an observer in the system. Taking quantum effects into account builds a more solid foundation for this connection, which shifts from being seen as a simple analogy to becoming a window into a quantum theory of gravity.\\
		The laws of black hole thermodynamics, as they came to be known, are articulated in four statements. The zeroth law states that, given a stationary black hole, the surface gravity $\kappa$ is uniform over the entire event horizon. For a black hole in asymptotically flat spacetime, stationarity implies the existence of a Killing field normal $\xi^\alpha$ to the event horizon, which makes the latter a Killing horizon as well \cite{Hawking1972_BHinGR}. This ensures that the surface gravity, defined through the relation $\nabla^\alpha(\xi^\beta \xi_\beta)=-2\kappa\xi^\alpha$, is well defined on it. Bardeen, Carter and Hawking in \cite{Bardeen1973_LawsTDmechanics} exploit the Einstein equation and the dominant energy condition to prove that $\kappa$ is constant over the horizon. Another version of the zeroth law is given by Carter in \cite{Carter1973_ZerothLawAxisymmetric}, in which he doesn't make use of the Einstein equation but assumes that the black hole under consideration is either static or stationary and axisymmetric with the "$t-\phi$ orthogonality property. In a later paper, R\'{a}cz and Wald generalise these conclusions on purely geometrical grounds \cite{Racz1996_KillingStationaryBH}.  
		The first law relates the changes in mass $M$, area $A$ and angular momentum $J$ of a stationary black hole that undergoes a quasi-static perturbation. The resulting state is a stationary black hole of mass $M+\delta M$, area $A+\delta A$ and angular momentum $J+\delta J$ where the perturbed quantities satisfy the relation
		\begin{equation}
			\delta M = \frac{\kappa}{8\pi}\delta A + \Omega_H \delta J,
		\end{equation}
		with $\Omega_H$ as the angular velocity of the horizon. This result is proved in \cite{Bardeen1973_LawsTDmechanics} by generalising the Smarr equation.
		The second law requires that under the null energy condition, the area of the event horizon of a black hole cannot decrease in time
		\begin{equation}
			\delta A \geq 0.
		\end{equation} 
		Also known as the area theorem, it is a consequence of the fact that the generators of the event horizon can never run into caustics, which implies that the expansion of the congruence of the generators must be non-negative on the horizon. It was proven by Hawking in 1971 and used in his work \cite{Hawking1971_AreaLaw} to deduce that, when two black holes coalesce, the area of the resulting event horizon is greater than the sum of the two initial ones. This implication of the theorem has been recently confirmed observationally with an analysis of GW150914 \cite{Isi2021_AreaLawGW}.
		The third law, postulated in \cite{Bardeen1973_LawsTDmechanics} and precisely formulated by Israel in \cite{Israel1986_ThirdLawTD}, states that the surface gravity of a black hole cannot be reduced to zero in a finite advanced time (or in other words a black hole cannot be made extremal), assuming that the stress energy tensor of the accreted matter is bounded and satisfies the weak energy condition.\\
		These relations bear a remarkable resemblance to the laws of thermodynamics, which prompts the interpretation of the surface gravity as a temperature and of the horizon area as an entropy. While there have been attempts to make sense of this picture in a classical framework \cite{Curiel2014_HotClassicalBH}, this identification is problematic on general relativity grounds. Temperature separates thermodynamic systems into equivalence classes based on thermal equilibrium. On the other hand, a black hole is never in equilibrium with its surroundings, with everything flowing into it, so the only temperature that could be assigned to it would be zero, if any. Consequently, the entropy assigned to this system could not be finite. Indeed, an observer outside the black hole sees matter falling inside as deformed by Lorentz contraction as it infinitely approaches the event horizon. To them, the black hole is a husk of progressively thinner and thinner layers composed of everything that ever fell in that encompasses the horizon. The resulting picture is very similar to the UV catastrophe of the black body problem: without a physically well-motivated lower bound to the height of the inner layer, there is infinite room to store information in the UV modes.\\
		This conundrum is resolved by taking into consideration quantum effects. As Hawking first proved in \cite{Hawking1975_ParticleCreationBH}, black holes radiate particles with a blackbody spectrum at the well-defined temperature
		\begin{equation}
			T=\frac{\hbar\kappa}{2\pi k_B c}.
		\end{equation}
		Hawking's original derivation uses the framework of quantum field theory in an asymptotically flat curved spacetime. He considers a free quantum field propagating from past infinity to future infinity. The lack of Poincar\'e symmetries in curved spacetime makes the concept of particle ill-defined and the mode decomposition of the field ambiguous, so that a field initially in the vacuum state has a non-zero particle content when probed at late times. These particles can be brought back to an energy flux coming out across the event horizon, or, in other words, to the evaporation of the black hole. The consequent decrease of the horizon area doesn't constitute a violation of the area theorem, as the assumption of the null energy condition no longer applies \cite{Hawking1975_ParticleCreationBH} and the second law of thermodynamics is restored when taking into consideration the entropy of the outgoing radiation. 
		These results can be obtained in a variety of frameworks. Wald \cite{Wald1975_BHParticleCreation}, using the same scenario as Hawking, computes the full density matrix of the radiation at infinity and confirms it to be identical to that of black body emission. Gibbons and Hawking \cite{Gibbons1977_EuclideanBH} consider the black hole geometry in Euclidean spacetime by performing a Wick rotation. Smoothness on the horizon then requires periodicity in the imaginary time, which corresponds to a thermal ensemble at a temperature equal to the inverse of the period. Robinson and Wilczek \cite{Robinson2005_HawkingfromAnomalies} obtain the Hawking radiation assuming that the horizon as a dynamical system captures all the relevant dynamics of the interior and requires the cancellation of gravitational anomalies tied to the ambiguous definition of time across the horizon. Parikh and Wilczek \cite{Parik2000_HawkingRadiationasTunnelling} model the Hawking radiation as the result of a quantum tunnelling process across the horizon. The need to enforce the conservation of energy brings them to consider a dynamical geometry, which introduces corrections that stir the resulting spectrum away from perfect thermality.\\
		Having established that a black hole does have a non-zero temperature significantly strengthens the possibility of going through with the thermodynamic interpretation and assigning a physical entropy to it. The idea of the entropy being proportional to the area of the horizon, rather than to a volume, is consistent with the picture of the black hole as an empty husk to the outside observer that was briefly discussed before and holds the seed of the concept of holography, when compared to the point of view of an observer that falls through. Bekenstein first proposed it in his PhD thesis and later in \cite{Bekenstein1972_BHSecondLaw} \cite{Bekenstein1973_BHentropy}, where he also determines the proportionality constant to be of order unity divided by $\hbar$. The value of the coefficient is set after the computation of the Hawking temperature as
		\begin{equation}
			S=\frac{A k_B c^3}{4G\hbar}.
		\end{equation}
		This result can be obtained in many frameworks, reflecting the variety of the derivations of the Hawking radiation. Gibbons and Hawking \cite{Gibbons1977_EuclideanBH}, in the context of Euclidean black hole geometry cited above, compute it from the partition function of the black hole as a canonical ensemble. Brown and York \cite{Brown1993_SBHMicrocanonical} repeat the calculation considering a more appropriate microcanonical ensemble. Bombelli, Koul, Lee and Sorkin \cite{Bombelli1986_BHEntanglementEntropy} and Callan and Wilczek \cite{Callan1994_BHGeometricEntropy} introduce the picture of the Bekenstein entropy as entanglement entropy, a measure of the correlations between the modes in the interior and those in the exterior of the black hole, that at first order is proportional to the surface of the boundary between the two regions. In the same vein, Srednicki \cite{Srednicki1993_EntropyAndArea} obtains an entropy proportional to the boundary area by tracing out the modes inside the horizon in the ground state matrix for a massless free field. A functionally equivalent interpretation is that of the Bekenstein entropy as that of the thermal atmosphere surrounding the black hole, modelled as a free massless gas. Both calculations require the introduction of a cutoff to avoid UV divergences. Wald \cite{Wald1993_BHEntropyAsNoetherCharge} recovers the Bekenstein entropy as a N\"{o}ther charge associated with the diffeomorphism invariance of the gravitational Lagrangian. Specific programs of quantum gravity attempt to trace back the Bekenstein entropy to an evaluation of the logarithm of the number of microstates that correspond to a given macrostate of the black hole. Strominger and Vafa \cite{Strominger1996_MicroBHEntropy} famously succeeded in doing so in string theory for extremal and near-extremal black holes in five dimensions. Ashtekar, Baez, Corichi and Krasnov \cite{Ashtekar1998_BHEntropyInLQG} and Rovelli \cite{Rovelli1996_BHEntropyInLQG} (with later revisions by Domagala and Lewandoski \cite{Domagala2004_BHEntropyLQGCorrections} and Meissner \cite{Meissner2004_BHEntropyLQGCorrections}) perform the computation in the context of Loop Quantum Gravity, getting a result that is proportional to the horizon area without the need for a UV cutoff, but crucially depends on the free Immirzi-Barbero parameter. The supergravity framework prompts Carlip \cite{Carlip1999_SBHfromCFT} \cite{Carlip2000_SBHfromCFT} \cite{Carlip2015_SBHfromCFTvsLQG} to exploit the AdS/CFT correspondence to get the asymptotic density of states from the conformal structure of certain black holes. Having computed the central charge of the associated Virasoro algebra, the density of states can be obtained via the Cardy formula.\\
		The plethora of models and frameworks that make up the landscape of black hole thermodynamics is a testimony to the uncertainties that still govern its interpretation. The quantities involved are far from being well understood in a definitive way and the forced cohabitation of general relativity and quantum field theory produces inconsistencies that hinder the construction of a complete picture (see for example \cite{Wald201_TDofBH} for a review of black hole thermodynamics and some of the open issues, or \cite{Curiel2025_InfoLoss} for an overview of the subtleties in the definition of the information loss paradox). On the other hand, paradoxes highlight the inconsistencies in the way in which we understand gravity and quantum physics, providing a window towards a theory of quantum gravity. The fact that the thermodynamic features of black holes can be recovered in a variety of frameworks (and agree even up to corrections \cite{Carlip2000_SBHCorrections}) seems to indicate that they are quite a general feature of gravity when combined with the quantum world, or at least of the way in which we model the two.
		Indeed, the laws of black hole thermodynamics have been generalized since their first formulation to include black holes in non-asymptotically flat spacetime using dynamical trapping horizons \cite{Hayward1994_NonFlatBHTD} and avoiding the need for a Killing field in the neighbourhood of the horizon with isolated horizons \cite{Ashtekar2002_IsolatedHorizons} \cite{Ashtekar2000_IsolatedHorizonsFirstLaws} \cite{Ashtekar2001_IsolatedHorizonRotating} \cite{Ashtekar2002_DynamicalHorizons}. They also have been shown to not be limited to black holes and to also hold for cosmological de Sitter horizons \cite{Gibbons1977_CosmologicalHorizons} and Rindler horizons \cite{Unruh1976_UnruhEffect}, where the thermality can be recovered directly from the structure of the inertial propagator \cite{Padmanabhan2019_RindlerTDfromPropagator}.

		\subsection{A brief history of the thermodynamic nature of gravity}

		While not clarified in many of its aspects (and far from any experimental feedback), the thermodynamics of causal horizons has gained broad acceptance. The discovery of Hawking radiation and all the consequent derivations relying on the consideration of quantum effects have contributed substantially to giving weight to horizon thermodynamics beyond the status of formal analogy. Still, the laws of black hole mechanics were formulated in a classical framework \cite{Bardeen1973_LawsTDmechanics} that couldn't account for a temperature or an entropy, but predicted the correct relations between the geometric quantities in their thermodynamic interpretation. It is this surprising capability of general relativity that opens the door to seeing the dynamics of the metric as being of a thermodynamic nature. The key is the identification between the entropy and the horizon area that, through the laws of thermodynamics, shapes spacetime and translates into the geometric constraints of general relativity.\\
		The gateway paper that introduces the idea of gravity as an emergent force is by Sakharov \cite{Sakharov1991_InducedGravity}. He considers a spacetime manifold with a quantum field theory on it and argues that the dynamics of the geometry is induced by field fluctuations, as the terms of the one-loop effective action bear the same relation to the curvature as the Lagrangian of general relativity. The proposal of gravity as something different from a fundamental force has been developed in many directions, some even regarding spacetime and the metric itself as emergent \cite{Visser2002_SakharovModern} \cite{Carlip2012_ChallengesEmergentGravity}.
		Jacobson \cite{Jacobson1995_EinsteinEqofState}, inspired by the laws of black hole mechanics, sees gravity specifically through a thermodynamic lens. Given an accelerated observer, a local Rindler horizon can be seen as the boundary of a system at the Unruh temperature in thermodynamic equilibrium with the environment. The passage of matter across the horizon can then be interpreted as heat transferring energy from the degrees of freedom outside the horizon to those of the system
		\begin{equation}
			\delta Q = \int_{\mathcal{H}} T_{\alpha\beta} (-\kappa\lambda l^\alpha) l^\beta d\lambda dA
		\end{equation}
		where $\kappa$ is the acceleration of the observer, $l^\alpha=dx^\alpha/d\lambda$ is the tangent vector to the horizon generators for affine parameter $\lambda$, $T_{\alpha\beta}$ is the energy-momentum tensor, and the integral is taken over the horizon $\mathcal{H}$.
		This process is governed by the Clausius relation $\delta Q = T dS$, which, upon request that the entropy is proportional to the horizon area, translates to energy flux resulting in focusing of the horizon generators. Using the Raychaudhuri equation to write the expansion $\theta$ of the congruence of the horizon generators in terms of the Ricci tensor, gives for small affine parameter $\lambda$ 
		\begin{equation}
			dS = \eta\delta A = \eta \int_{\mathcal{H}} \theta d\lambda d A = \eta \int_\mathcal{H} \lambda R_{\alpha\beta} l^\alpha l^\beta d\lambda d A,
		\end{equation}
		where $\eta$ is an undetermined proportionality constant. Given the Unruh temperature as $T=\hbar \kappa / 2\pi$ \cite{Unruh1976_UnruhEffect}, the thermodynamic equation $\delta Q = T dS$ alongside the request that $T_{\alpha\beta}$ is free from divergences then implies
		\begin{equation}
			R_{\alpha\beta}-\frac{1}{2}R g_{\alpha\beta}+\Lambda g_{\alpha\beta} = \frac{2\pi}{\hbar\eta} T_{\alpha\beta}
		\end{equation}
		for some constant $\Lambda$, which is the Einstein equation for $G=(4\hbar\eta)^{-1}$. The possibility of repeating this construction for any point of spacetime makes general relativity nothing but the geometrical counterpart of equilibrium thermodynamics for the specific choice $dS=\eta \delta A$ of the entropy functional, with $G$ changing its status from fundamental constant to a quantity that has a non-trivial $\hbar\rightarrow0$ limit. The passage of matter across the horizon is a one-way transfer of energy to the degrees of freedom of the system, which therefore happens as heat rather than work and results in an increase in entropy. What these degrees of freedom are remains unclear. Jacobson explicitly interprets the horizon entropy as entanglement entropy, due to the correlations between the modes inside and outside of the horizon. This forces the introduction of a minimum length cutoff to avoid divergences that enter into the calculations through $\eta$. Compatibility with Einstein's equation hints at this minimum length being of the order of the Planck length. The use of Rindler horizons makes the whole construction heavily observer dependent, with irreversibility being introduced by the existence of a limiting velocity. Jacobson's setup has been generalised in \cite{Eling2006_NonEquilibriumTD} to non-equilibrium thermodynamics by introducing corrections to the entropy functional that are polynomial in the Ricci scalar.\\
		Verlinde in \cite{Verlinde2011_EntropicGravity} takes the thermodynamic picture a step further. He proposes that spacetime and the metric are themselves emergent objects and that gravity should be interpreted as an entropic force, regarding as fundamental the information of the relative position and displacement of masses. This information should be thought of as stored in discrete bits on surfaces between masses, enforcing the holographic principle that is hinted at by black hole physics and by the AdS/CFT correspondence. The holographic screen acts like a stretched horizon: it separates the emerged space from the pre-geometric mere distribution of information and is endowed with an entropy and a temperature. Time is assumed to be well defined in the underlying but unspecified microscopic theory, to allow for a definition of energy and temperature.
		The idea is that the holographic screen encodes the information of the pre-geometric side and that of the newly emerged space, so that a test mass $m$ approaching it in the latter will influence its entropy. In analogy with osmosis across a semipermeable membrane, the test mass under a displacement $\Delta x$ will experience an effective force due to the entropy gradient as
		\begin{equation} \label{entropic force}
			F \Delta x = T \Delta S.
		\end{equation}
		To see how the boundary gets a temperature, essential to get a non-vanishing force; $T$ can be interpreted as the Unruh temperature seen by the test mass as it is accelerated by the entropic force. Assuming that the change in entropy $\Delta S$ is linear in $m$ and in the displacement $\Delta x$ with the appropriate proportionality coefficient
		\begin{equation}
			\Delta S = 2\pi k_B \frac{m c}{\hbar} \Delta x,
		\end{equation}
		one can then recover the second law of dynamics $F=m a$ in the emergent space from the above relation.
		The argument can be refined considering a closed, spherical holographic boundary. The number of bits of information $N$ it encodes is naturally assumed to be proportional to the area $A$ of the boundary, with dimensionality and the Bekenstein formula for entropy suggesting
		\begin{equation}
			N=\frac{c^3}{G\hbar} A,
		\end{equation}
		where $G$ is, at this stage, to be interpreted as a generic proportionality constant. Assuming that the total energy of the system $E$ is evenly divided over the $N$ bits, the temperature is defined as the average energy per bit by the equipartition rule
		\begin{equation}
			E=\frac{1}{2} N k_B T.
		\end{equation}
		This energy would emerge as a mass $M$ enclosed by the screen through $E=M c^2$. The temperature $T$ can then be expressed in terms of the emerged mass as $T= (2 M G \hbar) / (A c k_B)$. Substituting these quantities and the area of the spherical boundary $A=4\pi R^2$, the entropic force of equation (\ref{entropic force}) reduces to Newton's gravitational law
		\begin{equation}
			F=G\frac{m M}{R^2}.
		\end{equation}
		In this view, Newton's potential $\Phi$ acts as a coarse-grained descriptor of the information encoded on the holographic screen. Equipotential surfaces provide a foliation of space that is to be seen as nested holographic screens: as they shrink, the data encoded on them gets more coarse-grained and the lost information is replaced by emergent space. The gradient $\nabla\Phi$ therefore identifies the emerging holographic direction. This picture can be translated into a relativistic scenario by considering a static background with a global time-like Killing field $\zeta^\alpha$. The generalisation of Newton's potential then reads $\phi=\ln(\sqrt{-\zeta^\alpha \zeta_\alpha})$ and allows a foliation of space based on surfaces of constant redshift.
		In \cite{Verlinde2017_EntropicGravityDM} expands his model by adding a volume term to the entropy functional that overtakes the area dependence for large distances, claiming that this modification can account for the observed cosmological constant in place of dark matter. Hossenfelder in \cite{Hossenfelder2017_CovariantVerlinde} offers a Lagrangian, generally covariant, version of Verlinde's theory.
		The arguments presented by Verlinde propose a radical shift in the modellization of gravity and spacetime and are, by the author's own admission, quite heuristic. He introduces the idea of gravity as an entropic force through an analogy with the elasticity of a polymer chain immersed in a heat bath; however, it is not clear how exactly this example should translate to a gravitational system \cite{Gao2011_IsGravityEntropic}. The main criticality of Verlinde's program lies in the fact that gravity is a conservative force, an aspect which is very difficult to reconcile with an entropic character \cite{Dai2017_IncostistenciesInVerlinde}. Visser \cite{Visser2011_ConservativeEntropicForce} in particular derives the constraints that such a requirement places on the form of the entropy and the temperature functionals. He concludes that a consistent picture would require multiple temperatures and entropies, one for each ordered pair of particles in the gravitational system, whose physical interpretation is far from obvious. Several authors have also cast doubts on the effectiveness of the modified MoND-like entropy functional in reproducing the observational data attributed to dark matter effects \cite{Dai2017_IncostistenciesInVerlinde} \cite{Dai2017_CorrectionToHossenfelder}. Kobakhidze \cite{Kobakhidze2011_GravityNotEntropic} \cite{Kobakhidze2011_MoreGravityNotEntropic} points out that Verlinde's description, when applied to quantum mechanical systems, implies effects which are in contrast with experimental results \cite{Nesvizhevsky2002_GravitationalUltracoldNeutrons}.

		\subsection{Padmanabhan's theory of emergent gravity}

		Together with Jacobson's and Verlinde's approaches, there is a third main line of research that explores the thermodynamic nature of gravity: that of Padmanabhan. We shall outline it in this separate section, as it is the one that bears the closest connection to the results of the q-metric paradigm.\\
		Padmanabhan's point of view first stemmed from seeing spacetime as the emergent, long-wavelength description of a microscopic structure, akin to the continuum limit of a solid. Diffeomorphisms $x^\alpha \rightarrow x'^\alpha = x^\alpha+v^\alpha (x)$ for some vector field $v^\alpha$ are then the analogue of elastic deformations, and it should be possible to obtain the field equations of general relativity from the maximisation of an appropriate entropy functional \cite{Padmanabhan2004_GravityAsElasticity}. The key observation to construct this functional comes once again from horizons. In general relativity, different observers (defined by time-like congruences) have access to different regions of spacetime. Each observer should provide a description based only on the quantities they have access to, or, in other words, as if they lived in an effective manifold where the region beyond the horizon is removed and encoded in the boundary. The gravitational action should then be the result of a surface term and a bulk term, which also encapsulates the interplay between the horizon and the rest of the accessible spacetime. Both of these depend on the chosen foliation, but their sum should be covariant. It turns out that relating the surface term to the entropy per unit area completely fixes the gravitational dynamics in the bulk \cite{Padmanabhan2005_GravityAndTDHorizon}. This is related to the fact that the Ricci scalar, and therefore the Einstein-Hilbert Lagrangian, can be decomposed into terms that are quadratic in the first derivative of the metric and contributions in the second derivative of the metric. The first can always be made to vanish along a world line with the right choice of coordinate system. This makes gravity intrinsically holographic, in the sense that the theory is completely determined by the surface term.\\
		Having assessed the central role of horizons in determining the gravitational physics of the bulk, the next step is studying the thermodynamics of the system as matter crosses the horizon. This event can be seen as a deformation of the horizon to swallow the matter as it gets close enough to it, and the displacement associated can be evaluated in terms of an entropy functional. The evolution of the system should then be governed by the maximisation of such a functional \cite{Padmanabhan2010_TDAspectsOfGravity}. The studies on the thermodynamics of Rindler horizons suggest that the resulting relation should hold for any null surface, with the entropy functional being dependent on the observer. The total entropy should consider both the degrees of freedom associated to the deformation of spacetime and those related to the matter being engulfed. For the latter, the density of the energy flux $T_{\alpha\beta}l^\alpha l^\beta$ already used in Jacobson's work \cite{Jacobson1995_EinsteinEqofState}, suggests
		\begin{equation}
			S_{matter} = \int_{\mathcal{V}} d^D x \sqrt{g} T_{\alpha\beta} l^\alpha l^\beta
		\end{equation}
		where $g$ is the absolute value of the determinant of the metric, $T_{\alpha\beta}$ is the matter energy-momentum tensor, and $l^\alpha$ is the null vector to which the displacement field $v^\alpha$ becomes proportional on the horizon. The integral is taken over a small patch $\mathcal{V}$ of the stretched horizon interested by the displacement in D-dimensional spacetime. The structure of this entropy functional is still that of an energy flux divided by a temperature, but the latter is hidden in the integration in the time variable: by Wick rotating the geometry to an Euclidean one, thermality translates into a periodicity in imaginary time, with period $\beta=T^{-1}$. This gives an overall factor $T^{-1}$ that multiplies the energy flux in $D-1$ dimensional space.\\
		The gravitational entropy is constructed following the analogy with the elasticity of a solid \cite{Padmanabhan2007_EntropyNullSurfaces}. It should be a volume integral of a local entropy density to ensure extensivity, and it should only have quadratic terms in the first derivatives of the displacement field to preserve translational invariance. As the entropy is a scalar, balancing out tensorial indices requires the introduction of a tensor $P_{\alpha\beta}^{\gamma\delta}$ to construct
		\begin{equation}
			S_{grav} = -4\int_{\mathcal{V}} d^D x \sqrt{g} P_{\alpha\beta}^{\phantom{\alpha\beta}\gamma\delta} \nabla_\gamma l^\alpha \nabla_{\delta} l^\beta.
		\end{equation}
		Following the elasticity analogy, $P_{\alpha\beta\gamma\delta}$ can be thought of as representing the 'elasticity constant' of spacetime. If $P_{\alpha\beta\gamma\delta}$ is assumed to have the same symmetries as the Riemann tensor $R_{\alpha\beta\gamma\delta}$, it can be written as
		\begin{equation}\label{P condition 1}
			P_\alpha^{\phantom{\alpha}\beta\gamma\delta} = \frac{\partial L}{\partial R^\alpha_{\phantom{\alpha}\beta\gamma\delta}}
		\end{equation}
		for some scalar $L$. Mimicking the conservation law $\nabla_{\alpha}T^{\alpha\beta}=0$ leads to also require
		\begin{equation} \label{P condition 2}
			\nabla_{\alpha}P^{\alpha\beta\gamma\delta}=0.
		\end{equation}
		The total entropy to be extremised for variations of $l^\alpha$ is $S=S_{matter}+S_{grav}$ \cite{Padmanabhan2010_TDAspectsOfGravity}. The resulting equation of motion should shape the null surface that $l^\alpha$ identifies without changing its causal character; therefore, $l^\alpha$ must remain null throughout the variation. This is achieved by imposing the broader requirement of constancy of the norm of $l^\alpha$ and enforcing it by introducing a Lagrangian multiplier $\lambda$. Then the entropy variation reads
		\begin{equation}
			\delta S = 2\int_{\mathcal{V}} d^D \sqrt{g} [-4P_{\alpha\beta}^{\gamma\delta}\nabla_\gamma l^\alpha (\nabla_\delta \delta l^\beta) + T_{\alpha\beta} l^\alpha \delta l^\beta + \lambda(x) g_{\alpha\beta} l^\alpha \delta l^\beta].
		\end{equation}
		Integrating by parts and requiring that variations of $l^\alpha$ vanish on the boundary of $\mathcal{V}$ leaves the equation
		\begin{equation}
			(T_{\alpha\beta} + \lambda g_{\alpha\beta}) l^\alpha - 2 P_{\alpha\beta}^{\phantom{\alpha\beta}\gamma\delta} (\nabla_\gamma \nabla_\delta - \nabla_\delta \nabla_\gamma) l^\alpha = (T_{\alpha\beta} + \lambda g_{\alpha\beta}) l^\alpha - 2 P_{\alpha\beta}^{\phantom{\alpha\beta}\gamma\delta} R^\alpha_{\phantom{\alpha}\beta\gamma\delta} l^\beta = 0.
		\end{equation}
		Multiplying the whole expression by the metric, exploiting the symmetries of $P_{\alpha\beta\gamma\delta}$ and $R_{\alpha\beta\gamma\delta}$ and renaming dummy indices allows us to rearrange the above expression as
		\begin{equation}
			(2P_\beta^{\phantom{\beta}\mu\nu\sigma} R^\alpha_{\phantom{\alpha}\mu\nu\sigma} - T^\alpha_\beta + \lambda \delta^\alpha_\beta) l_\alpha = 0
		\end{equation}
		and calling $\mathcal{G}^\alpha_\beta = [P_\beta^{\phantom{\beta}\mu\nu\sigma} R^\alpha_{\phantom{\alpha}\mu\nu\sigma} - (1/2) L \delta^\alpha_\beta]$ with $L$ being the scalar of definition (\ref{P condition 1}) leads to
		\begin{equation} \label{EinsteinEqPadmanabhan}
			[2 \mathcal{G}^\alpha_\beta - T^\alpha_\beta + (L+\lambda)\delta^\alpha_\beta] l_\alpha = 0.
		\end{equation}
		Rather than a relation that determines the dynamics of $l_\alpha$, this equation of motion is to be interpreted as holding for all the null vectors $l^\alpha$ and therefore imposing a constraint on the geometry of spacetime. Taking the covariant derivative of the above expression shows that $(L+\lambda)$ needs to be a constant, which is renamed as $\Lambda$, as $\nabla_\alpha \mathcal{G}^\alpha_\beta = 0 = \nabla_\alpha T^\alpha_\beta$. The equation of motion is therefore satisfied for all $l_\alpha$ if \cite{Padmanabhan2010_TDAspectsOfGravity}
		\begin{equation} \label{extended GR equation}
			\mathcal{G}^\alpha_\beta = \frac{1}{2} T^\alpha_\beta + \Lambda \delta^\alpha_\beta.
		\end{equation}
		This expression is very familiar, and in fact reduces to the Einstein equation with a cosmological constant if the scalar $L$ used to define $P_{\alpha\beta\gamma\delta}$ is chosen to be the curvature $R$, which makes $\mathcal{G}_{\alpha\beta}=R_{\alpha\beta}-(1/2)R g_{\alpha\beta}$. The application is even broader when thinking of $L$ as a generic gravitational Lagrangian from which $P_{\alpha\beta\gamma\delta}$ is constructed. Under condition (\ref{P condition 2}), $L$ is the Lagrangian for Lanczos-Lovelock models, which are generalisations of general relativity that still yield second-order field equations but where horizon entropy is not proportional to the area (see, for example \cite{Padmanabhan2013_LanczosLovelockModels} for a review). Varying the associated action with respect to the metric results in an equation equal to (\ref{extended GR equation}), without the cosmological constant term that still needs to be added by hand.\\
		The conditions imposed on $P_{\alpha\beta\gamma\delta}$ and the precise form of $S_{grav}$ were crucial in arriving at the equations of (extended) general relativity (\ref{extended GR equation}). Their choice was guided by the known Lagrangian counterpart of the theory. 
		A peculiarity of the thermodynamic approach, focused on the variation of the null vectors $l^\alpha$, is that the requirement that $l^\alpha l_\alpha$ is fixed makes the cosmological constant emerge as an integration variable, rather than an ad hoc parameter added to adhere to observations. Avoiding treating $g_{\alpha\beta}$ as the dynamical variable also solves an asymmetry between the gravitational Lagrangian and the Lagrangians that give rise to matter field equations that, according to Padmanabhan \cite{Padmanabhan2015_EmergentGravityProgress}, is the true core of the cosmological constant problem. Gravitational dynamics in the conventional approach is sensitive to the addition of a constant to the Lagrangian, which instead leaves the matter equations of motion unchanged. This additional term can be interpreted as a shift to the energy-momentum tensor $T^\alpha_\beta \rightarrow T^\alpha_\beta + const \delta^\alpha_\beta$ and results in the equation of motion $\mathcal{G}^\alpha_\beta = T^\alpha_\beta + (const.) \delta^\alpha_\beta$. This skewed sensitivity questions the physical meaningfulness of the observed value of the cosmological constant, which would be altered by any arbitrary choice of a zero-point energy in the matter sector. Avoiding treating the metric as the dynamical variable in the variational principle fixes this issue, as the equation $\delta(S_{matt}+S_{grav})=0$ interpreted as a variation in $l^\alpha$ is not affected by a constant shift to $T_{\alpha\beta}$: it would only introduce a term proportional to $l^\alpha l_\alpha = 0$. The cosmological constant, therefore, cannot be introduced as an ad hoc low-energy parameter, but has to emerge as an integration constant \cite{Padmanabhan2010_TDAspectsOfGravity}.\\
        The introduction of the q-metric allowed Padmanabhan to clarify the meaning of the functional used, changing the thermodynamic interpretation of the gravitational field equation as well. Considering for simplicity Einstein's general relativity in four dimensions, the equation
        \begin{equation}
            -\frac{1}{8\pi L_P^2} R_{\alpha\beta} l^\alpha l^\beta + T_{\alpha\beta} l^\alpha l^\beta = 0
        \end{equation}
        can be interpreted as requiring a balance on null surfaces between the heat density of spacetime due to gravitational degrees of freedom and the heat density of matter \cite{Padmanabhan2019_GravityAndQG}. This relation leads to considering the functional
        \begin{equation}\label{NewPadmanabhanFunctional}
            Q[l_\alpha(x)] = \int dV F\bigg( T^\alpha_\beta(x) l_\alpha l^\beta- \frac{1}{8\pi L_P^2} R^\alpha_\beta(x) l_\alpha l^\beta \bigg),
        \end{equation}
        where $F$ is a scalar function and the integration is performed over a region of spacetime with the covariant measure $dV$. This functional can be traced back to the attempt and the calculation discussed before by noting that the expression
        \begin{equation}
            A = \int \frac{d^4 x}{L_P^4}\sqrt{g} (L_P^4 T_{\alpha\beta} l^\alpha l^\beta + P^{\alpha\beta}_{\gamma\delta}\nabla_\alpha l^\gamma \nabla_\beta l^\delta),
        \end{equation}
        with $P^{\alpha\beta}_{\gamma\delta}=\frac{L_P^2}{8\pi}(\delta^\alpha_\beta \delta^\beta_\delta - \delta^\beta_\gamma \delta^\alpha_\delta)$ for Einstein's general relativity, is equivalent to (\ref{NewPadmanabhanFunctional}) when considering a variational principle, as the term in $P^{\alpha\beta}_{\gamma\delta}$ can be split into an ignorable total divergence and a term in $R_{\alpha\beta}$ as
        \begin{equation}
        P^{\alpha\beta}_{\gamma\delta} \nabla_\alpha l^\gamma \nabla_\beta l^\delta = \delta_\alpha (P^{\alpha\beta}_{\gamma\delta} l^\gamma \nabla_\beta l^\delta) + \frac{L_P^2}{8\pi} R_{\alpha\beta} l^\alpha l^\beta.
        \end{equation}
        Varying either functional with respect to $l^\alpha$ and adding a Lagrangian multiplier to enforce the constraint $l^\alpha l_\alpha=0$ gives the gravitational field equation
        \begin{equation}
        G^\alpha_\beta = 8\pi L_P^2 T^\alpha_\beta + \Lambda \delta^\alpha_\beta,
        \end{equation}
        where all the considerations for the previous setup apply.
        The functional (\ref{NewPadmanabhanFunctional}) can be interpreted as the total entropy on the null surface resulting from the mesoscopic density of states of spacetime $\rho_{grav}$ and the density of states for matter $\rho_{matter}$  \cite{Padmanabhan2019_GravityAndQG} and can then be written as $S_{grav}+S_{matter}=\ln(\rho_{grav}\rho_{matter})$.
        The density of states of spacetime at a given event $P$ can be naturally assumed to be a function of the area associated to that "point" $\Sigma(P)$. As detailed in subsection \ref{subsection: qmetric effects}, such an area is to be understood as that of the equigeodesic surface constructed around $P$ in the $\sigma\rightarrow 0$ limit and is non-vanishing in the q-metric framework. Given two separate events $P$ and $Q$, the fact that the degrees of freedom should be multiplicative for the total system as $\rho_{grav}(P)\rho_{grav}(Q)$ while the areas are additive suggests a logarithmic relation of the kind $\ln\rho_{grav}(P)\propto \Sigma_q (P)$. To obtain a dimensionless density of states, this definition can be refined to consider the $\sigma\rightarrow 0$ limit of the area element on the equigeodesic surface $\sqrt{h_q} d^{D-1}x$ normalized by its value in flat spacetime $\sqrt{h_q^{flat}} d^{D-1}x$ as
        \begin{equation}
        \ln \rho_{grav}(P) \propto \lim_{\sigma\rightarrow 0} \frac{\sqrt{h_q(P,\sigma)}}{\sqrt{h_q^{flat}(P, \sigma)}} = \bigg[ 1-\frac{L_0^2}{6} R_{\alpha\beta} n^\alpha n^\beta \bigg],
        \end{equation}
        where the last equality, valid at leading order, comes from the expansion in powers of $L_0$ of the Van Vleck determinant. The effective minimum spacetime length $L_0$ can be expressed in terms of the Planck length as $L_0^2= 3L_P^2/4\pi$ in order to recover the known Newtonian limit. The overall expression ends up depending on the vector normal to the equigeodesic surface, here named $n_\alpha=\nabla_\alpha\sigma$, which in the previous section entered the disformal term in the q-metric. In the $\sigma\rightarrow 0$ limit, this vector becomes arbitrary, with the sole requirement of having a fixed norm (equal to one in the Euclidean case and null in Lorentzian spacetime). Such a degree of freedom is a relic of the effective discrete small scale structure of spacetime, a fluctuating variable governed by some probability distribution whose average $\langle n^\alpha \rangle$ corresponds to the vector $l^\alpha$ of equation (\ref{EinsteinEqPadmanabhan}). In other words, each event $P$ is associated to both a set of coordinates and an internal degree of freedom: the quantum state of the (so far unspecified) microscopic structure determines the mean value $\langle n^\alpha \rangle$, which in the continuum limit acquires a geometrical interpretation as the normal $l^\alpha (x^\mu)$ to the local Rindler horizon in $P$. The thermodynamics of such null membranes then shapes the geometry through the extremization of the total entropy of the system, which arises from the degrees of freedom of spacetime and those of the matter being engulfed. In the continuum limit, the first contribution reads
        \begin{equation}
            \langle \ln\rho_{grav}(x^\mu, n_{\alpha}) \rangle \propto \frac{1}{4} \bigg[ 1-\frac{L_P^2}{2\pi} R_{\alpha\beta}l^\alpha l^\beta \bigg],
        \end{equation}
        for each event while the latter can be inferred from the heat density of matter falling through a Rindler horizon $\mathcal{H}$ as
        \begin{equation}
        \langle \ln\rho_{matter} \rangle \propto L_P^4 \mathcal{H} = L_P^4 T_{\alpha\beta}l^\alpha l^\beta.
        \end{equation}
        Both of these expressions then need to be integrated over the surface of the null membrane.\\
		Padmanabhan also proposes a thermodynamic framework for cosmology, making a step towards the interpretation of spacetime itself as being emergent, rather than being limited to the dynamics of an assumed metric. He \cite{Padmanabhan2012_SpacetimeEmergence} argues that the expansion of our Universe, seen as the emergence of cosmic spacetime, can be driven by a difference between its bulk degrees of freedom $N_{bulk}$ and its surface degrees of freedom $N_{surf}$. The thermodynamics of spacetime suggests the existence of yet unknown, microscopic degrees of freedom. Considering a spherical horizon, the Bekenstein formula for entropy suggests that $N_{surf}$ will be proportional to its area, while the law of equipartition hints that $N_{bulk}$ will be fixed by the energy and temperature as $E=(1/2) T N_{bulk}$ \cite{Padmanabhan2010_HorizonEquipartition} \cite{Padmanabhan2010_HorizonEquipartitionDoF}. The time-independent metric can then be fixed by requiring the holographic equipartition principle $N_{surf}=N_{bulk}$ \cite{Padmanabhan2012_EmergentGravityDE}. If our Universe can be modelled as a spherical bubble of radius equal to the inverse of the Hubble parameter $H$, the degrees of freedom of its surface can be written as
		\begin{equation}
			N_{surf} = \frac{4\pi}{L_P^2 H^2}
		\end{equation}
		and the degrees of freedom of the bulk as
		\begin{equation}
			N_{bulk} = \frac{2|E|}{T}
		\end{equation}
		where $E$ is the Komar energy $E=(\rho+3 P) V$ inside the proper volume $V=4\pi/3 H^3$ and the absolute value is required to deal with the negative contribution of dark energy. The expansion of the Universe can then be interpreted as driven by the tendency towards the equilibrium condition $N_{surf}=N_{bulk}$ and expressed by a law \cite{Padmanabhan2012_EmergentGravityDE}
		\begin{equation}
			\frac{d V}{d t} = L^2_P (N_{sur}-N_{bulk}).
		\end{equation}
		Substituting the expressions for $V$, $N_{surf}$ and $N_{bulk}$ with the temperature of the Hubble horizon as $T=H/2\pi$ gives
		\begin{equation}
			(\dot{H}+H^2) = L_P^2 \bigg[-\frac{4\pi}{3} (\rho+3P)\bigg]
		\end{equation}
		which, using $(\dot{H}+H^2) = \ddot{a}/a$ with $a(t)$ being the scale factor of the FLRW metric, is the familiar equation for the acceleration of the expansion in the Friedmann model.

		\section{Conclusions} \label{Conclusions}

		In this review we have presented the q-metric framework as a phenomenological approach to implement quantum effects in a semiclassical theory of gravity.
		The motivation for this construction stems from noting that many quantum gravity frameworks share a common prediction: the existence of a minimal spacetime length. Indeed, this feature can be inferred from fairly general grounds. It is supported by thought experiments on the localization of a body \cite{Mead1964_HeisenbergMicroscopeQG}, on distance measurement \cite{Salecker1958_LimitDistanceMeasurements} and on clock synchronization \cite{Mead1964_HeisenbergMicroscopeQG}: the main idea is that probes carry a momentum, which results in a gravitational field that perturbs the system. Taking into account general relativity, not only increasing the momentum won't result in an improved resolution after a certain threshold, but the concentration of energy could cause the formation of a black hole in the probed region \cite{Scardigli1999_GUPMicroBH}. 
		The existence of a limit length is also implied by the generalized covariant entropy bound. Other thermodynamic bounds that do not necessarily demand the consideration of gravitational effects (like the Hod's bound to relaxation time, the Bekenstein bound to entropy or the Kovtun-Son-Starinet bound to the viscosity to entropy density ratio) require a relation between the extension of a system and its temperature to be satisfied and are ultimately protected by the Heisenberg uncertainty principle. Taking gravity into account instead fixes a minimum extension for the system that is independent from its state \cite{Pesci2022_InformationContent}.
		Finally, the existence of a minimal length, generally thought to be of the order of the Planck length, can also be obtained as an explicit result of specific theories of quantum gravity. In string theory it arises as the outcome of a generalized uncertainty principle that emerges from the 1-dimensional nature of strings, in loop quantum gravity it is compatible with the fact that the area operator for physical systems has a discrete spectrum, in asymptotically safe gravity it is a direct consequence of the gravitational coupling constant being asymptotically finite and it is at the basis of the non-commutative geometry construction.\\
		The effects of a minimal length can be implemented in a semiclassical theory that still makes use of a continuous geometry, but substitutes the metric tensor with a quantum metric-like bitensor $q_{\alpha\beta}(x, x')$ that has a zero point length $L$. This construction has the advantage of providing a phenomenological approach that doesn't depend on the details of a specific quantum gravity structure and that can still be constructed on a manifold with the familiar tools of differential geometry.
		The fact that all the information on the geometry of a given spacetime is encoded in the squared geodesic distance allows to trade the degrees of freedom of the metric for those of a non-local object, on which then a deformation can be imposed that requires it to be non vanishing in the coincidence limit \cite{Kothawala2014_EntropySpacetime} \cite{Stargen2015_VanVleckDeterminant} \cite{Pesci2019_QMetricForNull}. The resulting q-metric is tied to the original one by a disformal transformation, that reduces to a conformal relation on equigeodesic surfaces \cite{Kothawala2014_Finsleresque}. Given the expression of the q-metric for space-like, time-like and null intervals, its consequences can be explored. Most notably, the existence of a minimum length results in a minimal area (but not a finite minimal spacetime volume) which avoids the formation of caustics \cite{Pesci2018_NullRaychaudhuri} \cite{Chakraborty2019_ZeroPointRaychaudhuri}. 
		Another non-trivial result comes for the calculation of the equivalent of the Ricci scalar in the q-metric framework. The computation can be carried out exploiting the simplified relation between the q-metric and the original metric on equigeodesic surfaces and then reconstructing the full q-Ricci biscalar via the Gauss-Codazzi equation \cite{Kothawala2014_EntropySpacetime} \cite{Pesci2020_RicciForNull}. Taking the coincidence limit $x\rightarrow x'$ followed by the local limit $L\rightarrow 0$ one would expect to recover the usual Ricci scalar as the classical limit of the q-metric theory. Instead, the resulting expression encodes additional degrees of freedom and is proportional to the functional for the heat density of horizons.\\
		This result connects the q-metric with the theories of thermodynamic emergent gravity like those by Jacobson \cite{Jacobson1995_EinsteinEqofState}, Verlinde \cite{Verlinde2011_EntropicGravity} and, more specifically, Padmanabhan \cite{Padmanabhan2004_GravityAsElasticity}. The idea at the core of these theories, inspired by black hole thermodynamics, is viewing null surfaces as one-way membranes for information, to which an entropy can be assigned. The field equations for gravity then emerge by extremizing such entropy.
		This is precisely what happens when writing the Einstein-Hilbert Lagrangian as the classical limit of q-metric quantities: the minimization of the action, interpreted as a variational principle for the metric in general relativity, naturally translates to a requirement for the microscopic degrees of freedom that make up spacetime. The metric is no longer a dynamical variable and gravity emerges as the collective behaviour of "atoms" of spacetime that evolve to maximize an entropy. In this viewpoint, gravity is no longer a fundamental force, but rather akin to elasticity: a non perturbative effect of the underlying small scale structure. This shift in perspective would have profound implications on the problem of the quantization of gravity. The result of such procedure would be like phonons for elasticity: pseudo-particles without a physical meaning that wouldn't shed light on the structure of spacetime. Furthermore, gravity should be thought as being quantum in nature as it couldn't emerge with a strictly continuous background.\\
		Further work in this field is needed towards clarifying the details of the emergence of gravity and perhaps of the metric itself. In particular, an expression for the momentum-energy tensor in the q-metric tensor has not yet been achieved.
		A natural application for the q-metric is on black holes horizons, where quantum effects are significant and the thermodynamics well recognized. The work of one of the authors is currently focused on this topic. An especially interesting result suggests that the minimal length implies a quantization of the horizon area (as hinted to by the expression of the Bekenstein entropy), which in turn sets a threshold energy for a particle to be absorbed by the black hole. The effects of this reflectivity could be captured by gravitational waves, as they would fall within the frequency range of the currently available interferometers even if the minimal step of the area were of the order of the Planck length squared \cite{Foit:2016uxn, Cardoso:2019, Agullo:2021, Datta:2021row, Sago:2021iku, KriPer, Pesci2025_MinLengthHorizons}.\\
        
        \noindent
        {\it Acknowledgments.} This work was partially supported by INFN grant FLAG.

	\bibliography{MinLengthBH_biblio}{}

@article{Padmanabhan2015_EmergentGravityProgress,
	title={Emergent gravity paradigm: Recent progress},
	volume={30},
	ISSN={1793-6632},
	url={http://dx.doi.org/10.1142/S0217732315400076},
	DOI={10.1142/s0217732315400076},
	number={03n04},
	journal={Modern Physics Letters A},
	publisher={World Scientific Pub Co Pte Lt},
	author={Padmanabhan, T.},
	year={2015},
	month=jan, pages={1540007} }

@article{Pesci2019_QMetricForNull,
	title={Quantum metric for null separated events and spacetime atoms},
	volume={36},
	ISSN={1361-6382},
	url={http://dx.doi.org/10.1088/1361-6382/ab0a40},
	DOI={10.1088/1361-6382/ab0a40},
	number={7},
	journal={Classical and Quantum Gravity},
	publisher={IOP Publishing},
	author={Pesci, Alessandro},
	year={2019},
	month=mar, pages={075009} }

@article{Pesci2020_RicciForNull,
	title={Minimum-length Ricci scalar for null separated events},
	volume={102},
	ISSN={2470-0029},
	url={http://dx.doi.org/10.1103/PhysRevD.102.124057},
	DOI={10.1103/physrevd.102.124057},
	number={12},
	journal={Physical Review D},
	publisher={American Physical Society (APS)},
	author={Pesci, Alessandro},
	year={2020},
	month=dec }

@article{Pesci2022_InformationContent,
	title={Information content and minimum-length metric: A drop of light},
	volume={54},
	ISSN={1572-9532},
	url={http://dx.doi.org/10.1007/s10714-022-02960-1},
	DOI={10.1007/s10714-022-02960-1},
	number={7},
	journal={General Relativity and Gravitation},
	publisher={Springer Science and Business Media LLC},
	author={Pesci, Alessandro},
	year={2022},
	month=jul }

@article{Padmanabhan2010_TDAspectsOfGravity,
	title={Thermodynamical aspects of gravity: new insights},
	volume={73},
	ISSN={1361-6633},
	url={http://dx.doi.org/10.1088/0034-4885/73/4/046901},
	DOI={10.1088/0034-4885/73/4/046901},
	number={4},
	journal={Reports on Progress in Physics},
	publisher={IOP Publishing},
	author={Padmanabhan, T},
	year={2010},
	month=mar, pages={046901} }

@article{Padmanabhan2005_GravityAndTDHorizon,
	title={Gravity and the thermodynamics of horizons},
	volume={406},
	ISSN={0370-1573},
	url={http://dx.doi.org/10.1016/j.physrep.2004.10.003},
	DOI={10.1016/j.physrep.2004.10.003},
	number={2},
	journal={Physics Reports},
	publisher={Elsevier BV},
	author={Padmanabhan, T.},
	year={2005},
	month=jan, pages={49–125} }

@article{Kothawala2014_EntropySpacetime,
	title={Entropy density of spacetime as a relic from quantum gravity},
	volume={90},
	ISSN={1550-2368},
	url={http://dx.doi.org/10.1103/PhysRevD.90.124060},
	DOI={10.1103/physrevd.90.124060},
	number={12},
	journal={Physical Review D},
	publisher={American Physical Society (APS)},
	author={Kothawala, Dawood and Padmanabhan, T.},
	year={2014},
	month=dec }

@article{Kothawala2013_SmallScaleStructure,
	title={Minimal length and small scale structure of spacetime},
	volume={88},
	ISSN={1550-2368},
	url={http://dx.doi.org/10.1103/PhysRevD.88.104029},
	DOI={10.1103/physrevd.88.104029},
	number={10},
	journal={Physical Review D},
	publisher={American Physical Society (APS)},
	author={Kothawala, Dawood},
	year={2013},
	month=nov }

@article{Stargen2015_VanVleckDeterminant,
	title={Small scale structure of spacetime: The van Vleck determinant and equigeodesic surfaces},
	volume={92},
	ISSN={1550-2368},
	url={http://dx.doi.org/10.1103/PhysRevD.92.024046},
	DOI={10.1103/physrevd.92.024046},
	number={2},
	journal={Physical Review D},
	publisher={American Physical Society (APS)},
	author={Stargen, D. Jaffino and Kothawala, Dawood},
	year={2015},
	month=jul }

@article{Hossenfelder2013_MinLength,
	title={Minimal Length Scale Scenarios for Quantum Gravity},
	volume={16},
	ISSN={1433-8351},
	url={http://dx.doi.org/10.12942/lrr-2013-2},
	DOI={10.12942/lrr-2013-2},
	number={1},
	journal={Living Reviews in Relativity},
	publisher={Springer Science and Business Media LLC},
	author={Hossenfelder, Sabine},
	year={2013},
	month=jan }

@article{Garay1995_QG,
	title={Quantum Gravity and Minimum Length},
	volume={10},
	ISSN={1793-656X},
	url={http://dx.doi.org/10.1142/S0217751X95000085},
	DOI={10.1142/s0217751x95000085},
	number={02},
	journal={International Journal of Modern Physics A},
	publisher={World Scientific Pub Co Pte Lt},
	author={Garay, Luis J.},
	year={1995},
	month=jan, pages={145–165} }

@article{Pesci2019_SpacetimeAtomsExtrinsic,
	title={Spacetime atoms and extrinsic curvature of equi-geodesic surfaces},
	volume={134},
	ISSN={2190-5444},
	url={http://dx.doi.org/10.1140/epjp/i2019-12749-0},
	DOI={10.1140/epjp/i2019-12749-0},
	number={7},
	journal={The European Physical Journal Plus},
	publisher={Springer Science and Business Media LLC},
	author={Pesci, Alessandro},
	year={2019},
	month=jul }

@article{Pesci2018_NullRaychaudhuri,
	title={Effective Null Raychaudhuri Equation},
	volume={1},
	ISSN={2571-712X},
	url={http://dx.doi.org/10.3390/particles1010017},
	DOI={10.3390/particles1010017},
	number={1},
	journal={Particles},
	publisher={MDPI AG},
	author={Pesci, Alessandro},
	year={2018},
	month=oct, pages={230–237} }

@article{Chakraborty2019_ZeroPointRaychaudhuri,
	title={Raychaudhuri equation with zero point length},
	volume={797},
	ISSN={0370-2693},
	url={http://dx.doi.org/10.1016/j.physletb.2019.134877},
	DOI={10.1016/j.physletb.2019.134877},
	journal={Physics Letters B},
	publisher={Elsevier BV},
	author={Chakraborty, Sumanta and Kothawala, Dawood and Pesci, Alessandro},
	year={2019},
	month=oct, pages={134877} }

@article{Padmanabhan2015_DistributionAtoms,
	title={Distribution Function of the Atoms of Spacetime and the Nature of Gravity},
	volume={17},
	ISSN={1099-4300},
	url={http://dx.doi.org/10.3390/e17117420},
	DOI={10.3390/e17117420},
	number={11},
	journal={Entropy},
	publisher={MDPI AG},
	author={Padmanabhan, Thanu},
	year={2015},
	month=oct, pages={7420–7452} }

@article{Padmanabhan2019_GravityAndQG,
	title={Gravity and quantum theory: Domains of conflict and contact},
	volume={29},
	ISSN={1793-6594},
	url={http://dx.doi.org/10.1142/S0218271820300013},
	DOI={10.1142/s0218271820300013},
	number={01},
	journal={International Journal of Modern Physics D},
	publisher={World Scientific Pub Co Pte Lt},
	author={Padmanabhan, T.},
	year={2019},
	month=nov, pages={2030001} }

@article{Poisson2011_PointParticlesCurved,
	title={The Motion of Point Particles in Curved Spacetime},
	volume={14},
	ISSN={1433-8351},
	url={http://dx.doi.org/10.12942/lrr-2011-7},
	DOI={10.12942/lrr-2011-7},
	number={1},
	journal={Living Reviews in Relativity},
	publisher={Springer Science and Business Media LLC},
	author={Poisson, Eric and Pound, Adam and Vega, Ian},
	year={2011},
	month=sep }

@article{Kothawala2014_Finsleresque,
	title={Intrinsic and extrinsic curvatures in Finsler esque spaces},
	volume={46},
	ISSN={1572-9532},
	url={http://dx.doi.org/10.1007/s10714-014-1836-6},
	DOI={10.1007/s10714-014-1836-6},
	number={12},
	journal={General Relativity and Gravitation},
	publisher={Springer Science and Business Media LLC},
	author={Kothawala, Dawood},
	year={2014},
	month=nov }

@article{Bekenstein1993_PhysicalAndGravitationalGeometry,
	title={Relation between physical and gravitational geometry},
	volume={48},
	ISSN={0556-2821},
	url={http://dx.doi.org/10.1103/PhysRevD.48.3641},
	DOI={10.1103/physrevd.48.3641},
	number={8},
	journal={Physical Review D},
	publisher={American Physical Society (APS)},
	author={Bekenstein, Jacob D.},
	year={1993},
	month=oct, pages={3641–3647} }

@article{Kouretsis2013_RelativisticFinsler,
	title={Relativistic Finsler geometry},
	volume={37},
	ISSN={1099-1476},
	url={http://dx.doi.org/10.1002/mma.2919},
	DOI={10.1002/mma.2919},
	number={2},
	journal={Mathematical Methods in the Applied Sciences},
	publisher={Wiley},
	author={Kouretsis, A.P. and Stathakopoulos, M. and Stavrinos, P.C.},
	year={2013},
	month=aug, pages={223–229} }

@article{Carlip2012_ChallengesEmergentGravity,
	author = {Steven Carlip},
	doi = {10.1016/j.shpsb.2012.11.002},
	journal = {Studies in History and Philosophy of Science Part B: Studies in History and Philosophy of Modern Physics},
	number = {2},
	pages = {200--208},
	publisher = {Elsevier},
	title = {Challenges for Emergent Gravity},
	volume = {46},
	year = {2012}
}

@article{Ruse1931_TaylorTheorem,
	author = {Ruse, H. S.},
	title = {Taylor's Theorem in the Tensor Calculus},
	journal = {Proceedings of the London Mathematical Society},
	volume = {s2-32},
	number = {1},
	pages = {87-92},
	year = {1931},
	month = {01},
	issn = {0024-6115},
	doi = {10.1112/plms/s2-32.1.87},
	url = {https://doi.org/10.1112/plms/s2-32.1.87},
	eprint = {https://academic.oup.com/plms/article-pdf/s2-32/1/87/4240938/s2-32-1-87.pdf},
}

@article{DeWitt1960_RadiationDamping,
	title = {Radiation damping in a gravitational field},
	journal = {Annals of Physics},
	volume = {9},
	number = {2},
	pages = {220-259},
	year = {1960},
	issn = {0003-4916},
	doi = {https://doi.org/10.1016/0003-4916(60)90030-0},
	url = {https://www.sciencedirect.com/science/article/pii/0003491660900300},
	author = {Bryce S DeWitt and Robert W Brehme}
}

@article{Synge1931_Function,
	author = {Synge, J. L.},
	title = {A Characteristic Function in Riemannian Space and its Application to the Solution of Geodesic Triangles},
	journal = {Proceedings of the London Mathematical Society},
	volume = {s2-32},
	number = {1},
	pages = {241-258},
	doi = {https://doi.org/10.1112/plms/s2-32.1.241},
	url = {https://londmathsoc.onlinelibrary.wiley.com/doi/abs/10.1112/plms/s2-32.1.241},
	eprint = {https://londmathsoc.onlinelibrary.wiley.com/doi/pdf/10.1112/plms/s2-32.1.241},
	year = {1931}
}

@article{Synge1960_OpticalObservations,
	author = {Synge, J. L.},
	title = {Optical observations in general relativity},
	journal = {Rendiconti del Seminario Matematico e Fisico di Milano},
	volume = {30},
	number = {1},
	pages = {271-302},
	doi = {https://doi.org/10.1007/BF02923262},
	year = {1960}
}

@article{Visser1993_VanVleck,
	title = {van Vleck determinants: Geodesic focusing in Lorentzian spacetimes},
	author = {Visser, Matt},
	journal = {Phys. Rev. D},
	volume = {47},
	issue = {6},
	pages = {2395--2402},
	numpages = {0},
	year = {1993},
	month = {Mar},
	publisher = {American Physical Society},
	doi = {10.1103/PhysRevD.47.2395},
	url = {https://link.aps.org/doi/10.1103/PhysRevD.47.2395}
}

@article{Padmanabhan2020_GeodesicDistance,
	author = {Padmanabhan, T.},
	title = {Geodesic distance: A descriptor of geometry and correlator of pregeometric density of spacetime events},
	journal = {Modern Physics Letters A},
	volume = {35},
	number = {12},
	pages = {2030008},
	year = {2020},
	doi = {10.1142/S0217732320300086},
	URL = {https://doi.org/10.1142/S0217732320300086},
	eprint = {https://doi.org/10.1142/S0217732320300086}
}

@misc{Pesci2018_SpacetimeAtomsInLorentz,
	title={Looking at spacetime atoms from within the Lorentz sector}, 
	author={Alessandro Pesci},
	year={2018},
	eprint={1803.05726},
	archivePrefix={arXiv},
	primaryClass={gr-qc},
	url={https://arxiv.org/abs/1803.05726}, 
}

@article{VanVleck1928_CorrespondencePrinciple,
	ISSN = {00278424, 10916490},
	URL = {http://www.jstor.org/stable/85735},
	author = {J. H. Van Vleck},
	journal = {Proceedings of the National Academy of Sciences of the United States of America},
	number = {2},
	pages = {178--188},
	publisher = {National Academy of Sciences},
	title = {The Correspondence Principle in the Statistical Interpretation of Quantum Mechanics},
	urldate = {2025-03-13},
	volume = {14},
	year = {1928}
}

@article{Morette1951_ApproxFeynman,
	title = {On the Definition and Approximation of Feynman's Path Integrals},
	author = {Morette, C\'ecile},
	journal = {Phys. Rev.},
	volume = {81},
	issue = {5},
	pages = {848--852},
	numpages = {0},
	year = {1951},
	month = {Mar},
	publisher = {American Physical Society},
	doi = {10.1103/PhysRev.81.848},
	url = {https://link.aps.org/doi/10.1103/PhysRev.81.848}
}

@article{Padmanabhan2021_WordlinePropagator,
	title={World-Line Path Integral for the Propagator Expressed as an Ordinary Integral: Concept and Applications},
	volume={51},
	ISSN={1572-9516},
	url={http://dx.doi.org/10.1007/s10701-021-00447-8},
	DOI={10.1007/s10701-021-00447-8},
	number={2},
	journal={Foundations of Physics},
	publisher={Springer Science and Business Media LLC},
	author={Padmanabhan, T.},
	year={2021},
	month=mar }

@book{Poisson2004_RelativistToolkit, 
	place={Cambridge}, 
	title={A Relativist’s Toolkit: The Mathematics of Black-Hole Mechanics}, 
	publisher={Cambridge University Press}, 
	author={Poisson, Eric}, 
	year={2004}}

@misc{Visser2017_WickRotate,
	title={How to Wick rotate generic curved spacetime}, 
	author={Matt Visser},
	year={2017},
	eprint={1702.05572},
	archivePrefix={arXiv},
	primaryClass={gr-qc},
	url={https://arxiv.org/abs/1702.05572}, 
}

@article{Greensite1993_DynamicalOriginLorentzian,
	title={Dynamical origin of the lorentzian signature of spacetime},
	volume={300},
	ISSN={0370-2693},
	url={http://dx.doi.org/10.1016/0370-2693(93)90744-3},
	DOI={10.1016/0370-2693(93)90744-3},
	number={1–2},
	journal={Physics Letters B},
	publisher={Elsevier BV},
	author={Greensite, J.},
	year={1993},
	month=feb, pages={34–37} }

@article{Carlini1994_WhyLorentian,
	title={Why is spacetime Lorentzian?},
	volume={49},
	ISSN={0556-2821},
	url={http://dx.doi.org/10.1103/PhysRevD.49.866},
	DOI={10.1103/physrevd.49.866},
	number={2},
	journal={Physical Review D},
	publisher={American Physical Society (APS)},
	author={Carlini, A. and Greensite, J.},
	year={1994},
	month=jan, pages={866–878} }

@article{White2010_SignatureChange,
	title={Signature change events: a challenge for quantum gravity?},
	volume={27},
	ISSN={1361-6382},
	url={http://dx.doi.org/10.1088/0264-9381/27/4/045007},
	DOI={10.1088/0264-9381/27/4/045007},
	number={4},
	journal={Classical and Quantum Gravity},
	publisher={IOP Publishing},
	author={White, Angela and Weinfurtner, Silke and Visser, Matt},
	year={2010},
	month=jan, pages={045007} }

@article{Barbero1996_RealEuclidean,
	title={From Euclidean to Lorentzian general relativity: The real way},
	volume={54},
	ISSN={1089-4918},
	url={http://dx.doi.org/10.1103/PhysRevD.54.1492},
	DOI={10.1103/physrevd.54.1492},
	number={2},
	journal={Physical Review D},
	publisher={American Physical Society (APS)},
	author={G., J. Fernando Barbero},
	year={1996},
	month=jul, pages={1492–1499} }

@article{Chua2021_NoTime,
	title={No Time for Time from No-Time},
	volume={88},
	ISSN={1539-767X},
	url={http://dx.doi.org/10.1086/714870},
	DOI={10.1086/714870},
	number={5},
	journal={Philosophy of Science},
	publisher={Cambridge University Press (CUP)},
	author={Chua, Eugene Y. S. and Callender, Craig},
	year={2021},
	month=dec, pages={1172–1184} }

@article{Sakharov1984_CosmologicalTransitions,
	author = "Sakharov, A. D.",
	title = "{Cosmological Transitions With a Change in Metric Signature}",
	reportNumber = "SLAC-TRANS-0211",
	doi = "10.1070/PU1991v034n05ABEH002502",
	journal = "Sov. Phys. JETP",
	volume = "60",
	pages = "214--218",
	year = "1984"
}

@article{Weinfurtner2007_SignatureChangeBEC,
	title={Trans-Planckian physics and signature change events in Bose gas hydrodynamics},
	volume={76},
	ISSN={1550-2368},
	url={http://dx.doi.org/10.1103/PhysRevD.76.124008},
	DOI={10.1103/physrevd.76.124008},
	number={12},
	journal={Physical Review D},
	publisher={American Physical Society (APS)},
	author={Weinfurtner, Silke and White, Angela and Visser, Matt},
	year={2007},
	month=dec }

@article{Hawking1993_NoBoundary,
	title = {The no-boundary proposal and the arrow of time},
	journal = {Vistas in Astronomy},
	volume = {37},
	pages = {559-568},
	year = {1993},
	issn = {0083-6656},
	doi = {https://doi.org/10.1016/0083-6656(93)90096-3},
	url = {https://www.sciencedirect.com/science/article/pii/0083665693900963},
	author = {S.W. Hawking}
}

@mastersthesis{Perri2024_MinimumLength,
	title = {Minimum length metric and horizon area
	variation},
	url = {https://amslaurea.unibo.it/id/eprint/30867/},
	author = {Perri, Aldo},
	school = {Università di Bologna},
	year = {2024}
}

@article{Gemelli2002_nullGauss-Codazzi,
	title = {Observer-dependent Gauss–Codazzi formalism for null hypersurfaces in the space–time},
	journal = {Journal of Geometry and Physics},
	volume = {43},
	number = {4},
	pages = {371-383},
	year = {2002},
	issn = {0393-0440},
	doi = {https://doi.org/10.1016/S0393-0440(02)00025-6},
	url = {https://www.sciencedirect.com/science/article/pii/S0393044002000256},
	author = {Gianluca Gemelli}
}

@article{Wald201_TDofBH,
	title={The Thermodynamics of Black Holes},
	volume={4},
	ISSN={1433-8351},
	url={http://dx.doi.org/10.12942/lrr-2001-6},
	DOI={10.12942/lrr-2001-6},
	number={1},
	journal={Living Reviews in Relativity},
	publisher={Springer Science and Business Media LLC},
	author={Wald, Robert M.},
	year={2001},
	month=jul }

@article{Bardeen1973_LawsTDmechanics,
		author = {Bardeen, J. M. and Carter, B. and Hawking, S. W.},
		date = {1973/06/01},
		doi = {10.1007/BF01645742},
		id = {Bardeen1973},
		isbn = {1432-0916},
		journal = {Communications in Mathematical Physics},
		number = {2},
		pages = {161--170},
		title = {The four laws of black hole mechanics},
		url = {https://doi.org/10.1007/BF01645742},
		volume = {31},
		year = {1973}}

@article{Hawking1975_ParticleCreationBH,
author = {Hawking, S.  W. },
date = {1975/08/01},
doi = {10.1007/BF02345020},
id = {Hawking1975},
isbn = {1432-0916},
journal = {Communications in Mathematical Physics},
number = {3},
pages = {199--220},
title = {Particle creation by black holes},
url = {https://doi.org/10.1007/BF02345020},
volume = {43},
year = {1975}}

@article{Hawking1972_BHinGR,
	author = {Hawking, S.  W. },
	date = {1972/06/01},
	doi = {10.1007/BF01877517},
	id = {Hawking1972},
	isbn = {1432-0916},
	journal = {Communications in Mathematical Physics},
	number = {2},
	pages = {152--166},
	title = {Black holes in general relativity},
	url = {https://doi.org/10.1007/BF01877517},
	volume = {25},
	year = {1972}}

@article{Racz1996_KillingStationaryBH,
		title={Global extensions of spacetimes describing asymptotic final states of black holes},
		volume={13},
		ISSN={1361-6382},
		url={http://dx.doi.org/10.1088/0264-9381/13/3/017},
		DOI={10.1088/0264-9381/13/3/017},
		number={3},
		journal={Classical and Quantum Gravity},
		publisher={IOP Publishing},
		author={Rácz, István and Wald, Robert M},
		year={1996},
		month=mar, pages={539–552} }

@article{Carter1973_ZerothLawAxisymmetric,
		author = {Carter, Brandon},
		date = {2009/12/01},
		doi = {10.1007/s10714-009-0888-5},
		id = {Carter2009},
		isbn = {1572-9532},
		journal = {General Relativity and Gravitation},
		number = {12},
		pages = {2873--2938},
		title = {Republication of: Black hole equilibrium states},
		url = {https://doi.org/10.1007/s10714-009-0888-5},
		volume = {41},
		year = {2009}}

@article{Israel1986_ThirdLawTD,
	title = {Third Law of Black-Hole Dynamics: A Formulation and Proof},
	author = {Israel, W.},
	journal = {Phys. Rev. Lett.},
	volume = {57},
	issue = {4},
	pages = {397--399},
	numpages = {0},
	year = {1986},
	month = {Jul},
	publisher = {American Physical Society},
	doi = {10.1103/PhysRevLett.57.397},
	url = {https://link.aps.org/doi/10.1103/PhysRevLett.57.397}
}

@article{Hawking1971_AreaLaw,
	title = {Gravitational Radiation from Colliding Black Holes},
	author = {Hawking, S. W.},
	journal = {Phys. Rev. Lett.},
	volume = {26},
	issue = {21},
	pages = {1344--1346},
	numpages = {0},
	year = {1971},
	month = {May},
	publisher = {American Physical Society},
	doi = {10.1103/PhysRevLett.26.1344},
	url = {https://link.aps.org/doi/10.1103/PhysRevLett.26.1344}
}

@article{Isi2021_AreaLawGW,
	title = {Testing the Black-Hole Area Law with GW150914},
	author = {Isi, Maximiliano and Farr, Will M. and Giesler, Matthew and Scheel, Mark A. and Teukolsky, Saul A.},
	journal = {Phys. Rev. Lett.},
	volume = {127},
	issue = {1},
	pages = {011103},
	numpages = {4},
	year = {2021},
	month = {Jul},
	publisher = {American Physical Society},
	doi = {10.1103/PhysRevLett.127.011103},
	url = {https://link.aps.org/doi/10.1103/PhysRevLett.127.011103}
}

@misc{Curiel2014_HotClassicalBH,
	author = "Curiel, Erik",
	title = "{Classical Black Holes Are Hot}",
	eprint = "1408.3691",
	archivePrefix = "arXiv",
	primaryClass = "gr-qc",
	month = "8",
	year = "2014"
}

@article{Wald1975_BHParticleCreation,
	author = {Wald, Robert M. },
	date = {1975/02/01},
	doi = {10.1007/BF01609863},
	id = {Wald1975},
	isbn = {1432-0916},
	journal = {Communications in Mathematical Physics},
	number = {1},
	pages = {9--34},
	title = {On particle creation by black holes},
	url = {https://doi.org/10.1007/BF01609863},
	volume = {45},
	year = {1975}}

@article{Gibbons1977_EuclideanBH,
	title = {Action integrals and partition functions in quantum gravity},
	author = {Gibbons, G. W. and Hawking, S. W.},
	journal = {Phys. Rev. D},
	volume = {15},
	issue = {10},
	pages = {2752--2756},
	numpages = {0},
	year = {1977},
	month = {May},
	publisher = {American Physical Society},
	doi = {10.1103/PhysRevD.15.2752},
	url = {https://link.aps.org/doi/10.1103/PhysRevD.15.2752}
}

@article{Parik2000_HawkingRadiationasTunnelling,
	title = {Hawking Radiation As Tunneling},
	author = {Parikh, Maulik K. and Wilczek, Frank},
	journal = {Phys. Rev. Lett.},
	volume = {85},
	issue = {24},
	pages = {5042--5045},
	numpages = {0},
	year = {2000},
	month = {Dec},
	publisher = {American Physical Society},
	doi = {10.1103/PhysRevLett.85.5042},
	url = {https://link.aps.org/doi/10.1103/PhysRevLett.85.5042}
}

@article{Bekenstein1973_BHentropy,
	title = {Black Holes and Entropy},
	author = {Bekenstein, Jacob D.},
	journal = {Phys. Rev. D},
	volume = {7},
	issue = {8},
	pages = {2333--2346},
	numpages = {0},
	year = {1973},
	month = {Apr},
	publisher = {American Physical Society},
	doi = {10.1103/PhysRevD.7.2333},
	url = {https://link.aps.org/doi/10.1103/PhysRevD.7.2333}
}

@article{Bekenstein1972_BHSecondLaw,
	author = {Bekenstein, J.  D. },
	date = {1972/08/01},
	doi = {10.1007/BF02757029},
	id = {Bekenstein1972},
	isbn = {1827-613X},
	journal = {Lettere al Nuovo Cimento (1971-1985)},
	number = {15},
	pages = {737--740},
	title = {Black holes and the second law},
	url = {https://doi.org/10.1007/BF02757029},
	volume = {4},
	year = {1972}}

@article{Robinson2005_HawkingfromAnomalies,
	title = {Relationship between Hawking Radiation and Gravitational Anomalies},
	author = {Robinson, Sean P. and Wilczek, Frank},
	journal = {Phys. Rev. Lett.},
	volume = {95},
	issue = {1},
	pages = {011303},
	numpages = {4},
	year = {2005},
	month = {Jun},
	publisher = {American Physical Society},
	doi = {10.1103/PhysRevLett.95.011303},
	url = {https://link.aps.org/doi/10.1103/PhysRevLett.95.011303}
}

@article{Brown1993_SBHMicrocanonical,
	title = {Microcanonical functional integral for the gravitational field},
	author = {Brown, J. David and York, James W.},
	journal = {Phys. Rev. D},
	volume = {47},
	issue = {4},
	pages = {1420--1431},
	numpages = {0},
	year = {1993},
	month = {Feb},
	publisher = {American Physical Society},
	doi = {10.1103/PhysRevD.47.1420},
	url = {https://link.aps.org/doi/10.1103/PhysRevD.47.1420}
}

@article{Bombelli1986_BHEntanglementEntropy,
	title = {Quantum source of entropy for black holes},
	author = {Bombelli, Luca and Koul, Rabinder K. and Lee, Joohan and Sorkin, Rafael D.},
	journal = {Phys. Rev. D},
	volume = {34},
	issue = {2},
	pages = {373--383},
	numpages = {0},
	year = {1986},
	month = {Jul},
	publisher = {American Physical Society},
	doi = {10.1103/PhysRevD.34.373},
	url = {https://link.aps.org/doi/10.1103/PhysRevD.34.373}
}

@article{Callan1994_BHGeometricEntropy,
	title={On geometric entropy},
	volume={333},
	ISSN={0370-2693},
	url={http://dx.doi.org/10.1016/0370-2693(94)91007-3},
	DOI={10.1016/0370-2693(94)91007-3},
	number={1–2},
	journal={Physics Letters B},
	publisher={Elsevier BV},
	author={Callan, Curtis and Wilczek, Frank},
	year={1994},
	month=jul, pages={55–61} }

@article{Srednicki1993_EntropyAndArea,
	title={Entropy and area},
	volume={71},
	ISSN={0031-9007},
	url={http://dx.doi.org/10.1103/PhysRevLett.71.666},
	DOI={10.1103/physrevlett.71.666},
	number={5},
	journal={Physical Review Letters},
	publisher={American Physical Society (APS)},
	author={Srednicki, Mark},
	year={1993},
	month=aug, pages={666–669} }

@article{Wald1993_BHEntropyAsNoetherCharge,
	title={Black hole entropy is the Noether charge},
	volume={48},
	ISSN={0556-2821},
	url={http://dx.doi.org/10.1103/PhysRevD.48.R3427},
	DOI={10.1103/physrevd.48.r3427},
	number={8},
	journal={Physical Review D},
	publisher={American Physical Society (APS)},
	author={Wald, Robert M.},
	year={1993},
	month=oct, pages={R3427–R3431} }

@article{Strominger1996_MicroBHEntropy,
	title={Microscopic origin of the Bekenstein-Hawking entropy},
	volume={379},
	ISSN={0370-2693},
	url={http://dx.doi.org/10.1016/0370-2693(96)00345-0},
	DOI={10.1016/0370-2693(96)00345-0},
	number={1–4},
	journal={Physics Letters B},
	publisher={Elsevier BV},
	author={Strominger, Andrew and Vafa, Cumrun},
	year={1996},
	month=jun, pages={99–104} }

@article{Rovelli1996_BHEntropyInLQG,
	author = "Rovelli, Carlo",
	title = "{Black hole entropy from loop quantum gravity}",
	eprint = "gr-qc/9603063",
	archivePrefix = "arXiv",
	doi = "10.1103/PhysRevLett.77.3288",
	journal = "Phys. Rev. Lett.",
	volume = "77",
	pages = "3288--3291",
	year = "1996"
}

@article{Ashtekar1998_BHEntropyInLQG,
	title={Quantum Geometry and Black Hole Entropy},
	volume={80},
	ISSN={1079-7114},
	url={http://dx.doi.org/10.1103/PhysRevLett.80.904},
	DOI={10.1103/physrevlett.80.904},
	number={5},
	journal={Physical Review Letters},
	publisher={American Physical Society (APS)},
	author={Ashtekar, A. and Baez, J. and Corichi, A. and Krasnov, K.},
	year={1998},
	month=feb, pages={904–907} }

@article{Domagala2004_BHEntropyLQGCorrections,
	title={Black-hole entropy from quantum geometry},
	volume={21},
	ISSN={1361-6382},
	url={http://dx.doi.org/10.1088/0264-9381/21/22/014},
	DOI={10.1088/0264-9381/21/22/014},
	number={22},
	journal={Classical and Quantum Gravity},
	publisher={IOP Publishing},
	author={Domagala, Marcin and Lewandowski, Jerzy},
	year={2004},
	month=oct, pages={5233–5243} }

@article{Meissner2004_BHEntropyLQGCorrections,
	title={Black-hole entropy in loop quantum gravity},
	volume={21},
	ISSN={1361-6382},
	url={http://dx.doi.org/10.1088/0264-9381/21/22/015},
	DOI={10.1088/0264-9381/21/22/015},
	number={22},
	journal={Classical and Quantum Gravity},
	publisher={IOP Publishing},
	author={Meissner, Krzysztof A},
	year={2004},
	month=oct, pages={5245–5251} }

@article{Carlip1999_SBHfromCFT,
	title={Entropy from conformal field theory at Killing horizons},
	volume={16},
	ISSN={1361-6382},
	url={http://dx.doi.org/10.1088/0264-9381/16/10/322},
	DOI={10.1088/0264-9381/16/10/322},
	number={10},
	journal={Classical and Quantum Gravity},
	publisher={IOP Publishing},
	author={Carlip, S},
	year={1999},
	month=sep, pages={3327–3348} }

@article{Carlip2000_SBHfromCFT,
	title={Black hole entropy from horizon conformal field theory},
	volume={88},
	ISSN={0920-5632},
	url={http://dx.doi.org/10.1016/S0920-5632(00)00748-9},
	DOI={10.1016/s0920-5632(00)00748-9},
	number={1–3},
	journal={Nuclear Physics B - Proceedings Supplements},
	publisher={Elsevier BV},
	author={Carlip, S.},
	year={2000},
	month=jun, pages={10–16} }

@article{Carlip2015_SBHfromCFTvsLQG,
	title={A note on black hole entropy in loop quantum gravity},
	volume={32},
	ISSN={1361-6382},
	url={http://dx.doi.org/10.1088/0264-9381/32/15/155009},
	DOI={10.1088/0264-9381/32/15/155009},
	number={15},
	journal={Classical and Quantum Gravity},
	publisher={IOP Publishing},
	author={Carlip, S},
	year={2015},
	month=jul, pages={155009} }

@article{Carlip2000_SBHCorrections,
	title={Logarithmic corrections to black hole entropy, from the Cardy formula},
	volume={17},
	ISSN={1361-6382},
	url={http://dx.doi.org/10.1088/0264-9381/17/20/302},
	DOI={10.1088/0264-9381/17/20/302},
	number={20},
	journal={Classical and Quantum Gravity},
	publisher={IOP Publishing},
	author={Carlip, S},
	year={2000},
	month=sep, pages={4175–4186} }

@misc{Curiel2025_InfoLoss,
	author       = {Curiel, Eric},
	title        = {{[Lecture Slides] 'Information Loss' is said in many ways, the responses legion, not all cogent}},
	howpublished = {SIGRAV International School 2025 "Quantum effects in curved spacetiems and the thermodynamics of horizons"},
	month        = jun,
	year         = {2025},
	url          = {https://agenda.infn.it/event/43787/sessions/31546/attachments/131713/196510/Slides.pdf}
}

@article{Hayward1994_NonFlatBHTD,
	title = {General laws of black-hole dynamics},
	author = {Hayward, Sean A.},
	journal = {Phys. Rev. D},
	volume = {49},
	issue = {12},
	pages = {6467--6474},
	numpages = {0},
	year = {1994},
	month = {Jun},
	publisher = {American Physical Society},
	doi = {10.1103/PhysRevD.49.6467},
	url = {https://link.aps.org/doi/10.1103/PhysRevD.49.6467}
}

@article{Ashtekar2002_IsolatedHorizons,
	title={Geometry of generic isolated horizons},
	volume={19},
	ISSN={1361-6382},
	url={http://dx.doi.org/10.1088/0264-9381/19/6/311},
	DOI={10.1088/0264-9381/19/6/311},
	number={6},
	journal={Classical and Quantum Gravity},
	publisher={IOP Publishing},
	author={Ashtekar, Abhay and Beetle, Christopher and Lewandowski, Jerzy},
	year={2002},
	month=mar, pages={1195–1225} }

@article{Ashtekar2002_DynamicalHorizons,
	title={Dynamical Horizons: Energy, Angular Momentum, Fluxes, and Balance Laws},
	volume={89},
	ISSN={1079-7114},
	url={http://dx.doi.org/10.1103/PhysRevLett.89.261101},
	DOI={10.1103/physrevlett.89.261101},
	number={26},
	journal={Physical Review Letters},
	publisher={American Physical Society (APS)},
	author={Ashtekar, Abhay and Krishnan, Badri},
	year={2002},
	month=dec }

@article{Ashtekar2000_IsolatedHorizonsFirstLaws,
	title={Isolated horizons: Hamiltonian evolution and the first law},
	volume={62},
	ISSN={1089-4918},
	url={http://dx.doi.org/10.1103/PhysRevD.62.104025},
	DOI={10.1103/physrevd.62.104025},
	number={10},
	journal={Physical Review D},
	publisher={American Physical Society (APS)},
	author={Ashtekar, Abhay and Fairhurst, Stephen and Krishnan, Badri},
	year={2000},
	month=oct }

@article{Ashtekar2001_IsolatedHorizonRotating,
	title = {Mechanics of rotating isolated horizons},
	author = {Ashtekar, Abhay and Beetle, Christopher and Lewandowski, Jerzy},
	journal = {Phys. Rev. D},
	volume = {64},
	issue = {4},
	pages = {044016},
	numpages = {17},
	year = {2001},
	month = {Jul},
	publisher = {American Physical Society},
	doi = {10.1103/PhysRevD.64.044016},
	url = {https://link.aps.org/doi/10.1103/PhysRevD.64.044016}
}

@article{Gibbons1977_CosmologicalHorizons,
	title = {Cosmological event horizons, thermodynamics, and particle creation},
	author = {Gibbons, G. W. and Hawking, S. W.},
	journal = {Phys. Rev. D},
	volume = {15},
	issue = {10},
	pages = {2738--2751},
	numpages = {0},
	year = {1977},
	month = {May},
	publisher = {American Physical Society},
	doi = {10.1103/PhysRevD.15.2738},
	url = {https://link.aps.org/doi/10.1103/PhysRevD.15.2738}
}

@article{Unruh1976_UnruhEffect,
	title = {Notes on black-hole evaporation},
	author = {Unruh, W. G.},
	journal = {Phys. Rev. D},
	volume = {14},
	issue = {4},
	pages = {870--892},
	numpages = {0},
	year = {1976},
	month = {Aug},
	publisher = {American Physical Society},
	doi = {10.1103/PhysRevD.14.870},
	url = {https://link.aps.org/doi/10.1103/PhysRevD.14.870}
}

@article{Padmanabhan2019_RindlerTDfromPropagator,
	title={Thermality of the Rindler horizon: A simple derivation from the structure of the inertial propagator},
	volume={100},
	ISSN={2470-0029},
	url={http://dx.doi.org/10.1103/PhysRevD.100.045024},
	DOI={10.1103/physrevd.100.045024},
	number={4},
	journal={Physical Review D},
	publisher={American Physical Society (APS)},
	author={Padmanabhan, T.},
	year={2019},
	month=aug }

@article{Visser2002_SakharovModern,
	title={SAKHAROV’S INDUCED GRAVITY: A MODERN PERSPECTIVE},
	volume={17},
	ISSN={1793-6632},
	url={http://dx.doi.org/10.1142/S0217732302006886},
	DOI={10.1142/s0217732302006886},
	number={15n17},
	journal={Modern Physics Letters A},
	publisher={World Scientific Pub Co Pte Lt},
	author={VISSER, MATT},
	year={2002},
	month=jun, pages={977–991} }

@article{Sakharov1991_InducedGravity,
	doi = {10.1070/PU1991v034n05ABEH002498},
	url = {https://dx.doi.org/10.1070/PU1991v034n05ABEH002498},
	year = {1991},
	month = {may},
	publisher = {},
	volume = {34},
	number = {5},
	pages = {394},
	author = {Andrei D Sakharov},
	title = {Vacuum quantum fluctuations in curved space and the theory of gravitation},
	journal = {Soviet Physics Uspekhi}
}

@article{Jacobson1995_EinsteinEqofState,
	title={Thermodynamics of Spacetime: The Einstein Equation of State},
	volume={75},
	ISSN={1079-7114},
	url={http://dx.doi.org/10.1103/PhysRevLett.75.1260},
	DOI={10.1103/physrevlett.75.1260},
	number={7},
	journal={Physical Review Letters},
	publisher={American Physical Society (APS)},
	author={Jacobson, Ted},
	year={1995},
	month=aug, pages={1260–1263} }

@article{Eling2006_NonEquilibriumTD,
	title={Nonequilibrium Thermodynamics of Spacetime},
	volume={96},
	ISSN={1079-7114},
	url={http://dx.doi.org/10.1103/PhysRevLett.96.121301},
	DOI={10.1103/physrevlett.96.121301},
	number={12},
	journal={Physical Review Letters},
	publisher={American Physical Society (APS)},
	author={Eling, Christopher and Guedens, Raf and Jacobson, Ted},
	year={2006},
	month=mar }

@article{Verlinde2011_EntropicGravity,
	title={On the origin of gravity and the laws of Newton},
	volume={2011},
	ISSN={1029-8479},
	url={http://dx.doi.org/10.1007/JHEP04(2011)029},
	DOI={10.1007/jhep04(2011)029},
	number={4},
	journal={Journal of High Energy Physics},
	publisher={Springer Science and Business Media LLC},
	author={Verlinde, Erik},
	year={2011},
	month=apr }

@article{Verlinde2017_EntropicGravityDM,
	title={Emergent Gravity and the Dark Universe},
	volume={2},
	ISSN={2542-4653},
	url={http://dx.doi.org/10.21468/SciPostPhys.2.3.016},
	DOI={10.21468/scipostphys.2.3.016},
	number={3},
	journal={SciPost Physics},
	publisher={Stichting SciPost},
	author={Verlinde, Erik P.},
	year={2017},
	month=may }

@article{Hossenfelder2017_CovariantVerlinde,
	title={Covariant version of Verlinde’s emergent gravity},
	volume={95},
	ISSN={2470-0029},
	url={http://dx.doi.org/10.1103/PhysRevD.95.124018},
	DOI={10.1103/physrevd.95.124018},
	number={12},
	journal={Physical Review D},
	publisher={American Physical Society (APS)},
	author={Hossenfelder, Sabine},
	year={2017},
	month=jun }

@article{Dai2017_CorrectionToHossenfelder,
	title={Comment on “Covariant version of Verlinde’s emergent gravity”},
	volume={96},
	ISSN={2470-0029},
	url={http://dx.doi.org/10.1103/PhysRevD.96.108501},
	DOI={10.1103/physrevd.96.108501},
	number={10},
	journal={Physical Review D},
	publisher={American Physical Society (APS)},
	author={Dai, De-Chang and Stojkovic, Dejan},
	year={2017},
	month=nov }

@article{Visser2011_ConservativeEntropicForce,
	title={Conservative entropic forces},
	volume={2011},
	ISSN={1029-8479},
	url={http://dx.doi.org/10.1007/JHEP10(2011)140},
	DOI={10.1007/jhep10(2011)140},
	number={10},
	journal={Journal of High Energy Physics},
	publisher={Springer Science and Business Media LLC},
	author={Visser, Matt},
	year={2011},
	month=oct }

@article{Kobakhidze2011_GravityNotEntropic,
	title={Gravity is not an entropic force},
	volume={83},
	ISSN={1550-2368},
	url={http://dx.doi.org/10.1103/PhysRevD.83.021502},
	DOI={10.1103/physrevd.83.021502},
	number={2},
	journal={Physical Review D},
	publisher={American Physical Society (APS)},
	author={Kobakhidze, Archil},
	year={2011},
	month=jan }

@misc{Kobakhidze2011_MoreGravityNotEntropic,
	title={Once more: gravity is not an entropic force}, 
	author={Archil Kobakhidze},
	year={2011},
	eprint={1108.4161},
	archivePrefix={arXiv},
	primaryClass={hep-th},
	url={https://arxiv.org/abs/1108.4161}, 
}

@article{Dai2017_IncostistenciesInVerlinde,
	title={Inconsistencies in Verlinde’s emergent gravity},
	volume={2017},
	ISSN={1029-8479},
	url={http://dx.doi.org/10.1007/JHEP11(2017)007},
	DOI={10.1007/jhep11(2017)007},
	number={11},
	journal={Journal of High Energy Physics},
	publisher={Springer Science and Business Media LLC},
	author={Dai, De-Chang and Stojkovic, Dejan},
	year={2017},
	month=nov }

@Article{Gao2011_IsGravityEntropic,
	AUTHOR = {Gao, Shan},
	TITLE = {Is Gravity an Entropic Force?},
	JOURNAL = {Entropy},
	VOLUME = {13},
	YEAR = {2011},
	NUMBER = {5},
	PAGES = {936--948},
	URL = {https://www.mdpi.com/1099-4300/13/5/936},
	ISSN = {1099-4300},
	DOI = {10.3390/e13050936}
}

@article{Nesvizhevsky2002_GravitationalUltracoldNeutrons,
	author = {Nesvizhevsky, Valery V. and B{\"o}rner, Hans G. and Petukhov, Alexander K. and Abele, Hartmut and Bae{\ss}ler, Stefan and Rue{\ss}, Frank J. and St{\"o}ferle, Thilo and Westphal, Alexander and Gagarski, Alexei M. and Petrov, Guennady A. and Strelkov, Alexander V.},
	date = {2002/01/01},
	doi = {10.1038/415297a},
	id = {Nesvizhevsky2002},
	isbn = {1476-4687},
	journal = {Nature},
	number = {6869},
	pages = {297--299},
	title = {Quantum states of neutrons in the Earth's gravitational field},
	url = {https://doi.org/10.1038/415297a},
	volume = {415},
	year = {2002}}

@article{Padmanabhan2004_GravityAsElasticity,
	title={GRAVITY AS ELASTICITY OF SPACETIME: A PARADIGM TO UNDERSTAND HORIZON THERMODYNAMICS AND COSMOLOGICAL CONSTANT},
	volume={13},
	ISSN={1793-6594},
	url={http://dx.doi.org/10.1142/S0218271804006358},
	DOI={10.1142/s0218271804006358},
	number={10},
	journal={International Journal of Modern Physics D},
	publisher={World Scientific Pub Co Pte Lt},
	author={PADMANABHAN, T.},
	year={2004},
	month=dec, pages={2293–2298} }

@article{Padmanabhan2007_EntropyNullSurfaces,
	title={Entropy of null surfaces and dynamics of spacetime},
	volume={75},
	ISSN={1550-2368},
	url={http://dx.doi.org/10.1103/PhysRevD.75.064004},
	DOI={10.1103/physrevd.75.064004},
	number={6},
	journal={Physical Review D},
	publisher={American Physical Society (APS)},
	author={Padmanabhan, T. and Paranjape, Aseem},
	year={2007},
	month=mar }

@article{Padmanabhan2013_LanczosLovelockModels,
	title={Lanczos–Lovelock models of gravity},
	volume={531},
	ISSN={0370-1573},
	url={http://dx.doi.org/10.1016/j.physrep.2013.05.007},
	DOI={10.1016/j.physrep.2013.05.007},
	number={3},
	journal={Physics Reports},
	publisher={Elsevier BV},
	author={Padmanabhan, T. and Kothawala, D.},
	year={2013},
	month=oct, pages={115–171} }

@misc{Padmanabhan2012_SpacetimeEmergence,
	title={Emergence and Expansion of Cosmic Space as due to the Quest for Holographic Equipartition}, 
	author={T. Padmanabhan},
	year={2012},
	eprint={1206.4916},
	archivePrefix={arXiv},
	primaryClass={hep-th},
	url={https://arxiv.org/abs/1206.4916}, 
}

@article{Padmanabhan2010_HorizonEquipartition,
	title={EQUIPARTITION OF ENERGY IN THE HORIZON DEGREES OF FREEDOM AND THE EMERGENCE OF GRAVITY},
	volume={25},
	ISSN={1793-6632},
	url={http://dx.doi.org/10.1142/S021773231003313X},
	DOI={10.1142/s021773231003313x},
	number={14},
	journal={Modern Physics Letters A},
	publisher={World Scientific Pub Co Pte Lt},
	author={PADMANABHAN, T.},
	year={2010},
	month=may, pages={1129–1136} }

@article{Padmanabhan2010_HorizonEquipartitionDoF,
	title={Surface density of spacetime degrees of freedom from equipartition law in theories of gravity},
	volume={81},
	ISSN={1550-2368},
	url={http://dx.doi.org/10.1103/PhysRevD.81.124040},
	DOI={10.1103/physrevd.81.124040},
	number={12},
	journal={Physical Review D},
	publisher={American Physical Society (APS)},
	author={Padmanabhan, T.},
	year={2010},
	month=jun }

@article{Padmanabhan2012_EmergentGravityDE,
	title={Emergent perspective of gravity and dark energy},
	volume={12},
	ISSN={1674-4527},
	url={http://dx.doi.org/10.1088/1674-4527/12/8/003},
	DOI={10.1088/1674-4527/12/8/003},
	number={8},
	journal={Research in Astronomy and Astrophysics},
	publisher={IOP Publishing},
	author={Padmanabhan, T.},
	year={2012},
	month=aug, pages={891–916} }

@misc{Tong2012_StringTheory,
		title={Lectures on String Theory}, 
		author={David Tong},
		year={2012},
		eprint={0908.0333},
		archivePrefix={arXiv},
		primaryClass={hep-th},
		url={https://arxiv.org/abs/0908.0333}, 
	}

@article{Mukhi2011_StringUpdates,
		doi = {10.1088/0264-9381/28/15/153001},
		url = {https://dx.doi.org/10.1088/0264-9381/28/15/153001},
		year = {2011},
		month = {jun},
		publisher = {},
		volume = {28},
		number = {15},
		pages = {153001},
		author = {Mukhi, Sunil},
		title = {String theory: a perspective over the last 25 years},
		journal = {Classical and Quantum Gravity}
	}

@article{Reuter2012_AsymptoticSafeGravity,
		title={Quantum Einstein gravity},
		volume={14},
		ISSN={1367-2630},
		url={http://dx.doi.org/10.1088/1367-2630/14/5/055022},
		DOI={10.1088/1367-2630/14/5/055022},
		number={5},
		journal={New Journal of Physics},
		publisher={IOP Publishing},
		author={Reuter, Martin and Saueressig, Frank},
		year={2012},
		month=may, pages={055022} }

@article{Rovelli2008_LoopQuantumGravity,
	author = {Rovelli, Carlo},
	date = {2008/07/15},
	doi = {10.12942/lrr-2008-5},
	id = {Rovelli2008},
	isbn = {1433-8351},
	journal = {Living Reviews in Relativity},
	number = {1},
	pages = {5},
	title = {Loop Quantum Gravity},
	url = {https://doi.org/10.12942/lrr-2008-5},
	volume = {11},
	year = {2008}}

@article{Ashtekar2021_LQGUpdate,
		title={A short review of loop quantum gravity},
		volume={84},
		ISSN={1361-6633},
		url={http://dx.doi.org/10.1088/1361-6633/abed91},
		DOI={10.1088/1361-6633/abed91},
		number={4},
		journal={Reports on Progress in Physics},
		publisher={IOP Publishing},
		author={Ashtekar, Abhay and Bianchi, Eugenio},
		year={2021},
		month=mar, pages={042001} }

@article{Chamseddine2023_NonCummutativeUpdate,
	title={Noncommutativity and physics: a non-technical review},
	volume={232},
	ISSN={1951-6401},
	url={http://dx.doi.org/10.1140/epjs/s11734-023-00842-4},
	DOI={10.1140/epjs/s11734-023-00842-4},
	number={23–24},
	journal={The European Physical Journal Special Topics},
	publisher={Springer Science and Business Media LLC},
	author={Chamseddine, Ali H. and Connes, Alain and van Suijlekom, Walter D.},
	year={2023},
	month=may, pages={3581–3588} }

@article{Hinchliffe2004_NonCommutativePhenomenology,
	title={REVIEW OF THE PHENOMENOLOGY OF NONCOMMUTATIVE GEOMETRY},
	volume={19},
	ISSN={1793-656X},
	url={http://dx.doi.org/10.1142/S0217751X04017094},
	DOI={10.1142/s0217751x04017094},
	number={02},
	journal={International Journal of Modern Physics A},
	publisher={World Scientific Pub Co Pte Lt},
	author={HINCHLIFFE, I. and KERSTING, N. and MA, Y. L.},
	year={2004},
	month=jan, pages={179–204} }

@article{Mead1964_HeisenbergMicroscopeQG,
	title = {Possible Connection Between Gravitation and Fundamental Length},
	author = {Mead, C. Alden},
	journal = {Phys. Rev.},
	volume = {135},
	issue = {3B},
	pages = {B849--B862},
	numpages = {0},
	year = {1964},
	month = {Aug},
	publisher = {American Physical Society},
	doi = {10.1103/PhysRev.135.B849},
	url = {https://link.aps.org/doi/10.1103/PhysRev.135.B849}
}

@article{Scardigli1999_GUPMicroBH,
	title={Generalized uncertainty principle in quantum gravity from micro-black hole gedanken experiment},
	volume={452},
	ISSN={0370-2693},
	url={http://dx.doi.org/10.1016/S0370-2693(99)00167-7},
	DOI={10.1016/s0370-2693(99)00167-7},
	number={1–2},
	journal={Physics Letters B},
	publisher={Elsevier BV},
	author={Scardigli, Fabio},
	year={1999},
	month=apr, pages={39–44} }

@article{Salecker1958_LimitDistanceMeasurements,
	title = {Quantum Limitations of the Measurement of Space-Time Distances},
	author = {Salecker, H. and Wigner, E. P.},
	journal = {Phys. Rev.},
	volume = {109},
	issue = {2},
	pages = {571--577},
	numpages = {0},
	year = {1958},
	month = {Jan},
	publisher = {American Physical Society},
	doi = {10.1103/PhysRev.109.571},
	url = {https://link.aps.org/doi/10.1103/PhysRev.109.571}
}

@article{Pesci2007_BoussoBoundHydrodynamic,
	title={From Unruh temperature to the generalized Bousso bound},
	volume={24},
	ISSN={1361-6382},
	url={http://dx.doi.org/10.1088/0264-9381/24/24/005},
	DOI={10.1088/0264-9381/24/24/005},
	number={24},
	journal={Classical and Quantum Gravity},
	publisher={IOP Publishing},
	author={Pesci, Alessandro},
	year={2007},
	month=nov, pages={6219–6225} }

@inproceedings{Pesci2009_KSSBound,
		author = "Pesci, Alessandro",
		title = "{A Semiclassical approach to eta/s bound through holography}",
		booktitle = "{12th Marcel Grossmann Meeting on General Relativity}",
		eprint = "0910.0766",
		archivePrefix = "arXiv",
		primaryClass = "hep-th",
		doi = "10.1142/9789814374552_0469",
		pages = "2324--2326",
		month = "10",
		year = "2009"
	}

@article{Hod2007_RelaxationBound,
		title={Universal bound on dynamical relaxation times and black-hole quasinormal ringing},
		volume={75},
		ISSN={1550-2368},
		url={http://dx.doi.org/10.1103/PhysRevD.75.064013},
		DOI={10.1103/physrevd.75.064013},
		number={6},
		journal={Physical Review D},
		publisher={American Physical Society (APS)},
		author={Hod, Shahar},
		year={2007},
		month=mar }

@article{Amati1989_StringGUP,
	author        = "Amati, Daniele and Ciafaloni, Marcello and Veneziano,
	Gabriele",
	title         = "{Can space time be probed below the string size?}",
	reportNumber  = "CERN-TH-5207-88",
	journal       = "Phys. Lett. B",
	volume        = "216",
	pages         = "41-47",
	year          = "1989",
	url           = "https://cds.cern.ch/record/191788",
	doi           = "10.1016/0370-2693(89)91366-X",
}

@article{Konishi1989_StringGUPRenormalizationGroup,
	author = "Konishi, Kenichi and Paffuti, Giampiero and Provero, Paolo",
	title = "{Minimum Physical Length and the Generalized Uncertainty Principle in String Theory}",
	reportNumber = "IFUP-TH-46-89, GEF-TH-89-9",
	doi = "10.1016/0370-2693(90)91927-4",
	journal = "Phys. Lett. B",
	volume = "234",
	pages = "276--284",
	year = "1990"
}

@article{Amati1987_StringPlanckianScattering,
	author        = "Amati, Daniele and Ciafaloni, Marcello and Veneziano,
	Gabriele",
	title         = "{Superstring collisions at Planckian energies}",
	reportNumber  = "CERN-TH-4782-87",
	journal       = "Phys. Lett. B",
	volume        = "197",
	pages         = "81-88",
	year          = "1987",
	url           = "https://cds.cern.ch/record/179623",
	doi           = "10.1016/0370-2693(87)90346-7",
}

@article{Rovelli1995_LQGDiscreteArea,
	title={Discreteness of area and volume in quantum gravity},
	volume={442},
	ISSN={0550-3213},
	url={http://dx.doi.org/10.1016/0550-3213(95)00150-Q},
	DOI={10.1016/0550-3213(95)00150-q},
	number={3},
	journal={Nuclear Physics B},
	publisher={Elsevier BV},
	author={Rovelli, Carlo and Smolin, Lee},
	year={1995},
	month=may, pages={593–619} }

@article{Percacci2010_ASGMinimalLength,
	title={Asymptotic safety, emergence and minimal length},
	volume={27},
	ISSN={1361-6382},
	url={http://dx.doi.org/10.1088/0264-9381/27/24/245026},
	DOI={10.1088/0264-9381/27/24/245026},
	number={24},
	journal={Classical and Quantum Gravity},
	publisher={IOP Publishing},
	author={Percacci, Roberto and Vacca, Gian Paolo},
	year={2010},
	month=nov, pages={245026} }

@article{Pesci2025_MinLengthHorizons,
		doi = {10.1088/1742-6596/3017/1/012024},
		url = {https://dx.doi.org/10.1088/1742-6596/3017/1/012024},
		year = {2025},
		month = {jun},
		publisher = {IOP Publishing},
		volume = {3017},
		number = {1},
		pages = {012024},
		author = {Pesci, Alessandro},
		title = {Small-scale metric structure and horizons: Probing the nature of gravity},
		journal = {Journal of Physics: Conference Series}
	}

@article{Bousso1999_CovariantEntropy,
   title={A covariant entropy conjecture},
   volume={1999},
   ISSN={1029-8479},
   url={http://dx.doi.org/10.1088/1126-6708/1999/07/004},
   DOI={10.1088/1126-6708/1999/07/004},
   number={07},
   journal={Journal of High Energy Physics},
   publisher={Springer Science and Business Media LLC},
   author={Bousso, Raphael},
   year={1999},
   month=jul, pages={004–004} }

@article{Bousso2002_HolographicPrinciple,
   title={The holographic principle},
   volume={74},
   ISSN={1539-0756},
   url={http://dx.doi.org/10.1103/RevModPhys.74.825},
   DOI={10.1103/revmodphys.74.825},
   number={3},
   journal={Reviews of Modern Physics},
   publisher={American Physical Society (APS)},
   author={Bousso, Raphael},
   year={2002},
   month=aug, pages={825–874} }

@article{Padmanabhan2016_DimReduction,
   title={Spacetime with zero point length is two-dimensional at the Planck scale},
   volume={48},
   ISSN={1572-9532},
   url={http://dx.doi.org/10.1007/s10714-016-2053-2},
   DOI={10.1007/s10714-016-2053-2},
   number={5},
   journal={General Relativity and Gravitation},
   publisher={Springer Science and Business Media LLC},
   author={Padmanabhan, T. and Chakraborty, Sumanta and Kothawala, Dawood},
   year={2016},
   month=apr }

@inproceedings{Carlip2009_DimReduction,
   title={Spontaneous Dimensional Reduction in Short-Distance Quantum Gravity?},
   ISSN={0094-243X},
   url={http://dx.doi.org/10.1063/1.3284402},
   DOI={10.1063/1.3284402},
   booktitle={AIP Conference Proceedings},
   publisher={AIP},
   author={Carlip, Steven and Kowalski-Glikman, Jerzy and Durka, R. and Szczachor, M.},
   year={2009},
   pages={72–80} }

@article{KriPer,
    author = "N. V. Krishnendu and A. Perri and S. Chakraborty and A. Pesci",
    title = "Probing the existence of a minimal length through compact binary inspiral",
    eprint = "2505.22877",
    archivePrefix = "arXiv",
    primaryClass = "gr-qc",    
    journal = "",
    volume = "",
    pages = "",
    year = "arXiv:2505.22877[gr-qc], 2025"
}

@article{tHooft:1993dmi,
    author = "'t Hooft, Gerard",
    title = "{Dimensional reduction in quantum gravity}",
    eprint = "gr-qc/9310026",
    archivePrefix = "arXiv",
    reportNumber = "THU-93-26",
    journal = "Conf. Proc. C",
    volume = "930308",
    pages = "284--296",
    year = "1993"
}

@article{Susskind:1994vu,
    author = "Susskind, Leonard",
    title = "{The World as a hologram}",
    eprint = "hep-th/9409089",
    archivePrefix = "arXiv",
    reportNumber = "SU-ITP-94-33",
    doi = "10.1063/1.531249",
    journal = "J. Math. Phys.",
    volume = "36",
    pages = "6377--6396",
    year = "1995"
}

@article{Foit:2016uxn,
    author = "Foit, Valentino F. and Kleban, Matthew",
    title = "{Testing Quantum Black Holes with Gravitational Waves}",
    eprint = "1611.07009",
    archivePrefix = "arXiv",
    primaryClass = "hep-th",
    doi = "10.1088/1361-6382/aafcba",
    journal = "Class. Quant. Grav.",
    volume = "36",
    number = "3",
    pages = "035006",
    year = "2019"
}

@article{Cardoso:2019,
    author = "Cardoso, V and Foit, V F and Kleban, M",
    title = "{Gravitational wave echoes from black hole area quantization}",
    eprint = "1902.10164",
    archivePrefix = "arXiv",
    primaryClass = "hep-th",
    reportNumber = "",
    doi = "",
    journal = "J. Cosmol. Astropart. Phys.",
    volume = "08",
    pages = "006",
    year = "2019"
}

@article{Agullo:2021,
    author = "Agullo, Ivan and Cardoso, Vitor and del Rio, Adrian and Maggiore, Michele and Pullin, Jorge",
    title = "{Potential gravitational wave signatures of quantum gravity}",
    eprint = "2007.13761",
    archivePrefix = "arXiv",
    primaryClass = "gr-qc",
    reportNumber = "",
    doi = "",
    journal = "Phys. Rev. Lett.",
    volume = "126",
    pages = "041302",
    year = "2021"
}

@article{Datta:2021row,
    author = "Datta, Sayak and Phukon, Khun Sang",
    title = "{Imprint of black hole area quantization and Hawking radiation on inspiraling binary}",
    eprint = "2105.11140",
    archivePrefix = "arXiv",
    primaryClass = "gr-qc",
    reportNumber = "LIGO-P2100189",
    doi = "10.1103/PhysRevD.104.124062",
    journal = "Phys. Rev. D",
    volume = "104",
    number = "12",
    pages = "124062",
    year = "2021"
}

@article{Sago:2021iku,
    author = "Sago, Norichika and Tanaka, Takahiro",
    title = "{Oscillations in the extreme mass-ratio inspiral gravitational wave phase correction as a probe of a reflective boundary of the central black hole}",
    eprint = "2106.07123",
    archivePrefix = "arXiv",
    primaryClass = "gr-qc",
    reportNumber = "KUNS-2879, YITP-21-62, OCU-PHYS-542, AP-GR-170",
    doi = "10.1103/PhysRevD.104.064009",
    journal = "Phys. Rev. D",
    volume = "104",
    number = "6",
    pages = "064009",
    year = "2021"
}
	\bibliographystyle{plain}

\end{document}